\documentclass[%
 reprint,
 amsmath,amssymb,
 aps,
]{revtex4-2}

\usepackage{graphicx}% Include figure files
\usepackage{dcolumn}% Align table columns on decimal point
\usepackage{bm}% bold math
\usepackage{xcolor}
\usepackage{amsmath,amssymb}
\usepackage[colorlinks,linkcolor=black,citecolor=blue,urlcolor=blue ]{hyperref}

\begin{document}

\preprint{APS/123-QED}

\title{Combined $13\times2$-point analysis of the Cosmic Microwave Background and Large-Scale Structure: implications for the $S_8$-tension and neutrino mass constraints}

\author{Raphael Sgier}
 \email{raphael.sgier@phys.ethz.ch}
\author{Christiane S. Lorenz}%
 \email{chrlorenz@phys.ethz.ch}
\author{Alexandre Refregier}
 \email{alexandre.refregier@phys.ethz.ch}
\author{Janis Fluri,}
 \email{janis.fluri@phys.ethz.ch}
\author{Dominik Z\"{u}rcher}
 \email{dominik.zuercher@phys.ethz.ch}
\author{Federica Tarsitano}
 \email{federica.tarsitano@phys.ethz.ch}
 
\affiliation{Institute for Particle Physics and Astrophysics, Department of Physics, ETH Z\"urich, Wolfgang Pauli Strasse 27, 8093 Z\"urich, Switzerland}%

\date{\today}

\begin{abstract}
We present cosmological constraints for the flat $\Lambda$CDM model, including the sum of neutrino masses, by performing a multi-probe analysis of a total of 13 tomographic auto- and cross-angular power spectra. This is achieved by combining, at map level, the latest primary CMB and CMB-lensing measurements from the \textit{Planck} 2018 data release, as well as spectroscopic galaxy samples from BOSS DR12, and the latest Kilo-Degree Survey (KiDS-1000) tomographic weak lensing shear data release. Our analysis includes auto- and cross-correlations as well as calibration parameters for all cosmological probes, thus providing a self-calibration of the combined data sets. We find a good fit (reduced $\chi^2$=1.7) for the combined probes with calibration parameters only moderately different from their nominal value, thus giving a possible interpretation of the tension between the early- and late-Universe probes. The resulting value for the structure growth parameter is $S_8 = 0.754 \pm 0.016$ (68\% CL). We also obtain a $\sim$2.3$\sigma$ constraint on the neutrino mass sum of $\sum m_\nu = 0.51^{+0.21}_{-0.24}$ eV (68\% CL), which is compatible with current particle physics limits. We perform several tests by fixing the neutrino mass sum to a low value, considering narrower priors on the multiplicative bias parameters for cosmic shear, and by fixing all calibration parameters to their expected values. These tests result in worse fits compared to our fiducial run, especially for the case when all calibration parameters are fixed. This latter test also yields a lower upper limit of the neutrino mass sum. We discuss how the interplay between the cosmological and calibration parameters impact the $S_8$-tension and the constraints on the neutrino mass sum. Our analysis highlights how combined map-level analyses, including cross-correlations, offer powerful cosmological constraints already with current data, while providing stringent tests of systematics.
\end{abstract}

\maketitle

\tableofcontents

\section{\label{sec:introduction}Introduction}
A wide range of cosmological observations conducted in the last few decades led to the establishment of the spatially flat $\Lambda$CDM model. In this model, our Universe is composed of cold dark matter (CDM), dark energy (DE) responsible for the current accelerated expansion, baryons, radiation and neutrinos (e.g.~\cite{Planck2018I}). The $\Lambda$CDM model can successfully and independently describe measurements of various cosmological probes, such as the temperature fluctuations in the cosmic microwave background (CMB, \cite{Planck2018I}), the weak gravitational lensing of the CMB \cite{Planck2018VIII} and of background galaxies~\cite{Asgari2021, Troxel2018, Hamana2020}, baryon acoustic oscillations (BAO) and redshift-space distortions (RSD,~\cite{Alam2017, Alam2020}), the present-day acceleration rate measured with a distance ladder calibrated using Cepheid variables \cite{Riess2019} and the accelerated expansion rate measured with the distance-redshift relation of Type Ia supernovae (SNe,~\cite{Scolnic2018}).

Despite the remarkable success of the $\Lambda$CDM model, reported tensions between measurements of the precise value of the Hubble constant $H_0$ and the structure growth parameter $S_8 = \sigma_8 \sqrt{\Omega_\mathrm{m} / 0.3}$ hint at possible extensions of the $\Lambda$CDM model. Most prominently, differences of the order of $\sim4-5\sigma$ in the measurement of $H_0$ are found between the values inferred from CMB measurements~\cite{Aghanim:2018eyx,ACT:2020gnv} and from direct local measurements, such as the inverse distance ladder~\cite{Riess2019, Wong2020,Wong2019}. Similarly, differences are reported between the measured value of $S_8$ obtained with CMB and weak lensing measurements~\cite{Heymans2013, Troxel2018,Hikage2019, Heymans2021,Amon:2021kas}, such as a $\sim 3 \sigma$ tension found in the recent cosmic shear study based on the KiDS-1000 cosmic shear data~\cite{Asgari2021}.

The remarkable precision of future photometric and spectroscopic galaxy surveys, such as the Legacy Survey of Space and Time (LSST~\footnote{\url{http://www.lsst.org}}) conducted at the Vera C. Rubin Observatory, Euclid~\footnote{\url{http://sci.esa.int/euclid/}} and the Wide Field Infrared Telescope (WFIRST~\footnote{\url{http://wfirst.gsfc.nasa.gov}}) will hopefully shed light on the prevailing tensions between the early- and late-time probes of the Universe and test for possible extensions beyond a flat $\Lambda$CDM model~\cite{Abell2009, Laureijs2011, Abazajian2016cmbs4}.

A promising avenue to obtain new information retrieved from cosmological surveys is to consider the combination of different probes. This is possible since many surveys cover overlapping redshift ranges and regions in the sky and therefore probe the same matter density field in these shared regions. Considering the statistical correlation between cosmological probes is expected to provide tighter constraints on the cosmological parameters, and also offers a more stringent control over systematic effects~\cite{Weinberg2013}. Numerous studies conducted in the last years successfully detected cross-correlation signals and performed combined analyses. For example, cross-correlating the CMB temperature anisotropies with Large-Scale Structure (LSS) tracers from galaxy surveys allows the measurement of the Integrated Sachs-Wolfe effect (ISW)~\cite{Crittenden1996, Boughn1998, Fosalba2003, Cabre2006, Moura_Santos_2016}. Other detected cross-correlations include the correlations between the CMB observables themselves, such as CMB temperature and polarisation anisotropies and CMB lensing~\cite{Planck2018I, Han_2021}, or correlations between CMB lensing and lensing of galaxies~\cite{Hand2015, Kirk2016} and the positions of galaxies (e.g.~\cite{Hirata2008, Ho2008, Baxter2016, Giannantonio2016, Deraps2016, Bianchini2016}).

Joint analyses based on the real space correlation functions or power spectra have demonstrated the power of cross-correlations to obtain new information and thus stronger parameter constraints, as well as to cross-calibrate various calibrating nuisance parameters simultaneously. For example, combined analyses of cosmic shear, galaxy-galaxy lensing and angular clustering have been performed in~\cite{Uitert2017} for galaxies from the Galaxies and Mass Assembly (GAMA~\footnote{\url{http://www.gama-survey.org}}) survey and the Kilo-Degree Survey (KiDS~\footnote{\url{http://kids.strw.leidenuniv.nl}}), or in~\cite{Joudaki2017} for galaxies from the Baryon Oscillation Spectroscopic Survey (BOSS~\footnote{\url{https://www.sdss.org/surveys/boss/}}) and KiDS. In a fully joint analysis,~\cite{Nicola2016, Nicola2017} combined CMB temperature anisotropies, CMB lensing, weak lensing shear, galaxy positions and distance measurements from supernovae and direct measurements of the Hubble parameter $H_0$.

Incorporating the cross-correlation signal between different probes thus represents a promising way to test possible extensions beyond a flat $\Lambda$CDM model, such as the total mass of neutrinos $\sum m_\nu$~\cite{Lesgourgues:2007ix}. For example, cross-correlations between \textit{Planck} 2015 CMB lensing and spectroscopic samples from the Baryon Oscillation Spectroscopic Survey (BOSS) have been considered in~\cite{Doux2018}, resulting in an upper limit of $\sum m_\nu<0.28$ eV (68\% CL), 
%for a fixed equation of state parameter. 
compared to $\sum m_\nu < 0.21$ eV (95\% CL) for \textit{Planck} 2015 TT + lowP + BAO~\cite{Planck:2015fie}. 
Recent measurements by \textit{Planck}~\cite{Aghanim:2018eyx}, combining CMB data with distance measurements from BAO from the Sloan Digital Sky Survey (SDSS~\footnote{\url{https://www.sdss.org}}) and the 6dF galaxy survey~\footnote{\url{http://www.6dfgs.net/}} (i.e. \textit{Planck} 2018 TTTEEE+lowE+lensing+BAO)
provide an upper limit on the total neutrino mass sum of $\sum m_\nu < 0.12$ eV (95\% confidence limit (CL)).  Interestingly, as shown by various studies, combining late-Universe data with CMB data results often in loosening of the constraints of $\sum m_\nu$, compared to the constraints from using CMB data alone~\cite{Battye:2013xqa,Wyman:2013lza,Beutler:2014yhv,Poulin:2018zxs,Muir:2020puy}. %For example, combining data from the first year of observations of the Dark Energy Survey (DES) with CMB (\textit{Planck} 2015), BAO and type Ia supernovae data from the Joint Lightcurve Analysis (JLA) results in a 20\% weakened constraint from $\sum m_\nu < 0.22$ eV to $\sum m_\nu < 0.26$ eV (95\% CL)~\cite{Abbott:2017wau}. 
For example, a $3 \times 2$pt analysis of the DES data release 3~\cite{DES:2021wwk}, combining cosmic shear, galaxy clustering and galaxy-galaxy lensing, and additionally combined with \textit{Planck} CMB data, reported an upper limit on the sum of the neutrino masses of $\sum m_\nu<0.13$ eV (95\% CL). Similarly, a $3 \times 2$pt measurement from KiDS, BOSS and the spectroscopic 2-degree Field Lensing Survey (2dFLenS) reported an upper limit on the total mass of neutrinos of $\sum m_\nu < 1.76$ eV (95\% CL)~\cite{Troester2021}. 

Furthermore, the power of adding cross-correlations has been shown in various forecast analyses~\cite{Chen:2021vba,EUCLID:2020jwq,Euclid:2021qvm,Bermejo-Climent:2021jxf}. As shown in these studies, adding cross-correlations can significantly improve parameter constraints for extended cosmological models, for example for models including neutrino masses, time-varying dark energy and primordial non-Gaussianity. Morever, the inclusion of cross-correlations can help to constrain nuiscance parameters, as recently shown specifically with forecasts for Euclid~\cite{EUCLID:2020jwq,Euclid:2021qvm}. 

In this work, we perform a joint analysis, using a simulated multi-probe covariance matrix, by combining CMB temperature and polarisation anisotropies and CMB lensing from the \textit{Planck} 2018 data release with the spectroscopic galaxy samples from the 12$^\mathrm{th}$ BOSS and the latest KiDS-1000 weak lensing shear 'gold sample' data releases. 
We prepare and combine all data sets on a map level, which allows us to compute the auto- and cross-correlations between the different probes, as they share large overlapping regions in the sky. This map-based approach closely follows~\cite{Nicola2016, Nicola2017}, which has recently been used to perform a forecast analysis for a Stage-IV-like survey~\cite{Sgier2021}. We thus perform a tomographic $13\times2$-point combined analysis using a total of 10 auto- and 36 cross-spherical harmonic power spectra. We vary 6 cosmological and 9 calibration parameters simultaneously in the context of a flat $\Lambda$CDM model, including massive neutrinos for a degenerate mass neutrino hierarchy. Our statistical analysis is performed using a simulated multi-probe covariance matrix estimated using the lightcone generation code \textsc{UFalcon}~\cite{Sgier2019, Sgier2021}. \textsc{UFalcon} self-consistently generates full-sky maps (mocks) from the same simulated density field, which allows to consider cross-correlations between different probes. We use \textsc{PKDGrav3}~\cite{Stadel2001} as the underlying $N$-Body simulation code.

In Section \ref{sec:theory}, we describe the theoretical framework for our analysis, which uses the pseudo-$C_\ell$ method~\cite{Brown2005, Kogut2003} to take into account the finite sky coverage of the different cosmological probes. The different data sets from the \textit{Planck} 2018, BOSS DR12 and KiDS-1000 data releases and their preparation is detailed in Section \ref{sec:data}. The covariance matrix estimation process is presented in Section \ref{sec:cov}. In Section \ref{sec:inference} we describe the details of our statistical inference analysis and in Section~\ref{sec:constraints} we show the resulting parameter constraints, and discuss several consistency tests that we performed to cross-check our analysis. Lastly, we discuss our findings in Section \ref{sec:discussion}.

\section{\label{sec:theory}Theory}
\subsection{Analytical Predictions}
Various cosmological probes can be used to trace the large-scale matter distribution of the Universe. The projected field associated to a probe $X$ in the observed direction $\hat{n}$ can be related to the matter overdensity $\delta \equiv \delta \rho / \rho$ as
\begin{equation}
    X\left(\hat{n}\right) = \int \mathrm{d}z \frac{c}{H(z)} W^{X}(\chi(z)) \delta \left(\chi(z)\hat{n}, z \right)\, ,
\end{equation}
where $W^{X}(\chi(z))$ is the window function of the probe $X$, $H(z)$ is the Hubble parameter, $\chi(z)$ the comoving distance and $c$ the speed of light. The spherical harmonic cross-power spectrum as a function of multipole $\ell$ between two cosmological probes $X$ and $Y$ can be written as
\begin{eqnarray}
C_{\ell}^{XY} &= \int \mathrm{d}z \, \frac{c}{H(z)} \, \frac{W^{X} (\chi(z)) \, W^{Y} (\chi(z))}{\chi^{2} (z)} \\ &\times \, P^{\mathrm{nl}}_{\delta \delta} \left(k = \frac{\ell + 1/2}{\chi(z)}, z \right)\, , \nonumber
\label{cl_limber}
\end{eqnarray}
where we use the Limber approximation~\cite{Limber1953, Kaiser1992, Kaiser1998,LoVerde:2008re} in order to speed up the computation, which is valid for small angular scales ($\ell \gtrsim 30$) and broad redshift bins~\cite{Nicola2016, Peacock1999,Simon:2006gm,LoVerde:2008re}. The nonlinear power spectrum $P^{\mathrm{nl}}_{\delta \delta} \left(k, z \right)$ is computed using the halo-model fitting function presented in \cite{Mead2015, Mead2016}, which can predict power spectra including massive neutrinos accurately within a few percent up to $k \sim 10 h \mathrm{Mpc}^{-1}$.

In the present work we use a hybrid approach for the calculation of the analytical spherical harmonic power spectra: The power spectrum auto- and cross-correlations $C_{\ell}^{XY}$ for the probe-combinations $XY \in \{TT, TE, EE, \kappa_\mathrm{CMB}\kappa_\mathrm{CMB} \}$ are computed using the Cosmic Linear Anisotropy Solving System (\textsc{Class};~\cite{Lesgourgues2011, Blas2011}). The other combinations $XY \in \{\delta_g \delta_g, \gamma \gamma, \delta_g \gamma, \delta_g \kappa_\mathrm{CMB}, T \delta_g, T \gamma, T\kappa_\mathrm{CMB}\}$ are computed using \textsc{PyCosmo}~\cite{pycosmo2018, Tarsitano2020}, which is a Python-based framework to compute cosmological quantities and observables. The linear matter power spectrum used in \textsc{PyCosmo} is calculated using the Eisenstein \& Hu fitting formulas~\cite{Eisenstein1999}, describing the matter transfer functions, including massive neutrinos for a degenerate mass hierarchy. This transfer function fitting formula has been found to be accurate within $5\%$ for the transfer function ($10\%$ for the power spectrum) for moderate deviations from a pure $\Lambda$CDM model. It is not capable of fitting the acoustic oscillations created by a large baryon fraction, but accurately matches the underlying smooth function. The formula looses accuracy for baryon or massive neutrino fractions exceeding $30\%$ of the total matter density $\Omega_\mathrm{m}$ outside the range $0.06 \lesssim \Omega_\mathrm{m} h^2 \lesssim 0.4$. Within this range and for baryon fractions within $10\%$, the formula is accurate to better than $3\%$~\cite{Eisenstein1999}. We therefore do not expect to significantly bias our parameter constraints using this fitting formula for our inference analysis.

\begin{figure}
\centering
  \includegraphics[width=\linewidth]{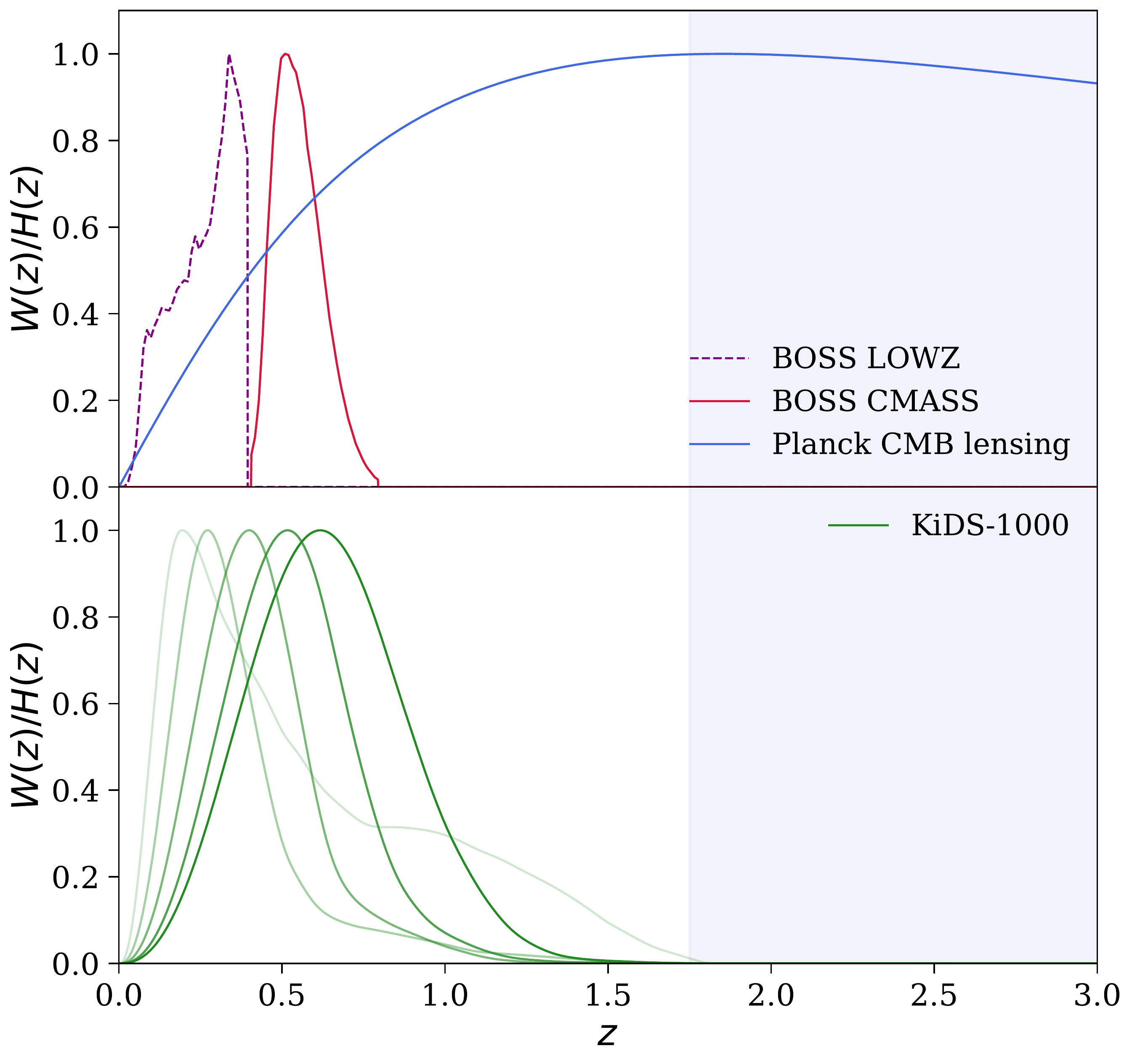}
  \caption{Normalised window functions multiplied by $1/H(z)$ for the galaxy overdensity, CMB lensing and weak lensing shear: $W^g$ using Eq. (\ref{wdelta}) for the LOWZ (pink, dashed line) and CMASS (red, solid line) $n(z)$, $W^{\kappa_\mathrm{CMB}}$ using Eq. (\ref{wkappa}) for the CMB lensing convergence (blue, solid line) and $W^\gamma$ using Eq. (\ref{wshear}) for weak lensing shear using the 5 $n(z)$'s of the tomographic bins from the KiDS-1000 cosmic shear gold sample (green). The blue shaded area starts at $z=1.75$ and indicates the range not covered by the \textsc{UFalcon} lightcone (see Sec. \ref{sec:cov_mat}). \label{wz_plot}}
\end{figure}

The window functions corresponding to the cosmological probes in this analysis are briefly described below and shown in Figure \ref{wz_plot}, using the fiducial cosmological parameters described in Section \ref{sec:nbody}. 
\\
\\
\textbf{Weak lensing shear.} The window function for weak lensing shear can be written as
\begin{equation}
W^{\gamma} (\chi(z)) = \frac{3}{2} \frac{\Omega_\mathrm{m} H_0^2}{c^2} \frac{\chi(z)}{a} \int^{\chi_\mathrm{h}}_{\chi(z)} \mathrm{d}z' n(z') \frac{\chi(z') - \chi(z)}{\chi(z')} \, ,
\label{wshear}
\end{equation}
where $\Omega_\mathrm{m}$ and $H_0$ denote the present day values of the matter density and the Hubble parameter, respectively. The comoving distance to the horizon is denoted by $\chi_\mathrm{h}$, and $n(z)$ is the normalised redshift selection function of the lensed galaxies.

The intrinsic ellipticities of galaxies are correlated with each other and with the large-scale structure. This effect, referred to as galaxy intrinsic alignment (IA), is one of the most important systematic effects in weak gravitational lensing~\cite{Heymans2006IA, Mandelbaum2006, Hirata2004, Hirata2007, Faltenbacher_2009}. The IA signal can be described by two components: The intrinsic-intrinsic (II) term, originating from the correlation between the ellipticities of the galaxies and the large-scale structure (i.e. the intrinsic ellipticities are aligned around massive objects), and the gravitational-intrinsic (GI) term, originating from the correlation between the intrinsic ellipticities of the foreground galaxies and the shear of background galaxies~\cite{Heavens2000}. We include the IA effect in the analytical code \textsc{PyCosmo} through the nonlinear alignment model (NLA;~\cite{Bridle2007, Hirata2004, Joachimi2011}). The NLA model has previously been used for the fiducial cosmic shear results of the KiDS-450~\cite{Hildebrandt2016} and the KiDS-1000~\cite{Asgari2021} analyses. It relates the nonlinear matter power spectrum to the two intrinsic alignment power spectra through
\begin{eqnarray}
    P_\mathrm{II}(k, z) = F^{2}(z) P(k, z) \\
    P_\mathrm{GI}(k, z) = F(z) P(k, z)\, ,
\end{eqnarray}
where $k$ represents the wavenumber. The cosmology and redshift dependent normalisation $F(z)$ is given by
\begin{equation}
    F(z) = - A_\mathrm{IA} C_1 \rho_\mathrm{crit} \frac{\Omega_m}{D_{+}(z)} \left( \frac{1+z}{1+z_0} \right)^{\eta} \left(\frac{\bar{L}}{L_0} \right)^{\beta}\, ,
\end{equation}
where $\rho_\mathrm{crit}$ is the critical density at $z=0$, $D_{+}(z)$ is the normalised linear growth factor satisfying $D_{+}(0)=1$ and $C_1 = 5 \times 10^{-14} h^{-2}M_{\odot} \mathrm{Mpc}^3$ is a normalisation constant. The free parameters of the NLA model are given by $\eta$ and $\beta$ and the intrinsic alignment amplitude is denoted by $A_\mathrm{IA}$. As for the fiducial results of the KiDS-450~\cite{Hildebrandt2016} and the KiDS-1000~\cite{Asgari2021} analyses, we choose the amplitude of the NLA model to be independent of redshift and luminosity, i.e. $\eta = \beta = 0$.\\
\\
\textbf{Galaxy overdensity.} The window function for the galaxy overdensity is given by
\begin{equation}
W^{\delta_g} (\chi(z)) = \frac{H(z)}{c} b(z) n(z) + W^{\delta_g}_\mathrm{lensing} (\chi(z))\, ,
\label{wdelta}
\end{equation}
where the second term $W^{\delta_g}_\mathrm{lensing} (\chi(z))$ is associated to the effects of gravitational lensing, which can be neglected for the two BOSS samples LOWZ and CMASS used in this work~\cite{Chisari2013}. We further use a constant, linear, scale- and redshift-independent galaxy bias $b(z)$, which is a valid assumption on large scales.

The local peculiar motion of galaxies along the line-of-sight alter the measured redshift and therefore distort their true comoving position, leading to redshift-space distortions (RSD) in the clustering of galaxies~\cite{Torre2012, Percival2011}. The effect of RSD leads to an increase in power on scales $\ell \lesssim 60$ for the galaxy clustering spherical harmonic power spectrum. Therefore, the galaxy overdensity window function implemented in \textsc{PyCosmo} needs to be modified to~\cite{Padmanabhan2008}
\begin{equation}
    W^{\delta_g}_{\mathrm{tot}} (\chi(z)) = W^{\delta_g} (\chi(z)) + W^{\delta_g}_\mathrm{RSD} (\chi(z))\, ,
\end{equation}
where the second term is given by
\begin{equation}
\begin{split}
W^{\delta_g}_\mathrm{RSD} &=  \frac{H(z)}{c} b(z) n(z) \left[ \beta \frac{(2 \ell^2 + 2 \ell - 1)}{(2 \ell + 3)(2 \ell - 1)} \right. \\
 & \left. - \beta \frac{\ell (\ell - 1)}{(2 \ell - 1)(2 \ell + 1)}  - \beta \frac{(\ell + 1)(\ell + 2)}{(2 \ell + 1)(2 \ell + 3)} \right] \, ,
\end{split}
\end{equation}
with the approximated redshift distortion parameter $\beta \approx \Omega_m^{0.6}/b(z)$.\\
\\
\textbf{CMB temperature and polarisation anisotropies.} The CMB temperature anisotropies $\Delta T_\ell (k)$ are related to the primordial power spectrum generated during inflation, $\mathcal{P}_{\delta \delta} (k)$, and to the evolution of matter perturbations through their spherical harmonic power spectrum. The resulting angular temperature power spectrum can be expressed as \cite{Dodelson2003}
\begin{equation}
C_{\ell}^{TT} = \frac{2}{\pi} \int \mathrm{d} k \,  k^2 \mathcal{P}_{\delta \delta} (k) \left| \frac{\Delta T_\ell (k)}{\delta (k)} \right| \, .
\end{equation}
A similar relation also applies to the amplitude of polarisation anisotropies $\Delta P_\ell (k)$, given by \cite{Dodelson2003}
\begin{equation}
C_{\ell}^{PP} = \frac{2}{\pi} \int \mathrm{d} k \,  k^2 \left|\Delta P_\ell (k) \right| \, .
\end{equation}

Secondary temperature anisotropies of the CMB can be generated through the ISW effect, leading to correlations between the CMB temperature anisotropies and tracers of LSS, such as weak lensing shear and galaxy overdensity \cite{SachsWolfe1967}. The corresponding spherical harmonic cross-power spectrum can be written as \cite{Crittenden1996}
\begin{eqnarray}
C_{\ell}^{XT} &=& T_\mathrm{CMB} \left( \frac{3 \Omega_m H_0^2}{c^2} \right) \frac{1}{(\ell + 1/2)^2} \\
 & \times & \int \mathrm{d}z \frac{\mathrm{d}}{\mathrm{d} z} \left[ D(z) (1 + z) \right] D(z) W^{X} (\chi (z)) \nonumber \\ & \times & P^{\mathrm{lin}}_{\delta \delta} \left(k = \frac{\ell + \frac{1}{2}}{\chi(z)}, 0 \right) \, , \nonumber
\label{iswcross}
\end{eqnarray}
where $T_\mathrm{CMB}$ denotes the mean CMB temperature at present time and $X \in \{ \delta_g, \gamma, \kappa_\mathrm{CMB}\}$. In the above equation, the linear matter power spectrum is separated into its redshift-dependent linear growth factor $D(z)$ and its scale-dependent part as $P^{\mathrm{lin}}_{\delta \delta} \left(k, z \right) = D(z) P^{\mathrm{lin}}_{\delta \delta} \left(k, 0 \right)$.\\
\\
\textbf{CMB lensing convergence.} The weak gravitational lensing of the CMB is caused by the deflection of CMB photons through potential gradients along the way from the surface of last scattering to us. Therefore, the observed, lensed CMB temperature field in the direction $\hat{n}$ corresponds to the unlensed field deflected by the deflection angle $\vec{\alpha}(\hat{n})$ as \cite{Lewis2006}
\begin{equation}
    \tilde{T}(\hat{n}) = T(\hat{n} + \vec{\alpha})\, .
\end{equation}
The deflection angle is given by
\begin{equation}
\vec{\alpha} (\hat{n}) = -2 \int_0^{z_\ast} \mathrm{d}z \frac{\chi(z_\ast) - \chi(z)}{\chi(z_\ast) \chi(z)} \nabla_{\hat{n}} \frac{\Phi (\chi \hat{n}, z)}{c^2}\, , 
\end{equation}
where recombination is approximated to be an instantaneous process at redshift $z_\ast \approx 1100$ and $\nabla_{\hat{n}}$ representes the divergence operator on the sky (see e.g. \cite{Lewis2006, Hu2000, BartelmannSchneider2001, Refregier2003}). The CMB lensing convergence is then defined as $\kappa_\mathrm{CMB} = -\frac{1}{2}\nabla_{\hat{n}} \vec{\alpha}$. Together with the relation $\nabla^2_{\hat{n}} \simeq \nabla^2$~\cite{Jain2000}, valid on small angular scales, we can express the CMB lensing convergence window function as the single-plane limit of the weak lensing shear window function
\begin{equation}
W^{\kappa_\mathrm{CMB}} (\chi(z)) = \frac{3}{2} \frac{\Omega_\mathrm{m} H_0^2}{c^2} \frac{\chi(z)}{a} \frac{\chi(z_\ast) - \chi(z)}{\chi(z_\ast)} \, .
\label{wkappa}
\end{equation}

\subsection{\label{sec:pseudo_theory}Pseudo-Spherical Harmonic Power Spectrum}
\begin{figure}
\centering
  \includegraphics[width=\linewidth]{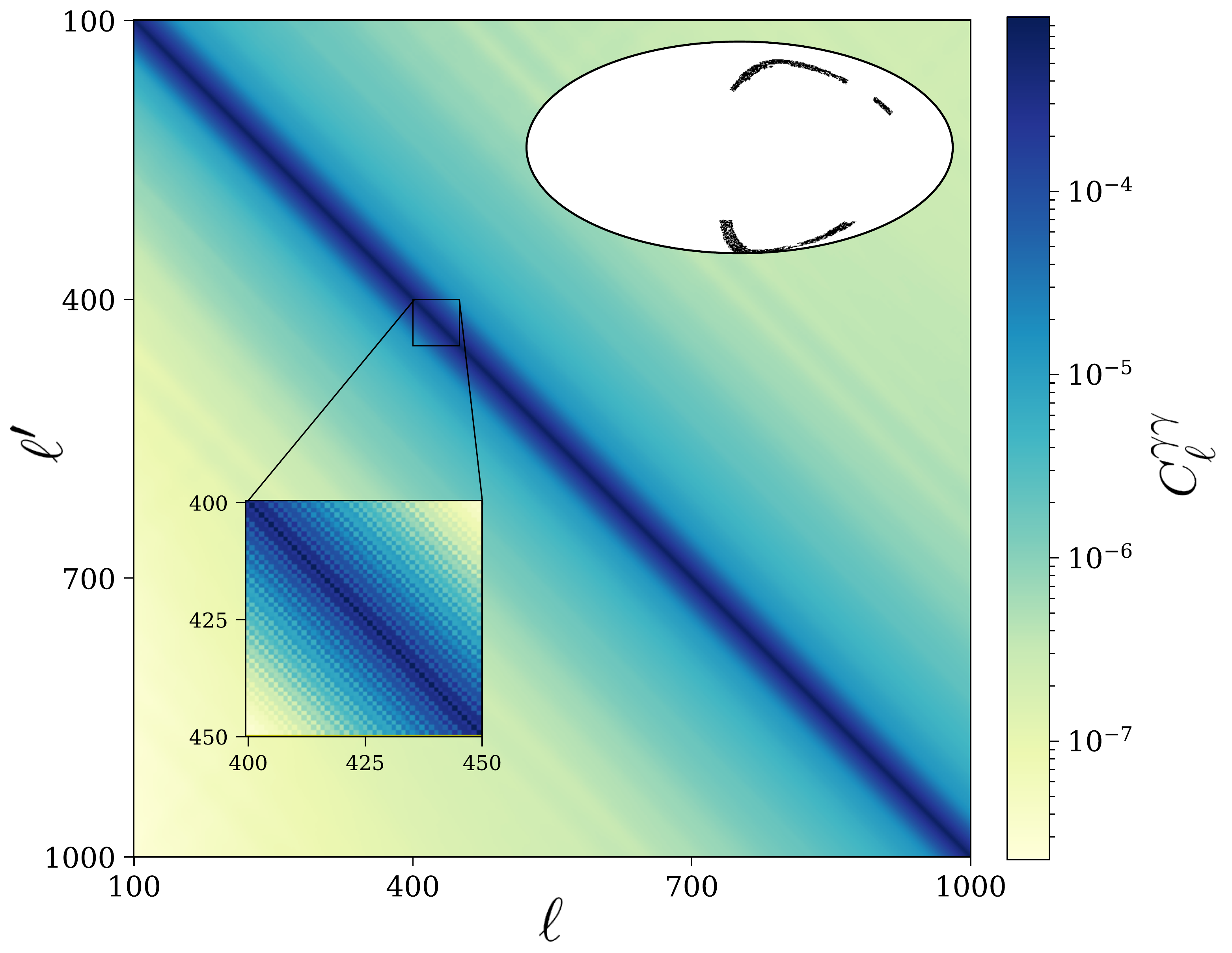}
  \caption{Mode-coupling matrix $M_{\ell \ell'}^{EE, EE}$ for the range $\ell=100$ to $1000$ in order to estimate the weak lensing shear pseudo-auto power spectra $C_\ell^{\gamma\gamma}$ using the KiDS-1000 mask (its mollweide projection is shown in the upper-right corner), transformed to galactic coordinates. The zoom-in region highlights the off-diagonal elements of the coupling matrix and therefore the coupling between different multipoles. \label{coupling_mat_kids}}
\end{figure}

We take into account the finite sky coverage of the considered surveys by using the pseudo-$C_\ell$ method originally based on~\cite{Brown2005, Kogut2003} and laid out for our purposes in Appendix D of~\cite{Sgier2021}. The effect of the mask is taken into account by the survey-specific window function $\mathcal{W}_{X}(\hat{n})$, which yields the value 1 if $\hat{n}$ lies in the observed region and 0 otherwise. The cut-sky field of any spin-0 or spin-2 type observable is then given by $\tilde{X}(\hat{n}) = \mathcal{W}_{X}(\hat{n}) X(\hat{n})$. The associated pseudo-spherical harmonic power spectrum between the fields $X$ and $Y$ are then related to their full-sky spectrum by
\begin{equation}
\tilde{C}_{\ell}^{XY} = \sum_{\ell'} M_{\ell \ell'} C_{\ell}^{XY}\, .
\label{cl_pseudo}
\end{equation}
The mode-coupling matrix $M$ takes into account the effect of the mask applied to the data, thereby the coupling between different multipoles. This becomes especially important for complicated sky cuts, where a simple rescaling of the power spectrum by the covered fraction of the sky $f_\mathrm{sky}$ is insufficient. The coupling matrices are given in terms of the Wigner-$3j$ symbols, which are given by Eq. (D.4) in \cite{Sgier2021}. In Figure \ref{coupling_mat_kids} we show the coupling matrix $M_{\ell \ell'}^{\gamma\gamma}$ used to estimate the weak lensing shear pseudo-auto power spectra using the KiDS-1000 footprint transformed to galactic coordinates within the $\ell$-range used in our analysis.

\section{\label{sec:data}Data}
In this section, we summarise relevant information about the data for the different cosmological probes obtained from the Planck 2018~\cite{Aghanim:2019ame,Aghanim:2018oex}, BOSS DR12~\cite{Reid2015, Alam2015} and KiDS-1000~\cite{Kuijken2019, Wright2020, Hildebrandt2020, Giblin2021} data releases. We extract the cosmological data from CMB temperature anisotropies, CMB polarisation, CMB lensing convergence, the galaxy overdensity field and weak lensing shear. Further, we describe the process of data selection and removal of systematics for each probe.

In this work, we perform a combined analysis of different cosmological probes at the map level. To this end, we create 2-dimensional spherical, pixelised maps with a \textsc{Healpix}~\footnote{\url{https://healpi.jpl.nasa.gov/}}~\cite{Gorski2005} resolution of $N_\mathrm{side} = 1024$ for all the probes considered.

The CMB probes from \textit{Planck} are given in galactic coordinates. We therefore transform the maps corresponding to the LSS probes, such as the galaxy overdensity and weak lensing shear, from equatorial (\texttt{RA}, \texttt{DEC}) to galactic (\texttt{l}, \texttt{b}) coordinates, using the \texttt{Rotator} class from \textsc{Healpix} to allow for a combination with the CMB maps. 

We choose to use the combined mask of two different probes when computing their spherical harmonic cross-power spectrum. Thereby we apply the combined mask, which constitutes of all pixels masked in at least one of the two maps, to both maps. The covered fractions of the sky of the combined mask for all considered cross-correlations, $f_\mathrm{sky}$, are listed in Table~\ref{cl_table}.

\begin{table}
\caption{Setup for the auto- and cross-power spectra considered in this work: covered sky fraction of the common mask, $f_\mathrm{sky}$, multipole range, and number of logarithmic bins within this range.}
\begin{tabular}{ llccc }
 \hline
 \\
 \multicolumn{2}{ l }{Power Spectrum} \quad & \quad $f_\mathrm{sky}$ \quad & $\ell$-range \quad & $N_\mathrm{log-bins}$\\
 \\
 \hline
 \hline
 \\
 \multicolumn{5}{ l }{Auto-Correlations}\\
 \\
 \hline\\
T &T & 0.746 & $\left[100, 1000\right]$ & 18\\[0.1cm]
E&E & 0.752 & $\left[100, 1000\right]$ & 18\\[0.1cm]
$\kappa_\mathrm{CMB}$ & $\kappa_\mathrm{CMB}$ &0.671 & $\left[50, 400\right]$ & 7\\[0.1cm]
$\delta_\mathrm{LOWZ}$ & $\delta_\mathrm{LOWZ}$ & 0.122 & $\left[50, 200\right]$ & 6\\[0.1cm]
$\delta_\mathrm{CMASS}$ & $\delta_\mathrm{CMASS}$ & 0.127 & $\left[50, 200\right]$ & 6\\[0.1cm]
$\gamma$ & $\gamma$ & 0.022 & $\left[100, 1000\right]$ & 5\\[0.1cm]
 \hline
 \\
 \multicolumn{5}{ l }{Cross-Correlations}\\
 \\
 \hline\\
T & E & 0.711 & $\left[100, 1000\right]$ & 18\\[0.1cm]
T & $\kappa_\mathrm{CMB}$ & 0.629 & $\left[50, 400\right]$ & 7\\[0.1cm]
T & $\delta_\mathrm{LOWZ}$ & 0.103 & $\left[50, 200\right]$ & 6\\[0.1cm]
T & $\delta_\mathrm{CMASS}$ & 0.109 & $\left[50, 200\right]$ & 6\\[0.1cm]
$\kappa_\mathrm{CMB}$ & $\delta_\mathrm{LOWZ}$ & 0.116 & $\left[50, 200\right]$ & 6\\[0.1cm]
$\kappa_\mathrm{CMB}$ & $\delta_\mathrm{CMASS}$ & 0.121 & $\left[50, 200\right]$ & 6\\[0.1cm]
$\gamma_{\mathrm{bin}\,i}$ & $\gamma_{\mathrm{bin}\,j}$ & 0.022 & $\left[100, 1000\right]$ & 5\\[0.1cm]
T & $\gamma$ & 0.021 & $\left[100, 1000\right]$ & 5\\[0.1cm]
$\kappa_\mathrm{CMB}$ & $\gamma$ & 0.021 & $\left[50, 400\right]$ & 7\\[0.1cm]
$\delta_\mathrm{LOWZ}$ & $\gamma$ & 0.001 & $\left[50, 200\right]$ & 6\\[0.1cm]
$\delta_\mathrm{CMASS}$ & $\gamma$ & 0.002 & $\left[50, 200\right]$ & 6\\[0.1cm]
 \hline
\end{tabular}
\label{cl_table}
\end{table}
\begin{figure}
\centering
  \includegraphics[width=\linewidth]{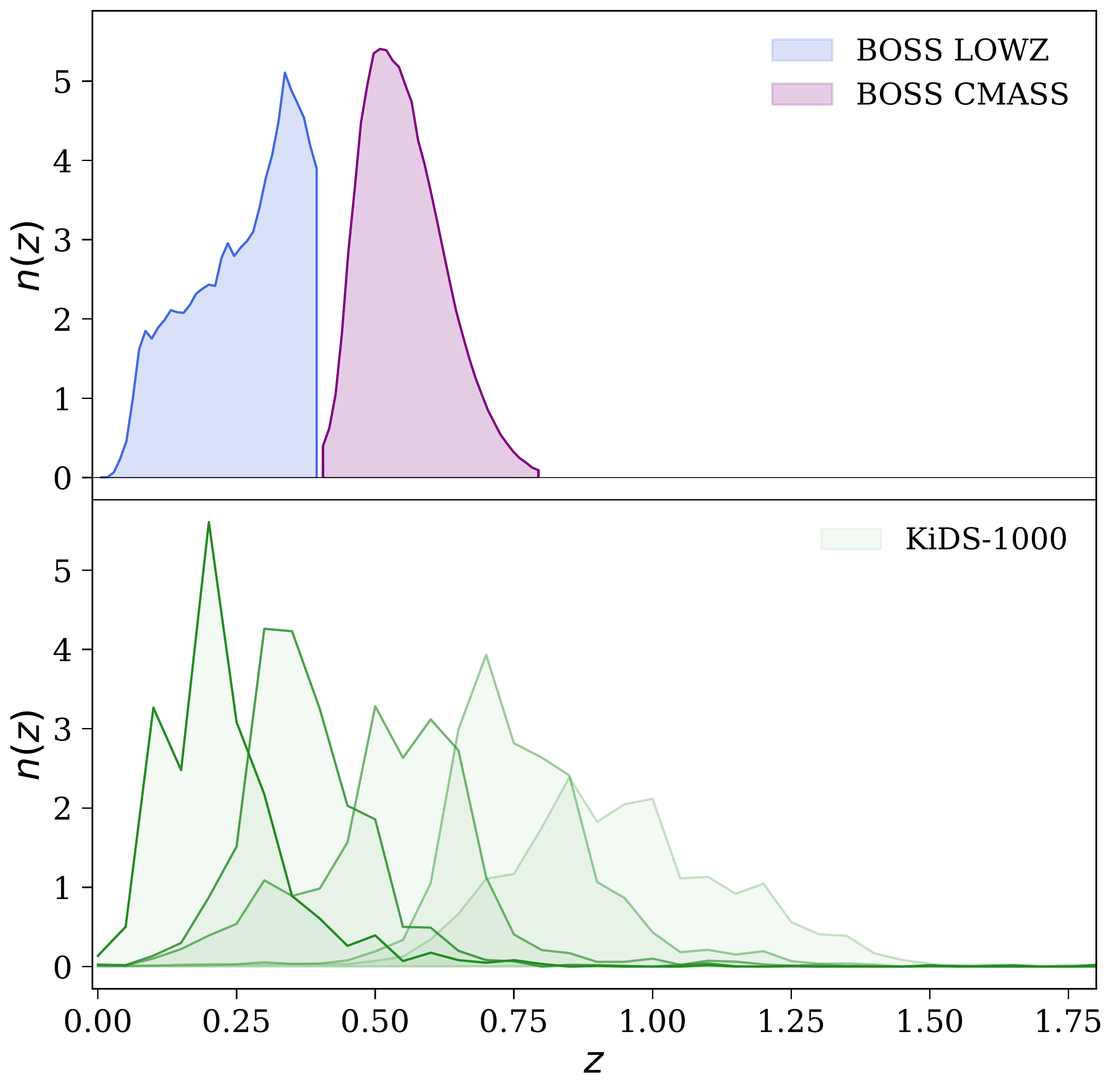}
  \caption{The galaxy redshift distributions of the used BOSS DR12 and KiDS-1000 data sets. The blue and purple histograms correspond to the LOWZ and CMASS samples, respectively. The green histograms correspond to the 5 KiDS-1000 tomographic samples.\label{nz_plot}}
\end{figure}

\begin{figure*}
\centering
  \includegraphics[width=9cm]{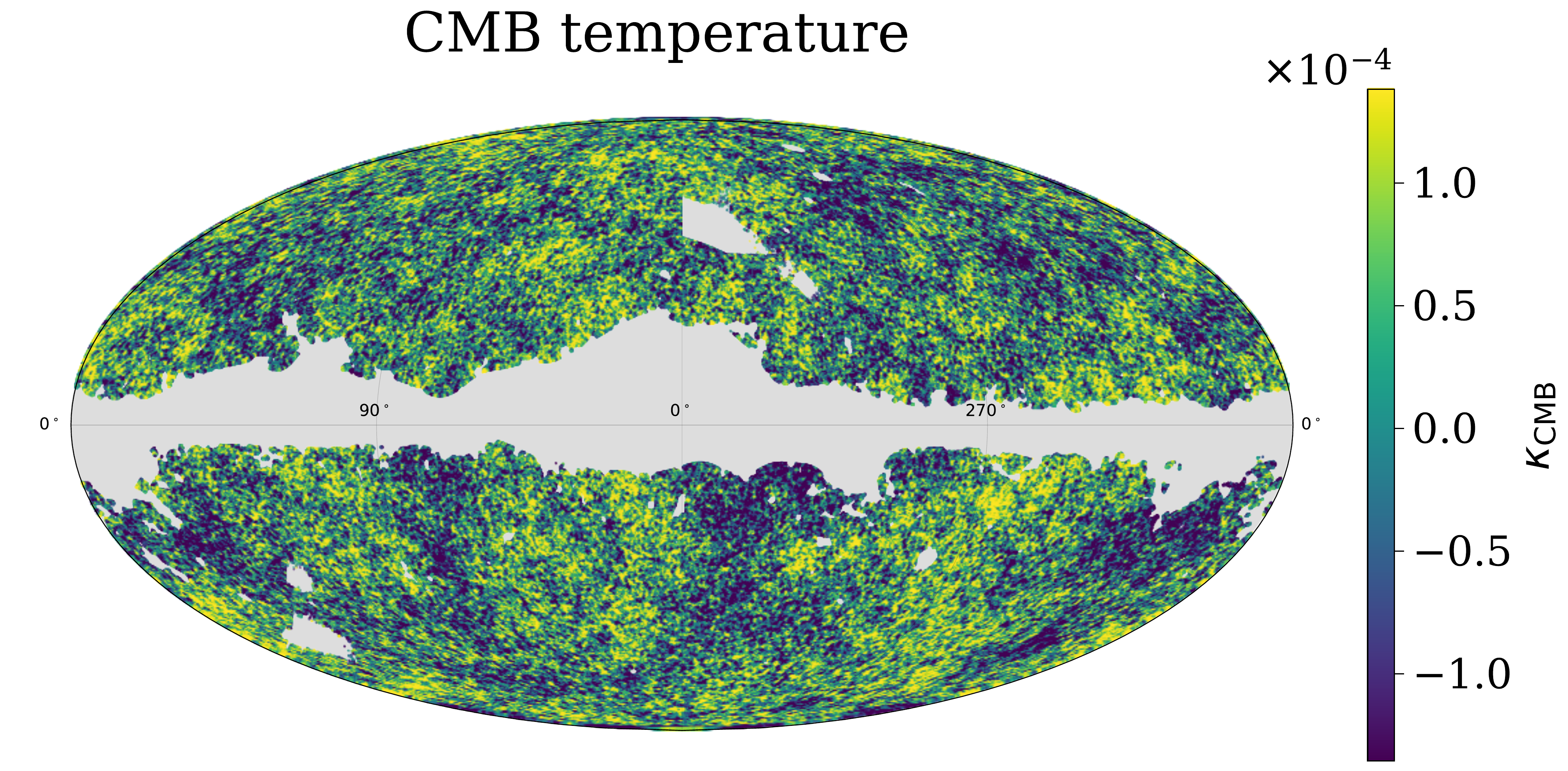}%
  \includegraphics[width=9cm]{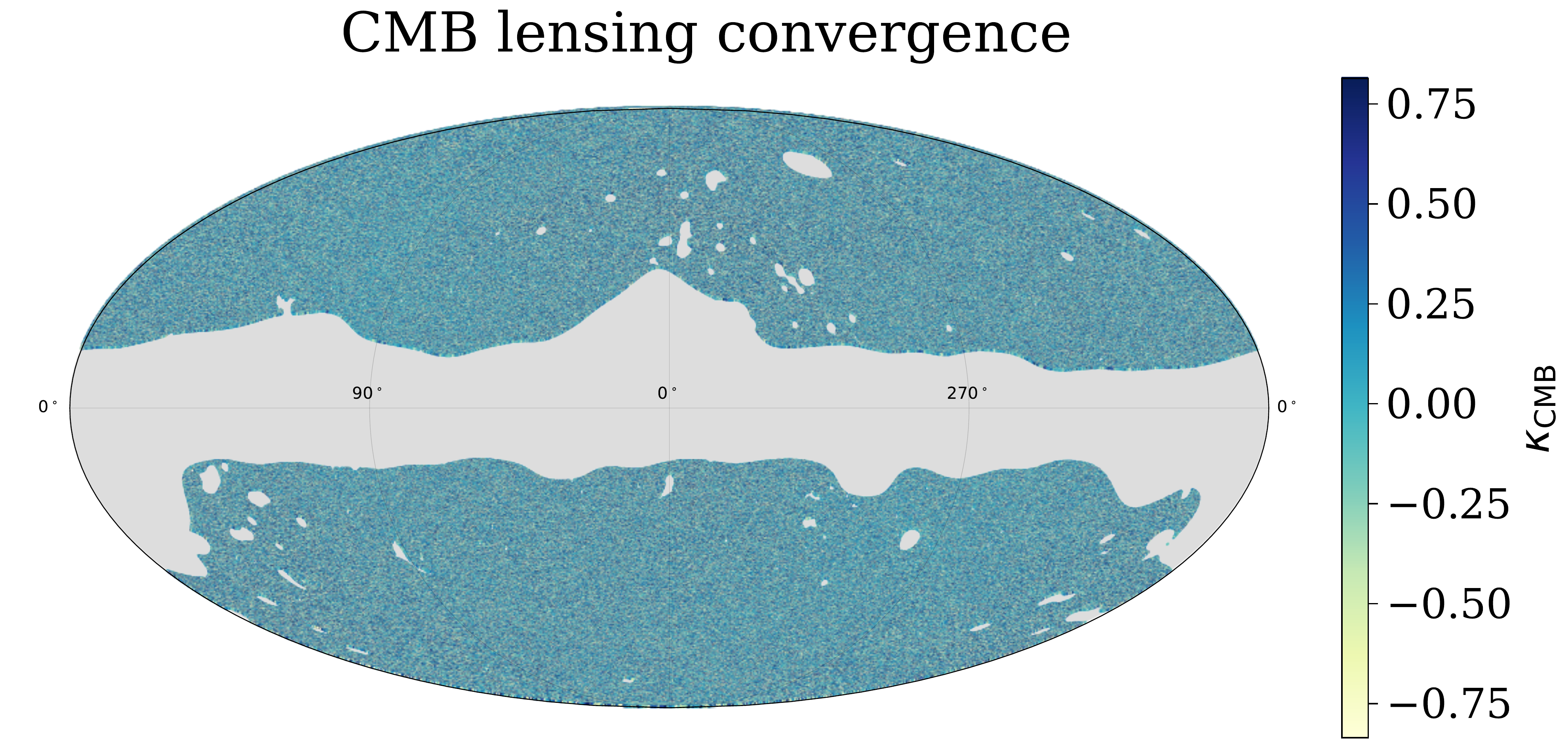}%
  \newline
  \includegraphics[width=9cm]{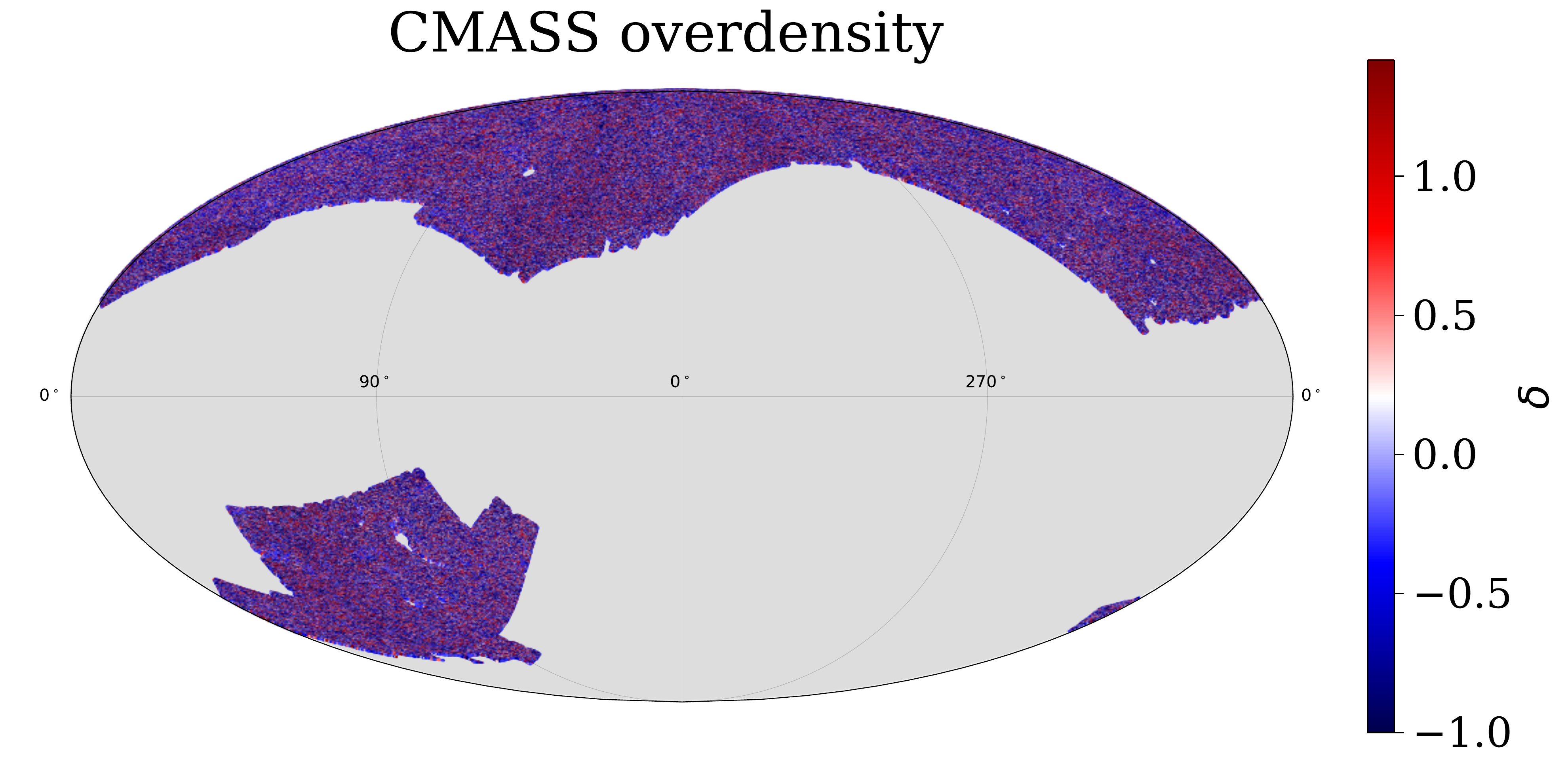}%
  \includegraphics[width=9cm]{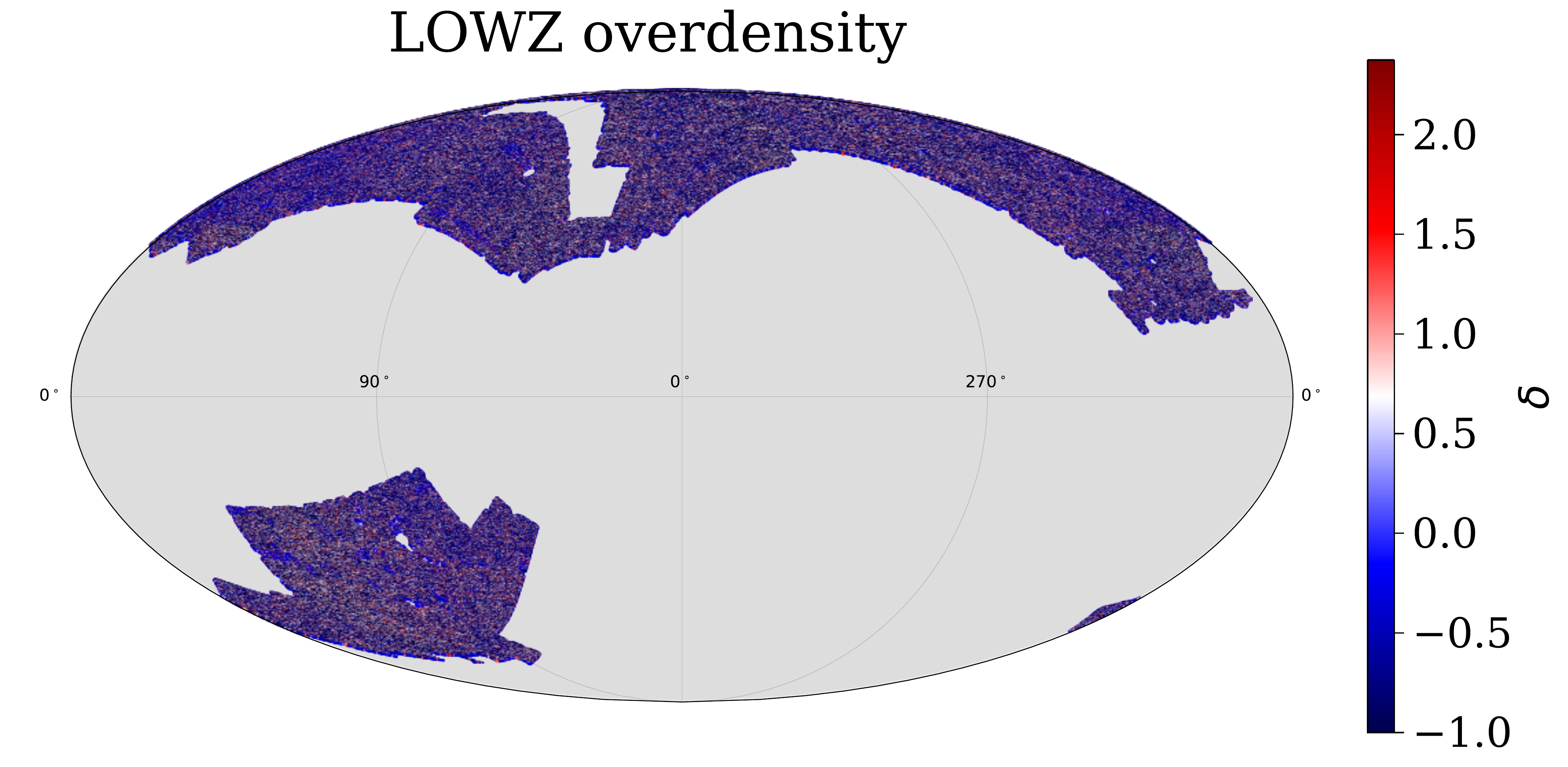}%
  \newline
  \includegraphics[width=9cm]{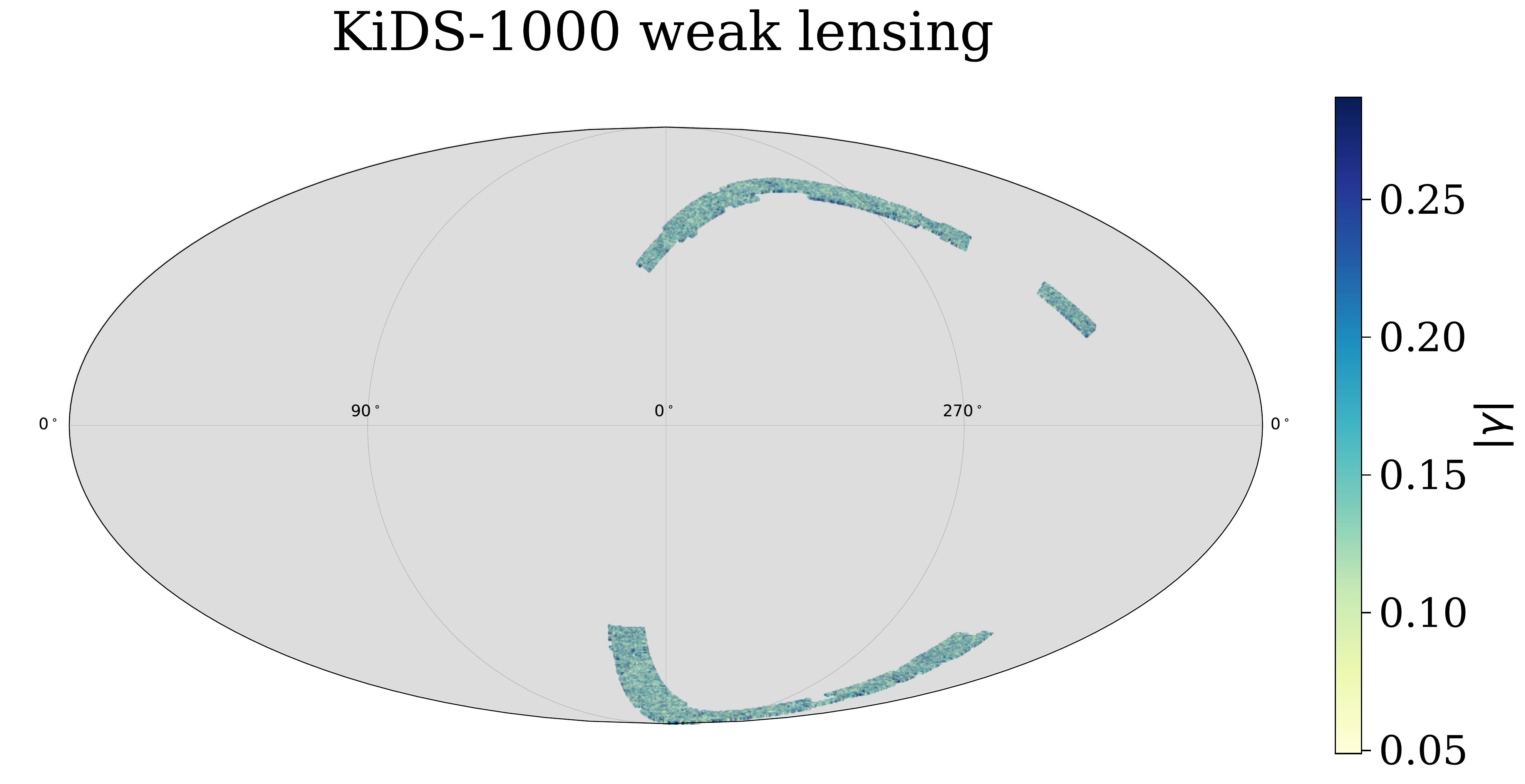}%
  \caption{Maps with resolution $N_\mathrm{side} = 1024$ of the different cosmological probes considered in this work: CMB temperature anisotropies map in units of $\mu$K (top left; the CMB polarisation map looks very similar); CMB lensing convergence map (top right); CMASS galaxy overdensity from BOSS (middle left); LOWZ galaxy overdensity from BOSS (middle right); KiDS-1000 weak lensing shear modulus $|\gamma|$ (bottom). Note that the KiDS-1000 and BOSS maps have been transformed to galactic coordinates. \label{all_maps}}
\end{figure*}

\subsection{\label{sec:data_planck}Planck 2018 Data}
In this work we use the CMB data provided by the \textit{Planck} collaboration \cite{Planck2018I} in their 2018 full-mission data release ("PR3"), which is available on the Planck Legacy Archive (PLA~\footnote{\href{http://pla.esac.esa.int/pla/}{http://pla.esac.esa.int/pla/}}).

\subsubsection{\label{sec:CMB}CMB Temperature and Polarisation}
As in earlier studies, the \textit{Planck} foreground-reduced CMB anisotropy maps have been derived from the nine frequency channel maps using four component-separation algorithms~\cite{Planck2018IV}: \texttt{Commander}, \texttt{NILC}, \texttt{SEVEM} and \texttt{SMICA}. For each method, the \textit{Planck} collaboration provides the full-mission CMB intensity and polarisation maps with corresponding confidence masks, inside which the CMB signal is trusted, and effective beam transfer functions. All maps are provided in Galactic coordinates at a \textsc{Healpix} resolution of $N_\mathrm{side} = 2048$ and at an approximate angular resolution of 5 arcmin full width at half maximum (FWHM). 

The estimation of the instrumental noise present in the \textit{Planck} observations is inferred from maps which are obtained by splitting the full-mission data sets in various ways \cite{Planck2018VII}. In this work, we use the half-mission maps, which are obtained by splitting the time-ordered data according to long time periods. More specifically, we use the half-mission half-sum (HMHS) maps, containing signal and noise, and the half-mission half-difference  (HMHD) maps, which provide an estimate for the combined impact of instrumental noise and residual systematic effects.

For the present analysis, we choose the \texttt{Commander} CMB maps together with the common temperature and polarisation masks, since it is the preferred algorithm of the \textit{Planck} collaboration to analyse CMB anisotropies at intermediate and large scales. The \texttt{Commander} CMB maps are derived by thresholding the standard deviation map, evaluated between each of the cleaned CMB maps based on the four algorithms. Furthermore, we use the temperature and polarisation masks for unobserved pixels for the half-mission data split, with $f_\mathrm{sky} \sim 0.96$ and $f_\mathrm{sky} \sim 0.961$, respectively. We downgrade all the CMB maps used in this analysis to a lower resolution of $N_\mathrm{side} = 1024$, using the following downgrading procedure \cite{Planck2018VII}: The high-resolution full-sky CMB maps are decomposed into their spherical harmonic coefficients at the input resolution of $N_\mathrm{side} = 2048$. These are then convolved to the new resolution, using their corresponding beam and pixel window function. The new coefficients are then synthesized into new maps at lower resolution of  $N_\mathrm{side} = 1024$. The HMHS CMB temperature anisotropy and polarisation maps at $N_\mathrm{side} = 1024$ are shown in the top panel of Fig. \ref{all_maps}.

\subsubsection{CMB Lensing Convergence}
\label{sec:cmb_lensing_data}
The weak gravitational lensing of the CMB photons causes secondary anisotropies in the CMB maps. The corresponding lensing potential, describing the deflection of the CMB photons through potential gradients along the line-of-sight, can be reconstructed using a quadratic estimator \cite{Okamoto2003}. The construction of the quadratic lensing estimator $\hat{\phi}$ for the \textit{Planck} 2018 data release follows \cite{Carron2017}. In this work, we use the CMB lensing convergence estimate from the \textit{Planck} 2018 data release \cite{Planck2018VIII}, which is given in the form of spherical harmonic coefficients $\hat{\kappa}_\mathrm{CMB, \ell ,m}$ after mean-field subtraction. The mean-field is estimated using the quadratic estimator mean over a set of the most faithful simulations and represents the map-level signal expected from the mask, noise and further anisotropic features in the absence of lensing \cite{Planck2018VIII}. The spherical harmonic coefficients are related to the CMB lensing potential estimate through
\begin{equation}
\hat{\kappa}_\mathrm{CMB, \ell ,m} = \frac{\ell (\ell + 1)}{2} \hat{\phi}_\mathrm{CMB, \ell ,m}\, ,
\label{klm_estimate}
\end{equation}
for a multipole range of $8 \leq \ell \leq 4096$. The above lensing potential estimate is obtained from foreground-cleaned temperature and polarisation maps, computed using the \texttt{SMICA} algorithm and combined into a minimum-variance estimator (MV). We use the coefficients given by Eq. (\ref{klm_estimate}) to create a map with resolution of $N_\mathrm{side} = 1024$ by using the \textsc{Healpix} routine \texttt{alm2map}, which we then use to compute cross-correlations between the CMB lensing convergence and the other probes considered. We further use the lens reconstruction analysis mask provided by the \textit{Planck} Collaboration with $N_\mathrm{side} = 2048$. We downgrade the resolution of the mask to $N_\mathrm{side} = 1024$, using the downgrading procedure based on \cite{Planck2018VII} and described above.

In order to obtain the auto-correlation of the CMB lensing convergence, we follow the steps detailed in \cite{Planck2018VIII}. First, we use the \textsc{Healpix} routine \texttt{alm2cl} to obtain the pseudo spherical harmonic power spectrum of the masked, reconstructed lensing map. The full-sky lensing power spectrum is then obtained by upweighting the pseudo power spectrum by $1/f_\mathrm{sky}$. Subsequently, additional bias terms have to be subtracted. 

Since the lensing auto-power spectrum estimator probes both the connected and the disconnected part of the four-point function, a lensing bias term $N^{(0)}$ taking into account the disconnected signal expected from Gaussian fluctuations (even in the absence of lensing) has to be subtracted. Furthermore, the $N^{(1)}$ term coming from non-primary couplings of the connected four-point function (non-Gaussian noise) and the point-source bias term $PS$, taking into account the contribution from the connected four-point function of shot-noise from unresolved point sources, have to be subtracted. We use the minimum variance (MV) reconstruction lensing spectrum biases $N^{(0)}$, $N^{(1)}$ and $PS$ provided by the \textit{Planck} Collaboration in the lensing simulation package, as defined in \cite{Planck2018VIII}. Finally, a multiplicative correction (the MC term), obtained through Monte Carlo simulations has to be applied to our lensing auto-power spectrum estimator, correcting for various isotropic and simplifying approximations made in the previous steps.

\subsection{\label{sec:data_boss}BOSS DR12 Data}
We use the spectroscopic galaxy catalogue from the BOSS $12^\mathrm{th}$ final public data release~\footnote{\url{https://data.sdss.org/sas/dr12/boss/}}.

The galaxy catalogue consists of 2.5 million objects of which around 1.5 million were classified as galaxies \cite{Reid2015}. The BOSS DR12 samples are divided into two galaxy catalogues, named LOWZ and CMASS, which have been created by the BOSS collaboration by applying colour-magnitude and colour-colour cuts to the SDSS photometric catalogue in order to generate galaxy target lists. We refer the reader to~\cite{Reid2015, Alam2015} for a more detailed description of the BOSS DR12 data set.

\subsubsection{The LOWZ and CMASS Samples}
In order to facilitate the comparison and reproducibility of our results, the preparation of the catalogues used in this work follows a procedure similar to that described in \cite{Doux2018}.

The LOWZ sample contains Luminous Red Galaxies (LRG) at low redshifts ($z \lesssim 0.4$), as an extension of the SDSS-I/II LRG Cut I sample \cite{Eisenstein2001}. The targets are selected by applying redshift-dependent magnitude cuts, which produces constant number density in the redshift range of this sample. For the present work we select galaxies within the redshift range $[0.0, 0.4]$, which results in $390\,200$ galaxies. The CMASS sample consists of galaxies at higher redshifts ($0.4 \lesssim z \lesssim 0.8$), with constant stellar mass, and was designed by extending the Cut-II LRGs from SDSS-I/II \cite{Eisenstein2001} to bluer and fainter objects. We select galaxies in the redshift range $[0.4, 0.8]$, which contains $823\,193$ galaxies. The corresponding redshift distributions are shown in the upper panel of Figure~\ref{nz_plot}.

\subsubsection{\label{sec:galaxy_maps}Galaxy Weights and Map Making}
\textbf{Total galaxy weights.} Clustering statistics calculated from the raw DR12 data contain biases due to various observational effects, such as fibre collisions and redshift failures. The BOSS collaboration provides a weighting scheme for each targeted galaxy in order to remove these biases~\cite{Reid2015, Anderson2014}. This weighting scheme consists of three components, associated to different observational effects: 
\begin{itemize}
\item The angular systematic weight $\omega_\mathrm{systot}$, which takes into account the effects of the stellar density with airmass, sky flux, reddening and further seeing conditions. This weight is simply given by the product of the weights $\omega_\mathrm{star}$ and $\omega_\mathrm{see}$, both provided by the BOSS collaboration.

\item $\omega_\mathrm{cp}$ accounting for the effect of fibre collisions, i.e., pairs of objects that cannot have both their spectra measured.
\item $\omega_\mathrm{noz}$ upweights galaxies, which are affected by redshift failure.
\end{itemize}

We use the above described components to combine them into a single weight for each galaxy:
\begin{equation}
\omega_\mathrm{tot} = \omega_\mathrm{star} \omega_\mathrm{see} \left(\omega_\mathrm{cp} + \omega_\mathrm{noz} - 1 \right)\, ,
\label{tot_galaxy_weight}
\end{equation}
where the default values for $\omega_\mathrm{cp}$ and $\omega_\mathrm{noz}$ are set to unity.\\
\\

\textbf{Completeness map.} The completeness of observation within a given region, i.e., the extent to which the redshifts of the galaxies were determined, is given by a continuous acceptance mask
\begin{equation}
C_\mathrm{BOSS} = \frac{N_\mathrm{obs} + N_\mathrm{cp}}{N_\mathrm{obs} + N_\mathrm{obs} + N_\mathrm{missed}}\, ,
\end{equation}
where $N_\mathrm{obs}$ is the number of spectroscopically observed objects (galaxies, stars and unclassified objects), $N_\mathrm{cp}$ represents the number of close-pair objects and $N_\mathrm{missed}$ is the number of targeted objects without spectra. Additionally, the BOSS collaboration provides several veto masks, which mark a given region as either good or bad, i.e. are given as binary maps. These masks remove regions vetoed for bad observational factors such as bad photometry, bright objects and stars, fibre centerpost collisions, collision priorities, seeing cuts and extinction cuts \cite{Reid2015}.

The acceptance and the veto masks are provided in the \textsc{Mangle}~\footnote{\href{https://space.mit.edu/~molly/mangle/}{https://space.mit.edu/~molly/mangle/}} format \cite{Swanson2008} and are converted to a single \textsc{Healpix} map with a resolution of $N_\mathrm{side} = 1024$. We then construct a binary mask function for the galaxy samples by assigning a value of 1 to pixels which pass a completeness cut of $C_\mathrm{BOSS} > 0.75$ and which are not marked as bad in any of the veto masks. This choice of completeness cut is based on \cite{Doux2018}.\\
\\

\textbf{Galaxy overdensity maps.} Based on the LOWZ and CMASS galaxy catalogues, we construct galaxy overdensity \textsc{Healpix} maps with a resolution of $N_\mathrm{side} = 1024$. First, we select galaxies from the data catalogues with redshifts within the ranges of $[0.0, 0.4]$ and $[0.4, 0.8]$ for LOWZ and CMASS, respectively. For both samples, we then compute the number of objects in each pixel $p$ weighted by the total galaxy weight (Eq. (\ref{tot_galaxy_weight})), given by $N_p = \sum_{i} \omega_{\mathrm{tot}, i}$. The final galaxy overdensity maps are then given by
\begin{equation}
  \delta_p =
    \begin{cases}
      \frac{N_p}{\bar{N}} - 1 & \text{if $C_{\mathrm{BOSS}, p} > 0.75$ and $p$ not vetoed} \\
      0 & \text{otherwise},
    \end{cases}       
\end{equation}
where $\bar{N} = \frac{1}{N_\mathrm{pix}} \sum_{p=1}^{N_\mathrm{pix}} N_p$ is the mean pixel count and $N_\mathrm{pix}$ denotes the total number of pixels. The resulting galaxy overdensity maps for both samples are shown in Fig. \ref{all_maps}.

However, by using only two broad redshift bins for this data set, we significantly compromise radial information. Following \cite{Asorey2012}, a spectroscopic survey with $z<1$ is allowed to have a thickest possible redshift bin size of $\Delta z = 0.05$ in order to retain sufficient radial information without suppression of the radial BAO signal.

\subsection{\label{sec:data_kids}KiDS-1000 Data}
We use the gold sample of weak lensing and photometric redshift measurements from the fourth data release of the Kilo-Degree Survey, covering 1006 $\mathrm{deg}^2$ of images~\cite{Kuijken2019, Wright2020, Hildebrandt2020, Giblin2021}, hereafter referred to as KiDS-1000. The used data products are publicly available on the KiDS website~\footnote{\url{http://kids.strw.leidenuniv.nl/DR4/}}. The KiDS survey is a public survey by the European Southern Observatory (ESO), mounted at the Cassegrain focus of the VLT Survey Telescope (VLT) operating in the four bands $ugri$. KiDS is specifically designed for weak lensing applications. Together with the infrared data from its partner survey VIKING (VISTA Kilo-degree INfrared Galaxy survey~\cite{Edge2013}), the observed galaxies have photometry in nine optical and near-infrared bans $ugriZYJHK_\mathrm{s}$. This allows for a better estimate of the photometric redshifts compared to using only the four optical bands observed by KiDS alone~\cite{Hildebrandt2020b}. Cosmological parameter constraints from KiDS-1000 have been presented in \cite{Asgari2021} from cosmic shear measurements, in \cite{Heymans2021} for a 3$\times$2pt analysis and in \cite{Troester2021} for beyond $\Lambda$CDM models, in all cases with the methodology presented in \cite{Joachimi2021}.

\subsubsection{Galaxy shear estimates}
In this work we perform a tomographic analysis of the gold sample by dividing the galaxies into five tomographic bins, according to their best-fitting photometric redshift $z_\mathrm{B}$. The photometric redshift of each galaxy $z_\mathrm{B}$ is estimated using the Bayesian photometric code \textsc{BPZ}~\cite{Benitez2000, Benitez2004}, whereas the self-organising map (SOM) method~\cite{Wright2020} is used to calibrate the redshift distribution of each tomographic bin. Thereby, galaxies are organised into groups based on their nine-band photometry and associated to matches within spectroscopic samples. We use the \textsc{BPZ} estimate of the redshift to split the galaxies into the same five tomographic bins as presented in \cite{Asgari2021}, whereas we only use galaxies with a redshift estimate between $z = 0.1$ and $z=1.2$. More information about the tomographic bins is given in Table~\ref{zbin_table}. Note that these estimated redshift ranges are based on different methods and may thus have varying accuracy. The redshift distributions for the five tomographic bins are shown in the lower panel of Figure~\ref{nz_plot}.

We further use the galaxy shear estimates produced using \textsc{Lensfit}~\cite{Miller2013, Conti2017}, described in \cite{Giblin2021}. \textsc{Lensfit} is based on a model of the point spread function of the individual exposures at the pixel level, and obtains the ellipticities of the galaxies using a likelihood-based method. The estimated shear value of each pixel is then calculated as
\begin{equation}
\gamma_\mathrm{pix} = \frac{\sum_{i \in \mathrm{pix}} w_i e_i}{\sum_{i \in \mathrm{pix}} w_i}\, ,
\label{shear_map}
\end{equation}
where $w_i$ is the \textsc{Lensfit} weight and $e_i$ is the measured ellipticity of each galaxy $i$. Note that the used \textsc{Lensfit} ellipticities are not corrected for multiplicative or additive shear bias, but are introduced as nuisance parameters in our parameter inference.

The estimated shear maps given by Eq. (\ref{shear_map}) both contain contributions from the cosmic shear signal and the shape noise of the galaxies, where the latter is attributed to intrinsic galaxy ellipticities. We generate a shape noise estimate by randomly rotating each galaxy in our sample \cite{Fluri2019}
\begin{equation}
    \gamma_\mathrm{pix}^\mathrm{noise} = \frac{\sum_{j \in \mathrm{pix}} \mathrm{exp}(\theta_j i) w_j \gamma_j}{\sum_{j \in \mathrm{pix}} w_j} \, ,
    \label{sn_map}
\end{equation}
where $\theta_j$ is drawn uniformly from $\left[0,2 \pi \right)$. This procedure breaks the spatial correlation between the galaxy shapes, leaving only the shape noise due to intrinsic galaxy ellipticities. 

For each redshift bin, we use the right ascension and declination of the galaxies (denoted as \texttt{RAJ2000} and \texttt{DECJ2000} in the catalogue) to infer the associated confidence mask used to compute to coupling matrix (see Section~\ref{sec:pseudo_theory}). The final weak lensing galaxy shear signal and noise maps and confidence masks are rotated from equatorial to galactic coordinates. The weak lensing shear modulus $| \gamma |$ for redshift bin 1 is shown in Figure~\ref{all_maps}.

\subsection{\label{sec:cls_data}Spherical Harmonic Power Spectra}
The spherical harmonic power spectrum measured from the maps $X$ and $Y$ can be calculated through
\begin{eqnarray}
%\hat{\tilde{C}}_{\ell}^{XY} = \frac{1}{2 \ell + 1} \sum_{m = - \ell}^{m = + \ell} \tilde{X}_{\ell m}^* %\tilde{Y}_{\ell m}^{}\, ,
\hat{\tilde{C}}_{\ell}^{XY, \mathrm{maps}} = \frac{1}{2 \ell + 1} \sum_{m = - \ell}^{m = + \ell} &&\left[ \operatorname{Re}(\tilde{X}_{\ell m})\operatorname{Re}(\tilde{Y}_{\ell m}) \right. \\
 &+& \left. \operatorname{Im}(\tilde{X}_{\ell m})\operatorname{Im}(\tilde{Y}_{\ell m}) \right] \, , \nonumber
\label{cl_measured}
\end{eqnarray}
where $\tilde{X}_{\ell m}$ and $\tilde{Y}_{\ell m}$ denote the pseudo-spherical harmonic coefficients. The maps derived from the different data sets include instrumental noise and only have partial sky coverage. This affects the above power spectrum estimate. Therefore, and after correction for the \textsc{Healpix} pixel window function and the beam window function, the measured power spectrum estimates can be written as
\begin{equation}
\left<\hat{\tilde{C}}_{\ell}^{XY}\right> = \sum_{\ell'} M_{\ell \ell'} C_\ell + \delta_{XY} \tilde{N}_{\ell}^{X}\, ,
\end{equation}
where $\left< \ldots \right>$ represents the ensemble average. On average, the measured power spectra are thereby related to the underlying full-sky analytical power spectra $C_\ell$ through the mode-coupling matrix $M_{\ell \ell'}$, taking into account the effect of the mask (see Section~\ref{sec:pseudo_theory}). By using the  Kronecker delta in the term $\delta_{XY} \tilde{N}_{\ell}^{X}$, we assume that different cosmological probes have uncorrelated noise~\cite{Hinshaw2003, Doux2018}. We therefore subtract the corresponding noise spectra for the auto correlations only. Following the approach in~\cite{Nicola2016, Nicola2017, Doux2018}, we resort to simulations to estimate the pseudo-noise power spectra $\tilde{N}^X_{\ell}$ for the cosmological probes $\kappa_\mathrm{CMB}, \delta_g$ and $\gamma$. The final power spectrum estimate is given by
\begin{equation}
\hat{\tilde{C}}_{\ell}^{XY} = \hat{\tilde{C}}_{\ell}^{XY, \mathrm{maps}} - \delta_{XY} \hat{\tilde{N}}_{\ell}^{X}\, .
\end{equation}

We measure the auto- and cross-pseudo-spectra between the maps $X,Y \in \{\mathrm{T}, \mathrm{E}, \kappa_\mathrm{CMB}, \delta_\mathrm{LOWZ}, \delta_\mathrm{CMASS}, \gamma_{\mathrm{bin} \, 1}, \ldots, \gamma_{\mathrm{bin} \, 5}\}$ with a resolution of $N_\mathrm{side} = 1024$, based on their respective combined mask using the \textsc{Healpix} routine \texttt{anafast}. All the obtained spectra are corrected for the \textsc{Healpix} pixel window function and the spectra involving the CMB temperature anisotropies or polarisation are further corrected for the effective beam window functions, where the latter are provided by the \textit{Planck} Collaboration. 

Most of the analytical predictions for the spectra considered in this work are based on \textsc{PyCosmo}~\cite{pycosmo2018,Tarsitano2020} (apart from the spectra $\mathrm{TT}, \mathrm{TE}, \kappa_\mathrm{CMB} \kappa_\mathrm{CMB}$, for which we use \textsc{Class}~\cite{Lesgourgues2011, Blas2011}), which uses the Limber approximation. Following~\cite{LoVerde2008, Ho2012}, the Limber approximation becomes accurate for angular scales larger than $\ell \sim 30$. We therefore choose a common minimum lower cut of $\ell_\mathrm{min} = 50$ for all spectra considered. As detailed in Appendix \ref{sec:wig_mat_check}, we do not expect a significant mixing of potentially miscalculated modes from the regime $\ell < 30$ with larger modes by applying the mode-coupling matrix. An overview of the spectra used in this analysis and the applied multipole cuts is given in Table~\ref{cl_table}. More details on the measurement of the power spectra and the noise estimates are presented below.
\begin{table}
\caption{Redshift ranges and number of objects for the used redshift bins in the BOSS DR12 and KiDS-1000 data sets.}
\begin{tabular}{ lccr }
 \hline
 \\
 Survey&Bin&z-range&number of objects\\[0.1cm]
 \hline
  \hline
  \\
BOSS DR12 \quad & LOWZ \quad & $\left[0.0, 0.4\right]$ \quad & 390'200\\[0.1cm]
BOSS DR12&CMASS & $\left[0.4, 0.8\right]$ & 823'193\\[0.1cm]
 \hline
 \\
KiDS-1000& 1 & $\left[0.1, 0.3\right]$ & 1'792'136\\[0.1cm]
KiDS-1000& 2 & $\left[0.3, 0.5\right]$ & 3'681'319\\[0.1cm]
KiDS-1000& 3 & $\left[0.5, 0.7\right]$ & 6'148'102\\[0.1cm]
KiDS-1000& 4 & $\left[0.7, 0.9\right]$ & 4'544'395\\[0.1cm]
KiDS-1000& 5 & $\left[0.9, 1.2\right]$ & 5'096'059\\[0.1cm]
 \hline
\end{tabular}
\label{zbin_table}
\end{table}
\subsubsection{Auto-Correlations}
\begin{figure*}[htbp!]
\centering
\includegraphics[width=0.8\paperwidth]{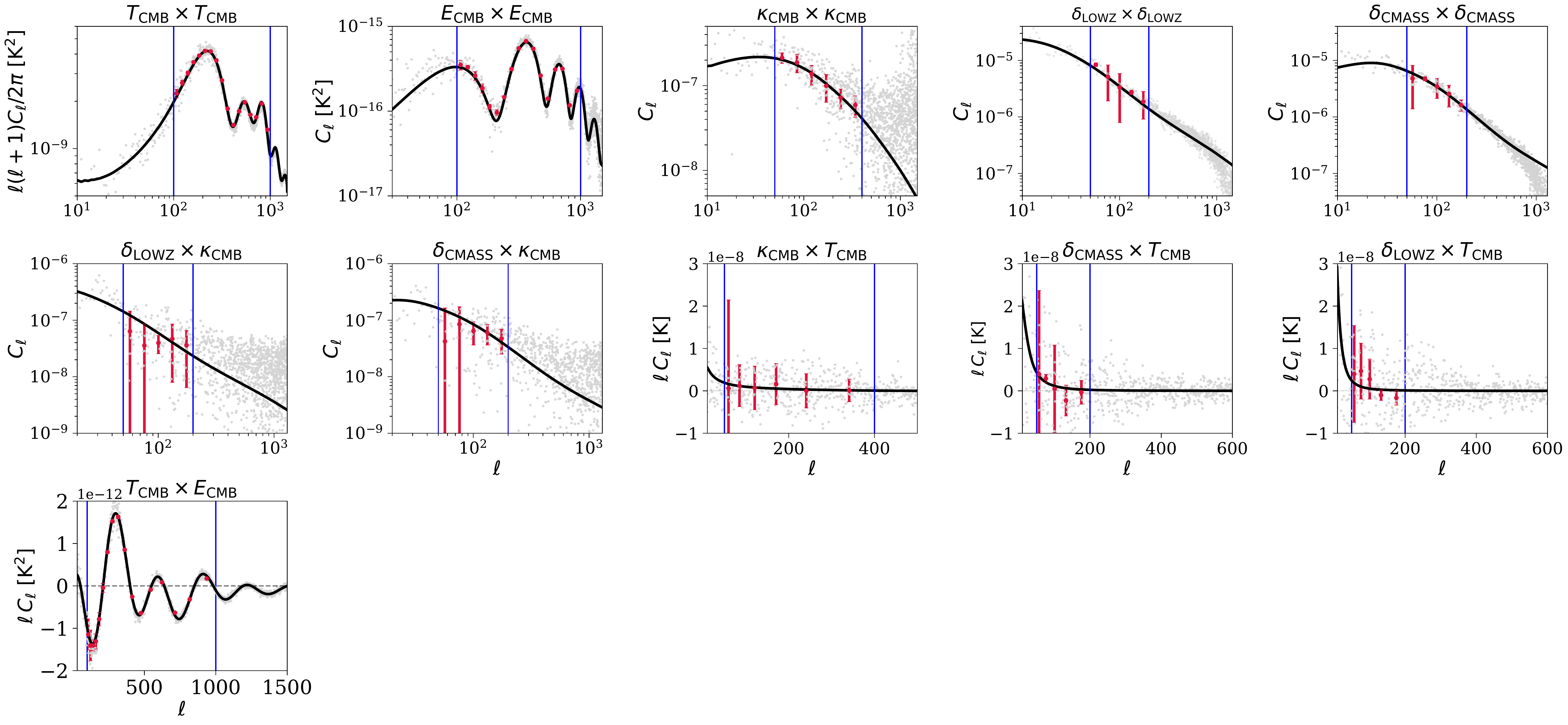}
\caption{Pseudo auto- and cross-spherical harmonic power spectra for CMB temperature and polarisation, CMB lensing convergence and the two BOSS galaxy samples LOWZ and CMASS. The observed spectra are shown as light grey dots, which have been binned according to Table \ref{cl_table} for our combined analysis (red dots). The errorbars represent the $1\sigma$-error and are obtained from our multi-probe covariance matrix. The best-fit model prediction for the pseudo-spectra from the joint analysis of auto- and cross-spectra are given by the solid black lines (the parameter values are listed in the last row of Table \ref{prior_post_table}). The two blue lines represent the lower and upper mutlipole cuts used in our analysis.}
\label{cls_data_A}
\end{figure*}

\textbf{CMB temperature and polarisation.} We estimate the CMB temperature and polarisation signal power spectra using the half-mission half-sum (HMHS) maps and subtract an estimate of the noise present in the HMHS maps using the half-mission half-difference (HMHD) maps
\begin{equation}
    \hat{\tilde{N}}_{\ell}^{X} = C_{\ell}^{X, \mathrm{HMHD}}\, ,
\end{equation}
for $X \in \{\mathrm{T}, \mathrm{E} \}$. We choose a rather conservative upper cut in angular multipole at $\ell = 10^3$ in order to reduce potential biases by residual foregrounds~\cite{Nicola2016} and a lower cut at $\ell_\mathrm{min} = 100$. The resulting power spectra for the CMB temperature and polarisation are shown in Fig. \ref{cls_data_A}.\\
\\
\textbf{CMB lensing convergence.} As described in Section~\ref{sec:cmb_lensing_data}, the calculation of the auto power spectrum of the CMB lensing convergence can be summarised as follows \cite{Planck2018VIII}: First, using the \textsc{Healpix} routine \texttt{alm2cl}, we obtain the pseudo power spectrum, which is subsequently upweighted by $1/f_\mathrm{sky}$ to obtain a full-sky power spectrum estimate:
\begin{equation}
    \hat{C}_\ell^{\hat{\kappa}_\mathrm{CMB}} = \frac{1}{(2\ell + 1) f_\mathrm{sky}} \sum_{m = - \ell}^{m = + \ell} \tilde{\kappa}_{\mathrm{CMB}_{\ell m}}^{*} \tilde{\kappa}_{\mathrm{CMB}_{\ell m}^{}}\, .
\end{equation}
Next, we subtract the lensing biases as in \cite{Planck2018VIII}
\begin{equation}
    \hat{C}_\ell^{\kappa_\mathrm{CMB}} = \hat{C}_\ell^{\hat{\kappa}_\mathrm{CMB}} - N_\ell^{(0)} - N_\ell^{(1)} - {PS}_\ell\, .
\end{equation}
In contrast to the \textit{Planck} 2015 release of the CMB lensing measurement, the 2018 release provides lensing maps to higher multipole $\ell_\mathrm{max} = 4096$. Even though \textit{Planck} performs a CMB lensing power spectrum meaurement using an \textit{aggressive} multipole range of $8 \leq \ell \leq 2048$, the most significant measurement at 40$\sigma$ is achieved using the MV estimate over the \textit{conservative} range $8 \leq \ell \leq 400$~\cite{Planck2018VIII}.
For the present analysis, we consider multipoles in the range of $50 \leq \ell \leq 400$ due to potential uncertainties in the lensing bias terms $N^{(0)}$ and $N^{(1)}$ becoming increasingly important at small scales and potential errors in modelling the survey anisotropies at large scales. The obtained power spectrum estimate for the CMB lensing convergence is shown in Fig. \ref{cls_data_A}.\\
\\
\textbf{Galaxy clustering.} We follow \cite{Nicola2016} to estimate the Poisson shot noise contribution of the galaxy overdensity power spectra. For both samples, LOWZ and CMASS, we randomise the positions of all galaxies inside the mask, keeping their corresponding weight. This breaks the spatial correlations between the galaxies corresponding to the galaxy clustering signal, leaving only an estimate of the Poisson shot noise. The Poisson noise maps are then constructed using the procedure described in Section ~\ref{sec:galaxy_maps}. We obtain a well-converged estimate of the pseudo-noise power spectrum by taking the mean of $N_\mathrm{P} = 1000$ Poisson noise spectra realisations
\begin{equation}
    \hat{\tilde{N}}_{\ell}^{X} = \frac{1}{N_\mathrm{P}} \sum_{i = 1}^{N_\mathrm{P}} \hat{\tilde{N}}_{\ell}^{X, i}\, ,
\end{equation}
where $X \in \{\mathrm{LOWZ}, \mathrm{CMASS} \}$. 

The linear, scale-independent galaxy bias $b(z)$ accurately relates the galaxy overdensity to the matter overdensity on large scales. On smaller scales, this relation is more complicated and the effects of the nonlinear galaxy bias are difficult to estimate. Since the galaxy clustering power spectrum obtains contributions from nonlinear structure formation on small scales, we restrict our analysis to angular scales where nonlinear contributions are small. We use \textsc{PyCosmo} to estimate the effect of nonlinear structure formation as a function of multipole. For this purpose, we compare the galaxy clustering power spectrum computed using the nonlinear to the one using the linear matter power spectrum. For both samples, we find that the difference reaches $5\%$ at $\ell \sim 200$. This upper multipole bound for the galaxy clustering power spectrum is in accordance with previous work~\cite{Ho2012, Nicola2016, Doux2018}. Moreover, using a linear, scale-independent bias approximation has been found to be accurate within $5\%$ down to scales of $20 h^{-1} \mathrm{Mpc}$~\cite{Cresswell2008, Torres2016}. We therefore use a conservative angular scale cut of $50 \leq \ell \leq 200$ for the galaxy clustering power spectra. The minimum multipole $\ell_\mathrm{min} = 50$ is chosen because the Limber approximation used in the \textsc{PyCosmo} calucaltions breaks down on larger scales. On similar scales, $\ell \lesssim 30$, the effects of RSD are expected to increase the power of the galaxy clustering power spectra~\cite{Padmanabhan2008}, caused through mixing of redshift and peculiar velocities of galaxies. Even though our analysis includes the effects of RSD in the analytical prediction code \textsc{PyCosmo} and in the lightcone generator \textsc{UFalcon} used for the covariance matrix estimation, we do not expect a significant impact due to our choice of $\ell_\mathrm{min}$.

The situation becomes more complicated in the presence of massive neutrinos, where the galaxy bias defined with respect to the total matter field also depends on the sum of the neutrino masses and becomes scale-dependent also on large scales~\cite{Villaescusa-Navarro:2017mfx,Giusarma:2018jei,Vagnozzi:2018pwo}. Not accounting for such a neutrino-induced scale-dependent bias for future surveys (especially for galaxy clustering surveys) could result in severe systematics and biased results for the sum of the neutrino masses and other cosmological parameters, as has been shown in~\cite{Vagnozzi:2018pwo} for a Euclid-like survey. For current surveys this has been shown not to be an issue~\cite{Raccanelli2018}. Given the sensitivity of the surveys analysed in this work, we neglect this effect here and leave it to be implemented in future analyses including galaxy clustering data.\\
\\
\textbf{Cosmic shear.} We compute the weak lensing shear auto-power spectrum for each of the five tomographic bins in the KiDS-1000 sample, which contains both cosmic shear signal and shape noise. We estimate the shape noise power spectrum from the mean of $N_\mathrm{SN}=200$ noise power spectra based on randomised noise maps as
\begin{equation}
    \hat{\tilde{N}}_{\ell}^{X} = \frac{1}{N_\mathrm{SN}} \sum_{i = 1}^{N_\mathrm{P}} \hat{\tilde{C}}_{\ell}^{\gamma_\mathrm{noise}, i}\, ,
\end{equation}
where $X \in \{\gamma_\mathrm{bin 1}, \ldots, \gamma_\mathrm{bin 5}\}$ and $\gamma_\mathrm{noise}$ is given by Eq. (\ref{sn_map}).
For the present analysis we choose angular scales of $100 \leq \ell \leq 1000$ for cosmic shear power spectra. The lower multipole cut is chosen to avoid large-scale regimes, where the Limber approximation breaks down and the effects of randomisation on the simulated \textsc{UFalcon} power spectra become dominant. With the currently used resolution settings for the \textsc{PkdGrav3} simulations, our \textsc{UFalcon} power spectra used for the covariance matrix estimation agree within $5\%$ with the analytical predictions up to $\ell \sim 1200$. Therefore, we use a conservative upper multipole cut of $\ell_\mathrm{max}=1000$ for all weak lensing shear power spectra.

As found in \cite{Asgari2021}, we set the intrinsic alignment amplitude to the maximum posterior value of the full multivariate distribution (MAP), $A_\mathrm{IA}=0.973$, inferred from the band power analysis. 

\begin{figure*}[htbp!]
\centering
\includegraphics[width=0.8\paperwidth]{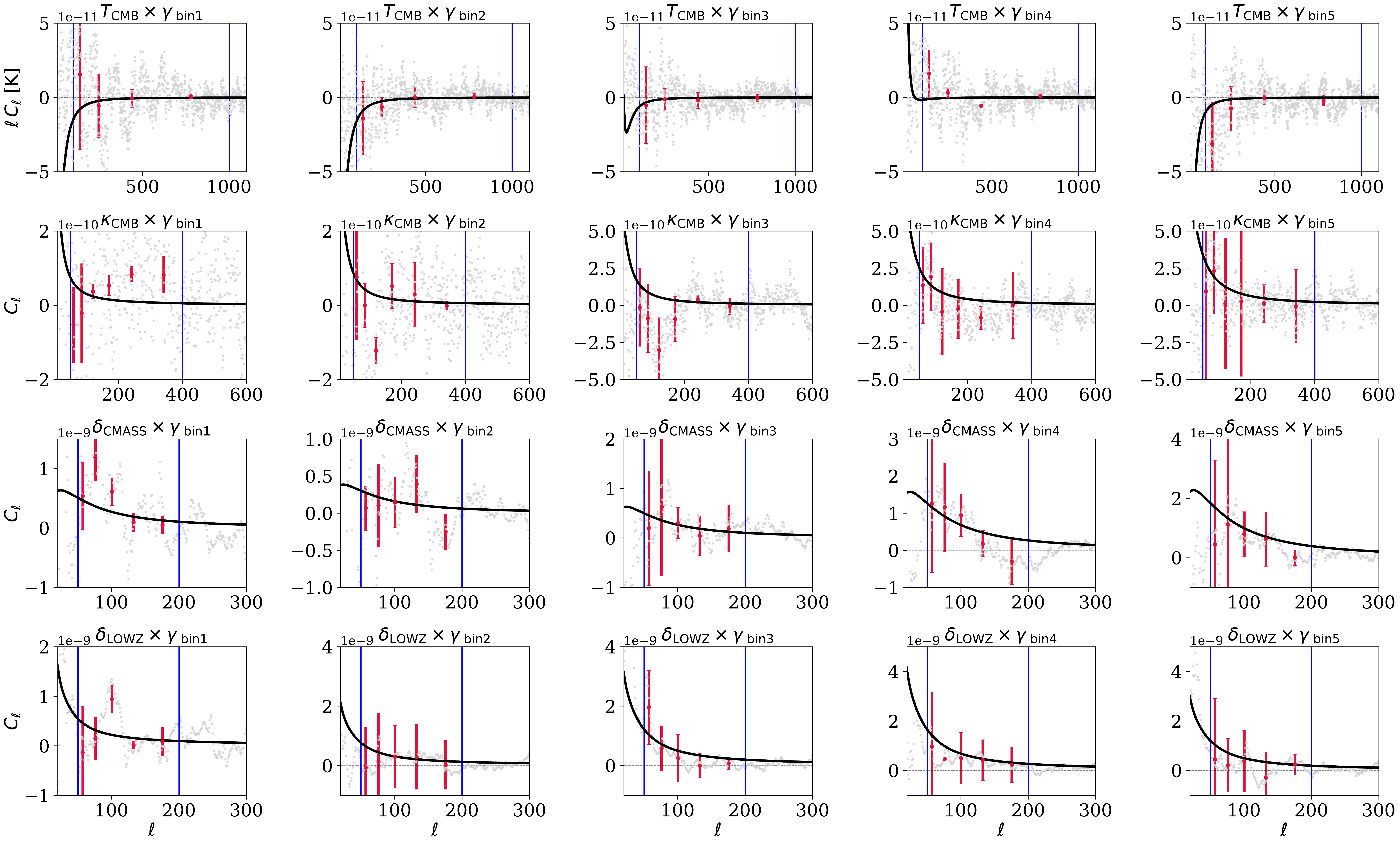}
\caption{Pseudo cross-spherical harmonic power spectra between the CMB temperature, CMB lensing convergence, LOWZ and CMASS galaxy samples and the five KiDS-1000 tomographic weak lensing shear bins. The observed spectra are shown as light grey dots, which have been binned according to Table \ref{cl_table} for our combined analysis (red dots). The errorbars represent the $1\sigma$-error and are obtained from our multi-probe covariance matrix. The best-fit model prediction for the pseudo-spectra from the joint analysis of auto- and cross-spectra are given by the solid black lines (the parameter values are listed in the last row of Table \ref{prior_post_table}). The two blue lines represent the lower and upper mutlipole cuts used in our analysis.}
\label{cls_data_B}
\end{figure*}
\begin{figure*}[htbp!]
\centering
\includegraphics[width=0.8\paperwidth]{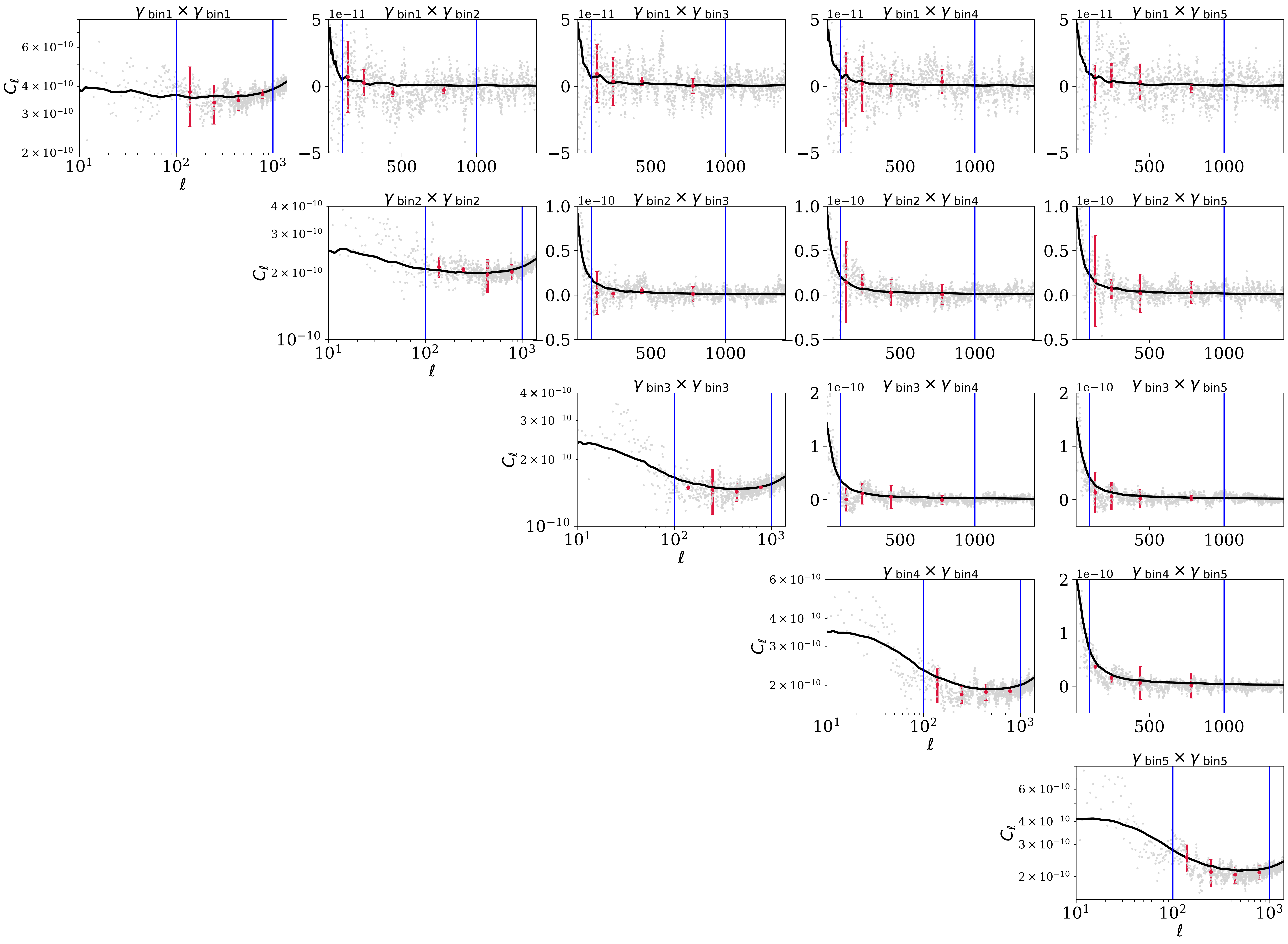}
\caption{Pseudo auto- and cross-spherical harmonic power spectra between the five KiDS-1000 tomographic weak lensing shear bins. The observed spectra are shown as light grey dots, which have been binned according to Table \ref{cl_table} for our combined analysis (red dots). The errorbars represent the $1\sigma$-error and are obtained from our multi-probe covariance matrix. The best-fit model prediction for the pseudo-spectra from the joint analysis of auto- and cross-spectra are given by the solid black lines (the parameter values are listed in the last row of Table \ref{prior_post_table}). The two blue lines represent the lower and upper mutlipole cuts used in our analysis}
\label{cls_data_C}
\end{figure*}

\subsubsection{Cross-Correlations}
\textbf{CMB cross-correlations.} We compute the cross-correlations between the CMB temperature, polarisation and lensing convergence using the maps explained in Section~\ref{sec:CMB} and the corresponding combined masks. The resulting spherical harmonic power spectra are shown in Figure~\ref{cls_data_A}. Note that the combination of our estimated spectra TT, TE, EE and KK would correspond to a modified \textit{Planck} 2018 "TT, TE, EE + lensing" data set with reduced multipole ranges (see Table \ref{cl_table} for our $\ell$-cuts) compared to the \textit{Planck} ranges given by $30 \leq \ell \leq 2508$ for TT, $30 \leq \ell \leq 1996$ for TE and EE and $8 \leq \ell \leq 400$ for CMB lensing. Note that using a reduced multipole range in this analysis leads to a shift in the obtained parameter constraints, compared to the results from the official \textit{Planck} 2018 analysis~\cite{Aghanim:2018eyx}. In Figure~\ref{figure_planck_consistency_TTTEEE} we compare our map-based results with the contours obtained when using \textsc{CosmoMC}~\cite{Bridle2007} for a multipole range of $100 \leq \ell \leq 1000$. We further include the cross-correlation between CMB temperature anisotropies and CMB lensing convergence T$\kappa$ in our analysis.\\
\\
\textbf{CMB $\times$ galaxy overdensity.} In the second row of Figure~\ref{cls_data_A}, we show the cross-correlations between the \textit{Planck} 2018 CMB temperature and lensing convergence and the two BOSS galaxy samples LOWZ and CMASS using the maps detailed in Sections~\ref{sec:CMB} and \ref{sec:data_boss} and their combined mask. The cross-spectra $\delta_\mathrm{LOWZ} \kappa$ and $\delta_\mathrm{CMASS} \kappa$ clearly reveal non-zero correlations. The cross-correlations between the CMB temperature and the galaxy overdensity show a relatively high noise level, such that a detection of the ISW effect is challenging. We nevertheless choose to include the $\delta_\mathrm{LOWZ}$T and $\delta_\mathrm{CMASS}$T spectra in our analysis.\\
\\
\textbf{CMB $\times$ weak lensing.} We compute the cross-correlations between the \textit{Planck} 2018 CMB temperature and lensing convergence and the KiDS-1000 weak lensing shear using the maps presented in Sections \ref{sec:CMB} and \ref{sec:data_kids} and the probe-specific combined mask. The resulting spherical harmonic power spectra are shown in the two top rows of Figure \ref{cls_data_B}. As can be seen from the figure, the high noise level in the cross-spectra T$\gamma_i$ and $\kappa \gamma_i$ for all KiDS-1000 tomographic bins $i$ complicate a detection of the ISW effect.\\
\\
\textbf{Galaxy overdensity $\times$ weak lensing.} The last two rows in Figure \ref{cls_data_B} show the cross-correlations between the two BOSS galaxy samples LOWZ and CMASS and the 5 KiDS-1000 weak lensing shear tomographic bins using the maps detailed in Sections \ref{sec:data_boss} and \ref{sec:data_kids}.\\
\\

\subsection{\label{sec:systematics}Systematic Effects}
Systematic uncertainties associated with the different cosmological observations should be taken into account in order to obtain unbiased parameter constraints. We take into account potential sources of systematic effects through nine different nuisance parameters. These parameters are explained in more detail below.\\
\\
\textbf{CMB temperature}. As pointed out in \cite{Planck2013XV}, the foreground-reduced CMB temperature anisotropy maps are potentially contaminated by unresolved extragalactic sources. According to \cite{Planck2013XII}, this systematic effect can be taken into account by considering the power spectra of these foregrounds. These power spectra are given by an unclustered Poisson component and a clustered component, which become significant at higher multipoles. According to~\cite{Planck2013XII}, this residual foreground subtraction starts to have a significant impact on the CMB temperature power spectrum starting at $\ell \gtrsim 1500$. Furthermore, the effect of these residual foregrounds on the parameter constraints has been analysed in \cite{Nicola2017}, where no significant impact has been found using a conservative upper multipole cut of $\ell_\mathrm{max} = 610$. By choosing an upper multipole cut of $\ell_\mathrm{max} = 1000$ for the CMB temperature and polarisation power spectra, our analysis is potentially affected by such systematic effects. We therefore introduce a scalar multiplicative bias parameter $m_\mathrm{T}$ for the estimated temperature field, $\hat{\mathrm{T}} = (1 + m_\mathrm{T}) \mathrm{T}$, in order to take into account any systematic effects caused by residual foregrounds.\\
\\
\textbf{CMB lensing convergence}. The CMB lensing estimator $\hat{\phi}$ (see also Section \ref{sec:cmb_lensing_data}) depends on the cosmological parameters chosen for an input CMB lensing potential. The \textit{Planck} 2018 lensing analysis thus corrects for this cosmology-dependent bias by normalising the CMB lensing estimator assuming a fiducial cosmological model \cite{Planck2018VIII}. Therefore, the normalisation of the CMB lensing estimator is cosmology-dependent and should be varied alongside the cosmological parameters in the inference process.

Furthermore, the \textit{Planck} analysis varies the phenomenological parameter $A_\mathrm{lens}$, which scales the modelled lensing potential spectrum as $C_\ell^{\psi \psi} \rightarrow A_\mathrm{lens}C_\ell^{\psi \psi}$ and therefore changes the amplitude of the lensing-induced smoothing of the acoustic peaks and transfer of power to the damping tail of the CMB spectra~\cite{Calabrese2008}. The expected value for a base $\Lambda$CDM model is $A_\mathrm{lens} = 1$. Measurements by the \textit{Planck} collaboration however prefer positive values $A_\mathrm{lens} > 1$~\cite{Aghanim:2018eyx, Planck2015XIII, Planck2013XVI}, which is commonly referred to as the $A_\mathrm{lens}$ anomaly~\cite{Motloch:2018pjy}.

Following \cite{Nicola2017}, we take into account the uncertainty in the normalisation by introducing a multiplicative bias parameter $m_\kappa$, which rescales the estimated CMB lensing convergence field as $\hat{\kappa}_\mathrm{CMB} = (1 + m_\kappa) \kappa_\mathrm{CMB}$. The parameter $m_\kappa$ thus only rescales the auto- and cross-power spectra including the CMB lensing convergence and does not affect the CMB temperature and polarisation spectra. Note that the expected value is $m_\kappa = 0$, whereas it is $A_\mathrm{lens}=1$ for the $A_\mathrm{lens}$ parameter.\\
\\
\textbf{Galaxy overdensity.} On sufficiently large scales, the galaxy overdensity $\delta_g$ can be related to the underlying dark matter density field $\delta$ through the relation
\begin{equation}
\delta_g (\chi \hat{n}, z) = b(z) \delta (\chi \hat{n}, z)\, ,
\end{equation}
where the function $b(z)$ represents the linear galaxy bias. In this analysis we assume constant, linear and scale-independent galaxy bias parameters $b_\mathrm{LOWZ}$ and $b_\mathrm{CMASS}$ for both BOSS samples LOWZ and CMASS respectively. As discussed in Section \ref{sec:cls_data}, we expect that using this galaxy bias model is sufficiently accurate for the large angular scales considered in our analysis. As detailed in Section~\ref{sec:data_boss}, we apply the provided veto masks to the galaxy overdensity maps in order to remove observational and sky systematics. Therefore, we do not include further nuisance parameters for the galaxy overdensity field in our analysis.\\
\\
\textbf{Weak lensing shear.} The observed shapes of galaxies can be altered by various observational systematic effects, which is typically modelled using the point spread function (PSF) \cite{Bernstein2002}. In addition to such observational systematics, there are also potential systematics introduced in the shear reduction pipeline, such as errors in the noise estimation~\cite{Hirata2004b, Refregier2012}. For example, errors in the estimation of the size or the ellipticity of the PSF can introduce a scalar multiplicative shear bias $m$. The observed shear of a single galaxy can thus be modelled as~\cite{Heymans2006}
\begin{equation}
    \hat{\gamma} = m \gamma \, ,
\end{equation}
where $\gamma$ denotes the actual shear. We incorporate potential calibration uncertainties using a multiplicative bias parameter $m_{\gamma i}$ for each tomographic bin $i \in \left[1, \ldots ,5 \right]$ through $\hat{\gamma}_i = (1 + m_{\gamma i}) \gamma_i$.

Photometric redshift uncertainty represents another source of systematic effects in cosmic shear studies. The redshifts of galaxies can only be measured photometrically in weak lensing surveys, which can result in errors or failures in the measurement process. Such systematic effects can alter the redshift distributions of the source galaxies, which in turn might introduce errors in the parameter inference process \cite{Hildebrandt2020b, Huterer2006}. The KiDS weak lensing ``gold" sample used for our analysis provides redshifts distributions based on the Bayesian photometric redshift (BPZ) estimates with reliable mean redshifts. The high accuracy of the redshift calibration attained for this catalogue is mainly due to the nine-band photometry of the KiDS galaxy images, which allows to break potential degeneracies of the galaxy spectral energy distribution when calibrating the data with spectroscopic samples performed in lower-dimensional colour spaces  \cite{Hildebrandt2020b}. For the present analysis, we use the best-fit $n(z)$ distributions based on the calibration method and do not take into account the mean shifts in the distributions. Therefore, we do not include potential photometric redshift uncertainties in the parameter inference and leave the implementation of such shift parameters for each tomographic redshift distribution to future work.
\section{\label{sec:cov}Covariance Matrix}
In this section, we describe the covariance matrix used for the joint cosmological analysis in this work and the different stages in the estimation procedure. Our multi-probe covariance matrix estimation is based on computing the pseudo-spherical harmonic power spectra from an ensemble of dark matter (DM)-only $N$-Body simulation outputs and closely follows the work presented in \cite{Sgier2021}. The underlying $N$-Body simulation code \textsc{PKDGrav3} and the lightcone generation code \textsc{UFalcon} are described in the following sections.

\subsection{\label{sec:nbody} N-Body Simulations}
In order to fully include the non-linear and non-Gaussian nature of the matter density field, we resort to $N$-Body simulations to estimate the relevant statistics. We use the DM-only $N$-Body code \textsc{PKDGrav3}~\cite{Stadel2001}, which scales linearly $\mathcal{O}(N)$ in the number $N$ of simulated particles and accurately computes the forces between the particles due to its underlying Fast Multipole Method (FMM) implementation. Furthermore, the usage of \textsc{PKDGrav3} accelerated when run with graphics processing unit (GPU) support. 
\\
We ran 20 \textsc{PKDGrav3} simulations with different random seeds for a fixed flat $\nu \Lambda$CDM cosmological model with fiducial parameter values given by
\begin{eqnarray}
\boldsymbol{\theta}_\mathrm{fid} &=& \{ h, \Omega_m, \Omega_b, n_s, \sigma_8, T_\mathrm{CMB} \}\label{fid_params} \\
&=& \{ 0.6736, 0.3152, 0.0493, 0.9649, 0.811, 2.276\, \mathrm{K}\} \, , \nonumber
\end{eqnarray}
which corresponds to the parameter values listed in Table 7 in \cite{Planck2018I} inferred from \textit{Planck} CMB temperature, polarisation and lensing power spectra. We additionally use a degenerate neutrino-hierarchy with a fiducial neutrino mass sum of $\sum m_{\nu} = 0.06$ eV. \textsc{PKDGrav3} treats neutrinos as a relativistic fluid following the treatment described in \cite{Tram_2019}. Similarly, it handles general relativistic effects to linear order. To do so, \textsc{PKDGrav3} uses a lookup table of transfer functions that was generated prior to the runs using \textsc{Class} \citep{Lesgourgues2011,Blas2011} and transformed into the $N$-body gauge using the utilities from \textsc{$\nu$CO$N$CEPT} \citep{Dakin_2019}. The same lookup table was also used to generate the initial conditions of the simulations. 
Otherwise, we use a similar simulation setup as described in Section 3.1 in \cite{Sgier2021}, which we briefly summarise below. 

The pseudo-power spectra used for the covariance matrix estimation should accurately resolve angular scales from $\ell = 50$ to 1000, in order to cover all the multipole cuts for all considered auto- and cross-power spectra (see Table \ref{cl_table}). A sufficiently high resolution is achieved by choosing a simulation volume of $V_\mathrm{sim} = (1150\, \mathrm{Mpc}/h)^3$ with $N_p = 1024^3$ particles. This setup also provides a large enough simulation volume to incorporate large-scale perturbation modes and to minimise super-sample covariance effects. Such effects originate from missing modes larger than the simulated volume and their coupling to small-scale modes within the volume. This can lead to systematic errors on the power spectrum and introduces an additional term in the covariance matrix~\cite{Li2014} (we refer the reader to~\cite{Sgier2021} for a discussion thereof for the current simulation setup). We ran a total of 20 \textsc{PKDGrav3} simulations with different random seeds for the initial conditions set at an initial redshift of $z_\mathrm{init}=99$. Each simulation uses 100 timesteps between $z_\mathrm{init}$ and $z_\mathrm{final}=0.0$, with more timesteps for lower redshifts, and outputs a snapshot every second timestep. We subsequently store the 40 snapshot outputs between redshift $z=0.0$ and $1.75$ and post-process them using the lightcone generation code \textsc{UFalcon}, which is described in detail in the next section.
%This configuration leads to a walltime-runtime of $\sim 8$ hours per simulation with GPU-support on the Piz Daint computer cluster, which corresponds to $\sim 64$ node-hours.

\subsection{Lightcone Generation using \textsc{UFalcon}}
We use the Ultra-Fast Lightcone (\textsc{UFalcon}; \cite{Sgier2019,Sgier2021}) code to post-process the $N$-Body simulation outputs. The pipeline replicates the underlying simulation snapshot output six times in the $x$, $y$ and $z$ direction from redshift $z=0$ up to a final redshift of $1.75$ by concentrically stacking pixelised shells with a resolution of $N_\mathrm{side} = 1024$ at different redshifts of the replicated density field around $z=0$. For all full-sky maps considered in this analysis, the same underlying matter field is used to compute the cross-power spectra between different probes.

The stacked shells are subsequently weighted and projected in radial direction in order to generate full-sky maps for galaxy overdensity, CMB lensing convergence and weak lensing shear. Full-sky maps for the temperature anisotropies from the ISW effect are obtained by first interpolating the density field and then integrating the quantity along the line-of-sight. The corresponding equations to generate the full-sky maps of the probes $\Delta T_\mathrm{CMB}$, $\kappa_\mathrm{CMB}$, $\kappa_{n(z)}$ and $\delta_g$ are given in Appendix \ref{sec:lightcones}. Note that the \textsc{UFalcon} full-sky weak lensing convergence maps $\kappa_{n(z)}$ are converted to weak lensing shear maps using \cite{Wallis2017}
\begin{equation}
 _{_2} \gamma_{\ell m} = \frac{-1}{\ell (\ell + 1)} \sqrt{\frac{(\ell + 2)!}{(\ell - 2)!}} \kappa_{\ell m} \, .
 \label{kappa2gamma}
\end{equation}
We increase the number of quasi-independent realisations by a factor of 50 using the shell randomisation procedure implemented in \textsc{UFalcon}, leading to a total number of 1000 quasi-independent realisations. Random operations consisting of rotations by $90^\circ$, translations and parity flips are applied coherently to the particle positions or the overdensity field of all snapshots lying within the redshift shells bundled together to have a thickness of $\mathrm{d}z \sim 0.1$. Such a tessellation preserves the continuity of the density field within a shell bundle and therefore correlations between shells within such a bundle are not broken. Each shell-bundle of the lightcone is randomised using another random seed and are therefore uncorrelated. This procedure therefore avoids the repetition of the same structures of the density field along the line-of-sight and generates a new quasi-independent realisation of the lightcone. 

\subsection{\label{sec:cov_mat}Multi-Probe Covariance Matrix}

In this section, we describe the computation of the multi-probe covariance matrix used in our analysis.\\
\\
As the amplitude of the radiation perturbations is small at all times, they evolve linearly from the initial fluctuations and preserve Gaussianity. Therefore, the CMB signal can be characterised through the spherical harmonic power spectrum \cite{Dodelson2003}. The pseudo-power spectra estimates for the probe-combinations TT, TE and EE are obtained using Gaussian fields as follows: First we use the \textsc{Healpix} routine \texttt{synfast} to generate 1000 synthetic Gaussian CMB temperature and polarisation maps from theoretical \textsc{Class} power spectra. We then add the associated HMHD maps (see Section \ref{sec:CMB}) as a noise estimate to each synthetic CMB map, apply the provided temperature and polarisation masks and compute the power spectra using \texttt{anafast}. We follow \cite{Nicola2016} and do not randomise the HMHD maps for each realisation, as this would break significant correlations within the noise map. This approach slightly underestimates the noise for the CMB temperature and polarisation maps. However, the noise in the CMB temperature and polarisation power spectra are dominated by cosmic variance on the angular scales considered. We therefore do not expect this effect to have a significant impact on our results.

In general, we estimate all the remaining pseudo-auto- and cross-power spectra based on the probes $X,Y \in \{\Delta T_\mathrm{ISW}, \kappa_\mathrm{CMB}, \delta_g, \gamma\}$ from the 1000 quasi-independent realisations, as long as the product of their window functions $W^{X} (\chi(z)) \, W^{Y} (\chi(z))$ lie within the redshift range of $z=0.0-1.75$ covered by the \textsc{UFalcon} lightcone. Therefore, a hybrid approach for the estimation of the CMB lensing auto-power spectrum has to be adopted. Weak lensing of the CMB takes into account variations in the gravitational potential from recombination at $z_\ast \approx 1100$ to $z=0$. Even though the CMB lensing convergence window function peaks at $z \sim 2$, higher redshifts up to $z \sim 11$ still significantly contribute to the CMB lensing signal \cite{Carbone2008}. We thus use a hybrid approach for the estimation of the $\kappa_\mathrm{CMB}$ pseudo-auto power spectrum, covering a large redshift range: The simulation-based \textsc{UFalcon} full-sky map $\kappa_\mathrm{CMB}^\mathrm{UFalcon}$ covers the lower redshift range $z=0.0 - 1.75$, which fully includes the non-Gaussianities from structure formation. For redshifts within the range $z=1.75 - z_\ast$, we approximate the CMB lensing signal with Gaussian statistics. We generate a synthetic Gaussian map $\kappa_\mathrm{CMB}^\mathrm{Gaussian}$ using \texttt{synfast} applied to a theoretical \textsc{Class} prediction for $z=1.75 - z_\ast$ and for the fiducial cosmology $\boldsymbol{\theta}_\mathrm{fid}$. The resulting CMB lensing convergence map is given as the sum of the two maps $\kappa_\mathrm{CMB} = \kappa_\mathrm{CMB}^\mathrm{UFalcon} + \kappa_\mathrm{CMB}^\mathrm{Gaussian}$. We thereby approximate that the two redshift ranges, $z=0.0 - 1.75$ and $z=1.75 - z_\ast$, are uncorrelated. All full-sky maps corresponding to one realisation are based on the same underlying matter density field, including the CMB temperature anisotropies from the ISW effect $\Delta T_\mathrm{ISW}$. This enables us to compute cross-correlations between the CMB temperature anisotropies and tracers of the LSS, even though the CMB temperature auto-power spectra are obtained independently from Gaussian fields.

We subsequently add realistic noise maps to the simulated maps, which is done as follows: For the galaxy overdensity fields associated to the LOWZ and CMASS sample, we follow the procedure outlined in Section \ref{sec:cls_data}. We therefore randomise the positions of all galaxies inside the corresponding mask and add one resulting pixelised Poisson shot noise map realisation to each simulated galaxy overdensity map. For the weak lensing shear fields, we randomly rotate each galaxy in each KiDS-1000 tomographic bin according to Eq. (\ref{sn_map}) to generate shape noise maps. We then add one random shape noise realisation to each simulated weak lensing shear map. We use the lensing simulation package provided by the \textit{Planck} collaboration and used for the \textit{Planck} CMB lensing analysis in \cite{Planck2018VIII} to estimate noise maps for the CMB lensing convergence. This package consists of the 300 \texttt{FFP10} simulations of observed, minimum variance-estimated spherical harmonic coefficients (without mean-field subtraction), together with the spherical harmonic coefficients of the input CMB lensing convergence to each simulation. The spherical harmonic coefficients for the noise are obtained by subtracting the mean-field coefficients and the input coefficients. We then cycle through the 300 simulations and add one noise realisation to each CMB lensing convergence map. As pointed out in \cite{Nicola2017}, this approach only represents an approximation as it assumes linearity in signal and noise. The differences between the estimated noise power spectra and the true spectra have been found to be at most $3.5\%$, which is an acceptable accuracy for the computation of the covariance matrix. Furthermore, simulated CMB lensing noise estimates mainly consist of Gaussian noise from the disconnected part of the CMB temperature 4-point function and therefore depend on the cosmological model chosen. The \texttt{FFP10} simulations used in this work use the same cosmological model parameters as in the previous \texttt{FFP8} release \cite{Planck2018III}, which are given in \cite{Planck2015XII}. Since the fiducial cosmological parameters are very close to the parameters used for the \texttt{FFP10} simulations, we expect the error due to using different cosmological models for noise and $N$-Body simulation to be small.

The auto- and cross-power spectra between the different probes are obtained using \texttt{anafast}, after application of the corresponding combined mask (see Table \ref{cl_table}). We further correct the resulting spectra for the \textsc{Healpix} window function and subtract the probe-specific shot noise contribution, which we briefly describe here. Particle simulations contain a certain amount of Poisson shot noise due to the finite number of particles used. This noise contribution has to be subtracted from the derived power spectra in order to obtain the cosmological signal from the simulations. We estimate the noise for the \textsc{UFalcon} spherical harmonic power spectra by replacing the non-linear power spectrum in Eq. (\ref{cl_limber}) with the shot noise contribution to the matter power spectrum as $P^{\mathrm{nl}}_{\delta \delta} \rightarrow V_\mathrm{sim} / N_p$. A comparison performed in \cite{Sgier2021} showed that the shot noise estimation using the Limber approximation and using the Born approximation (i.e. by constructing a shot noise-lightcone, see Appendix B in~\cite{Sgier2021}) provided results of the same order of magnitude. We therefore resort to the faster method of using the Limber approximation for the shot noise estimation. More details about the Born approximation and the lightcone construction performed in \textsc{UFalcon} is given in Appendix \ref{sec:lightcones}.

\begin{figure*}[htbp!]
\centering
\includegraphics[width=0.8\paperwidth]{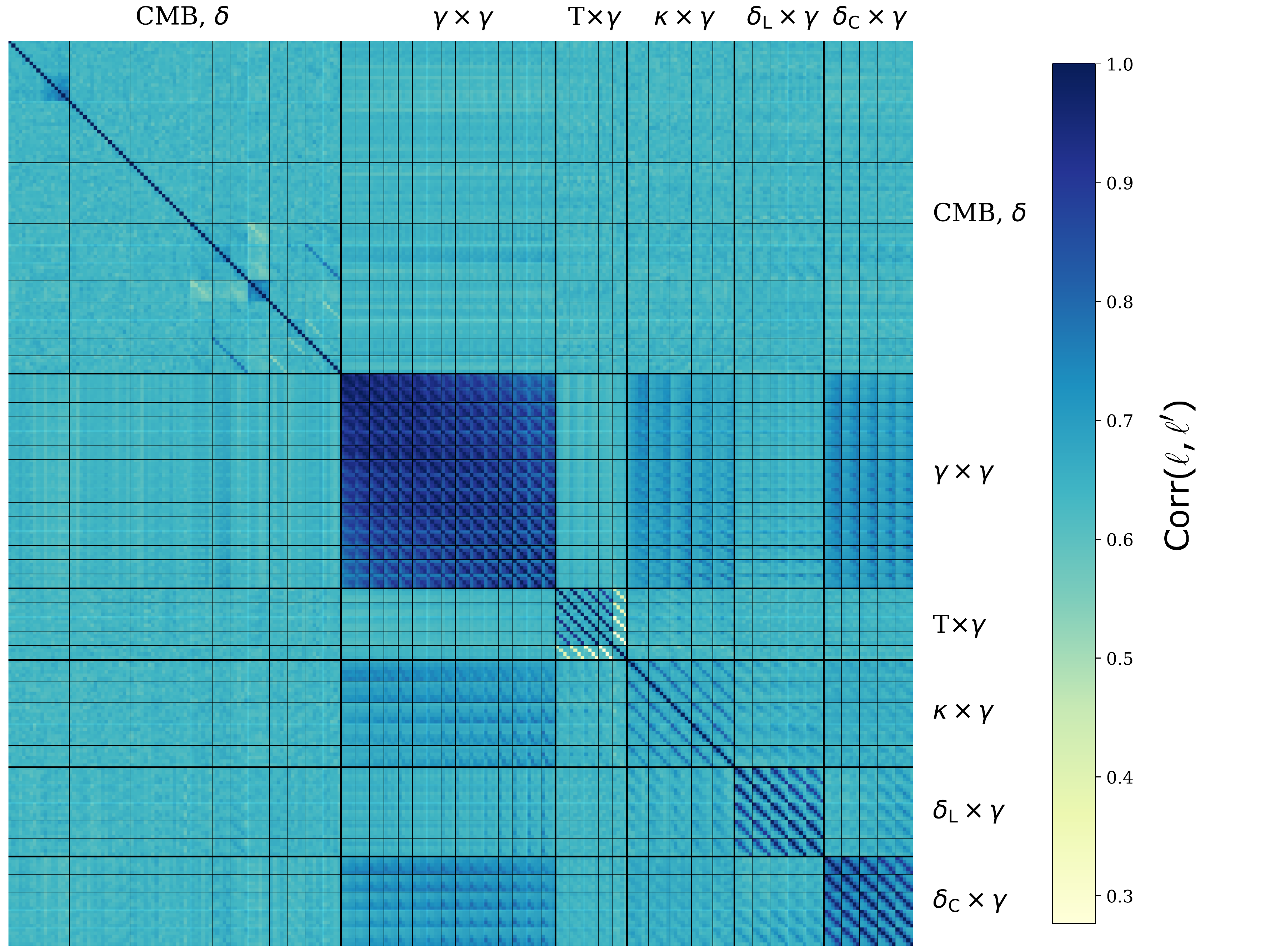}
\caption{Multi-probe covariance correlation matrix for the spherical harmonic power spectra estimated from 1000 quasi-independent \textsc{UFalcon} realisations: Each spectrum is logarithmically binned according to Table \ref{cl_table}. Each block constitutes the estimated spectra  $\hat{C}_\ell^{X,Y}$, where $X,Y$ reads (in this order): CMB temperature/polarisation and LOWZ/CMASS galaxy samples. $\boldsymbol{\mathrm{CMB}\, ,\delta}:$  TT, EE, TE, $\kappa\kappa$, $\delta_\mathrm{L}\delta_\mathrm{L}$, $\delta_\mathrm{C}\delta_\mathrm{C}$, $\kappa$T, $\delta_\mathrm{C}$T, $\delta_\mathrm{L}$T, $\kappa\delta_\mathrm{L}$, $\kappa\delta_\mathrm{C}$; weak lensing shear bins $\boldsymbol{\gamma \times \gamma}:$ 11, 12, 13, 14, 15, 22, 23, 24, 25, 33, 34, 35 , 44, 45, 55; CMB temperature and weak lensing shear bins $\boldsymbol{\mathrm{T} \times \gamma}:$ T1, T2, T3, T4, T5; CMB lensing convergence and weak lensing shear bins $\boldsymbol{\kappa \times \gamma}:$ $\kappa$1 $\kappa$2, $\kappa$3, $\kappa$4, $\kappa$5; LOWZ galaxy sample and weak lensing shear bins $\boldsymbol{\delta_\mathrm{L} \times \gamma}:$ $\delta_\mathrm{L}$1, $\delta_\mathrm{L}$2, $\delta_\mathrm{L}$3, $\delta_\mathrm{L}$4, $\delta_\mathrm{L}$5 and CMASS galaxy sample and weak lensing shear bins $\boldsymbol{\delta_\mathrm{C} \times \gamma}:$ $\delta_\mathrm{C}$1, $\delta_\mathrm{C}$2, $\delta_\mathrm{C}$3, $\delta_\mathrm{C}$4, $\delta_\mathrm{C}$5.}
\label{corr_cross_plot}
\end{figure*}

\begin{figure}[htbp!]
\centering
\includegraphics[width=\linewidth]{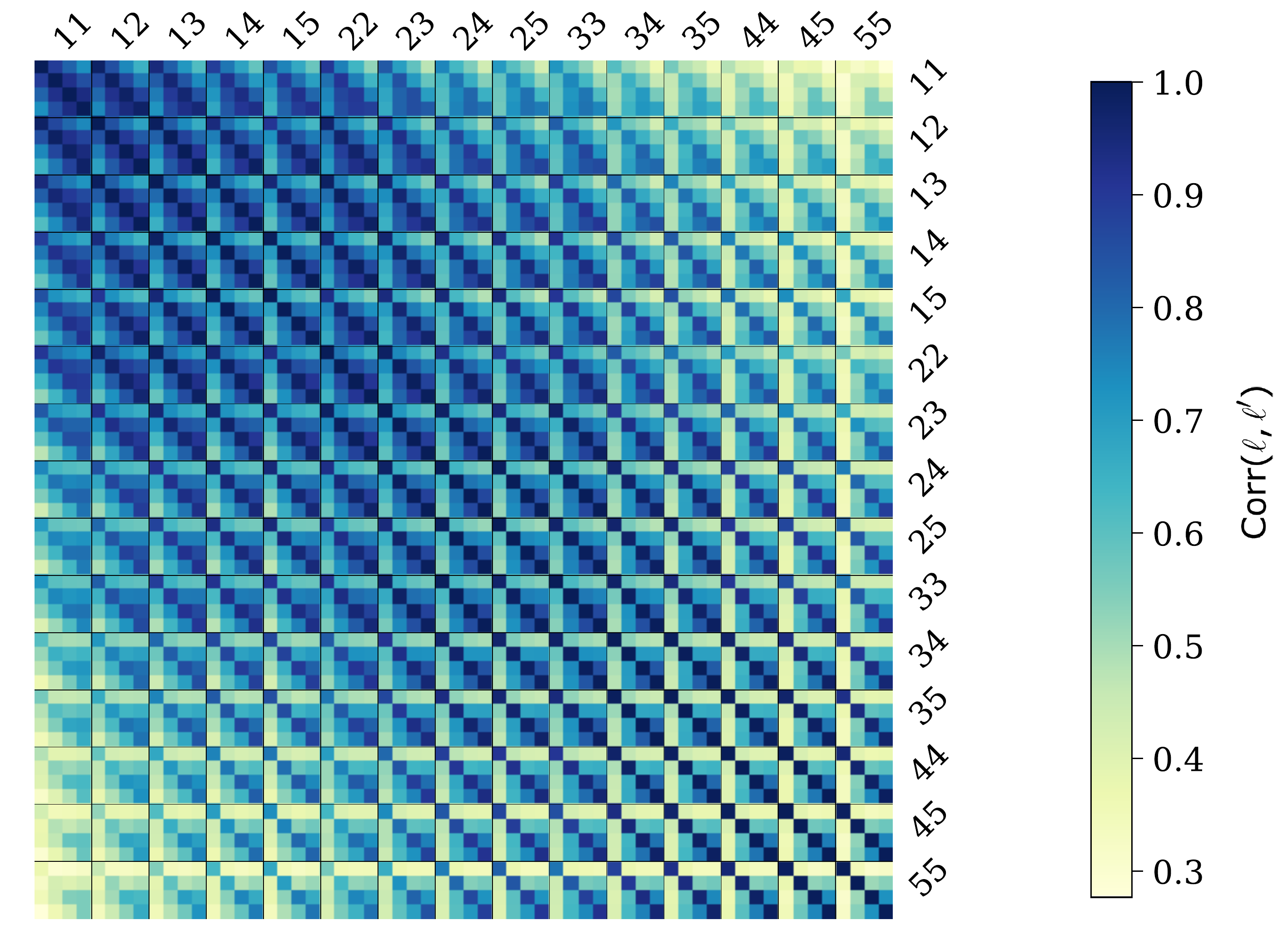}
\caption{Covariance correlation matrix for the spherical harmonic power spectra estimated from 1000 quasi-independent weak lensing shear \textsc{UFalcon} realisations (block $\boldsymbol{\gamma \times \gamma}$ in Figure \ref{corr_cross_plot}). Each spectrum is logarithmically binned according to Table \ref{cl_table}. Spectra based on tomographic bins which are closer together in redshift are more strongly correlated than spectra based on bins that are farther appart.}
\label{figure_corr_kids}
\end{figure}

The final multi-probe covariance matrix used in this analysis is computed from $N_\mathrm{sim} = 1000$ quasi-independent \textsc{UFalcon} realisations using the sample covariance estimator
\begin{equation}
\hat{\Sigma}_{\ell, \ell '} = \frac{1}{N_s - 1} \sum_{k = 1}^{N_s} \left[\hat{C}_{k}^{ij}(\ell) - \bar{C}_{k}^{ij}(\ell) \right] \left[\hat{C}_{k}^{i'j'}(\ell') - \bar{C}_{k}^{i'j'}(\ell') \right] \, ,
\label{cov2}
\end{equation}
where the different cosmological probes are denoted by $i,j,i',j' \in \{ T_\mathrm{ISW}, \kappa_\mathrm{CMB}, \delta_g, \gamma_1 \}$ and $\bar{C}_{k}^{ij}(\ell)$ is the mean over all realizations. The strength of the correlation between the power spectra at different multipoles can be quantified through the corelation coefficient
\begin{equation}
\mathrm{Corr} (\ell, \ell ') = \frac{\hat{\Sigma}_{\ell, \ell '}}{\sqrt{\hat{\Sigma}_{\ell, \ell} \hat{\Sigma}_{\ell', \ell '}}} \, ,
\end{equation}
which is normalised to unity for $\ell = \ell'$ and implies strong correlation for $\mathrm{Corr} \rightarrow 1$, no correlation for $\mathrm{Corr} \rightarrow 0$ or strong anti-correlation for $\mathrm{Corr} \rightarrow -1$. In Figure \ref{corr_cross_plot} we show the correlation coefficient of our full joint covariance matrix used for our analysis. The correlation coefficient matrix is built from six blocks, each of them labelled by the corresponding cosmological probes used to estimate the auto- and cross-power spectra. Each power spectrum estimate is logarithmically binned within its associated multipole range, which is detailed in Table \ref{cl_table}. The block corresponding to the auto- and cross-correlations between the five weak lensing shear bins is additionally shown in Figure \ref{figure_corr_kids}.


\section{\label{sec:inference}Statistical Inference}
In this section, we describe the underlying cosmological model, the setup for the statistical analysis and the resulting cosmological parameter constraints from the combined fit to the 10 auto- or 46 auto- and cross-spherical harmonic power spectra described in Section \ref{sec:cls_data}.

\subsection{Cosmological Model}
Using the likelihood given by Eq.~\eqref{likelihood}, we derive parameter constraints in the context of a flat ($\Omega_k = 0$) $\Lambda$CDM cosmological model with a dark energy equation of state parameter $w = -1$. Our baseline parameters therefore consist of $\{h, \Omega_\mathrm{m}, \Omega_\mathrm{b}, n_s, \sigma_8\}$, where $h$ is the dimensionless Hubble parameter, $\Omega_m$ and $\Omega_b$ are the fractional total matter and baryon density today, respectively, $n_s$ represents the scalar spectral index, $\sigma_8$ is the root mean square of the linear matter density fluctuations in spheres of radius $R = 8 h^{-1}\mathrm{Mpc}$. We further include massive neutrinos, which are parametrised by the effective number of neutrinos in the relativistic limit $N_\mathrm{eff} = 3.044$ \cite{Froustey_2020,Bennett_2021} and an effective temperature of $T_\nu / T_\gamma = 0.71611$, with $T_\gamma$ denoting the photon temperature. The squared mass differences between the neutrino mass eigenstates $m_{\nu,1}$, $m_{\nu,2}$ and $m_{\nu,3}$ have been tightly measured with neutrino oscillation experiments~\cite{10.1093/ptep/ptaa104}, allowing both a \textit{normal} or \textit{inverted} neutrino hierarchy, depending on the sign of $\Delta m_{\nu,13}^2$~\cite{Lesgourgues:2006nd}. Similarly as the official \textit{Planck} 2018 analysis~\cite{Aghanim:2018eyx}, we choose here a degenerate neutrino mass hierarchy. For large neutrino masses, the difference between the neutrino hierarchies is small. For small neutrino masses, it has been shown that current cosmological surveys cannot yet determine the neutrino mass hierarchy~\cite{Choudhury:2018byy,Vagnozzi:2017ovm,Gerbino:2016ehw,Gariazzo:2018pei,Jimenez:2010ev,Archidiacono:2020dvx}. A total neutrino mass sum below $\sum m_\nu<0.1$ eV though would rule out the inverted hierarchy. The total fractional matter density today is thus given by $\Omega_\mathrm{m} = \Omega_\mathrm{c} + \Omega_\mathrm{b} + \Omega_\nu$, where $\Omega_\mathrm{c}$ and $\Omega_\nu$ denote the fractional cold dark matter and massive neutrino density today respectively.

\subsection{Parameter Inference Setup}
In this section, we describe the setup for the statistical inference used to derive constraints from the combined fit of the 10 auto- or 46 auto- and cross-spherical harmonic power spectra described in Section \ref{sec:cls_data}. Assuming a Gaussian likelihood function is a good approximation for the case when an analytical covariance matrix is used. However, since our analysis is based on a simulated covariance matrix, we need to take into account the intrinsic uncertainty of the latter. The resulting likelihood is no longer Gaussian distributed, but is instead given by a modified version of a t-distribution~\cite{Sellentin2016, Sellentin2017} \begin{equation}
\mathcal{L} (D | \theta, \hat{\Sigma}, N_s) \propto \left[ 1 + \frac{\left(\boldsymbol{C}_{\ell}^\mathrm{diff}\right)^{T} \hat{\Sigma}^{-1}\boldsymbol{C}_{\ell}^\mathrm{diff}}{N_s -1} \right]^{- \frac{N_s}{2}} \, ,
\label{likelihood}
\end{equation}
where $N_s = 1000$ is the number of quasi-independent power spectra realisations used for the covariance matrix estimation (described in Section \ref{sec:cov_mat}). Note that we assume the covariance matrix $\hat{\Sigma}$ to be cosmology independent, i.e. the $N$-Body simulations used for the calculation of the covariance matrix are based on the fixed fiducial parameter values $\boldsymbol{\theta}_\mathrm{fid}$, outlined in Eq.~\eqref{fid_params}. It has been shown that this is a valid approximation~\cite{Kodwani:2018uaf}. 

We have defined the difference between the observed power spectrum vector and its corresponding analytical prediction as
\begin{equation}
    \boldsymbol{C}_{\ell}^\mathrm{diff} = \boldsymbol{\hat{\tilde{C}}}_{\ell}^\mathrm{obs} - \boldsymbol{\tilde{C}}_\ell^\mathrm{theory}\, ,
\end{equation}
where we used  Eq.~(\ref{cl_pseudo}) to calculate the analytical pseudo-spherical harmonic power spectrum vector $\boldsymbol{\tilde{C}}_\ell^\mathrm{theory}$. The order of the cosmological probes entering the observed auto-power spectrum vector is defined as follows
\begin{eqnarray*}
XX &=& \left( \mathrm{TT}\,\, \mathrm{EE}\,\, \kappa \kappa\,\, \delta_\mathrm{L}\delta_\mathrm{L}\,\, \delta_\mathrm{C}\delta_\mathrm{C}\,\, \gamma_\mathrm{1}\gamma_\mathrm{1}\,\, \gamma_\mathrm{2}\gamma_\mathrm{2}\,\, \gamma_\mathrm{3}\gamma_\mathrm{3}\,\, \gamma_\mathrm{4}\gamma_\mathrm{4}\,\, \gamma_\mathrm{5}\gamma_\mathrm{5} \right) \, .
\label{datavec_auto}
\end{eqnarray*}
The order adopted for the 46 auto- and cross-correlations between the probes entering the combined observed power spectrum vector is given by
\begin{eqnarray*}
XY &=& \left( \mathrm{TT}\,\,\mathrm{EE}\,\,\mathrm{TE}\,\,\kappa \kappa\,\,\delta_\mathrm{L}\delta_\mathrm{L}\,\,\delta_\mathrm{C}\delta_\mathrm{C}\,\, \kappa\mathrm{T}\,\, \delta_\mathrm{C}\mathrm{T}\,\, \delta_\mathrm{L}\mathrm{T}\,\, \kappa\delta_\mathrm{L}\,\, \kappa\delta_\mathrm{C} \right. \\
&& \gamma_\mathrm{1}\gamma_\mathrm{1}\,\, \gamma_\mathrm{1}\gamma_\mathrm{2}\,\, \gamma_\mathrm{1}\gamma_\mathrm{3}\,\, \gamma_\mathrm{1}\gamma_\mathrm{4}\,\, \gamma_\mathrm{1}\gamma_\mathrm{5}\,\, \gamma_\mathrm{2}\gamma_\mathrm{2}\,\, \gamma_\mathrm{2}\gamma_\mathrm{3}\,\, \gamma_\mathrm{2}\gamma_\mathrm{4}\,\, \gamma_\mathrm{2}\gamma_\mathrm{5}\\
&& \gamma_\mathrm{3}\gamma_\mathrm{3}\,\, \gamma_\mathrm{3}\gamma_\mathrm{4}\,\, \gamma_\mathrm{3}\gamma_\mathrm{5}\,\, \gamma_\mathrm{4}\gamma_\mathrm{4}\,\, \gamma_\mathrm{4}\gamma_\mathrm{5}\,\, \gamma_\mathrm{5}\gamma_\mathrm{5}\,\, \mathrm{T}\gamma_\mathrm{1}\,\, \mathrm{T}\gamma_\mathrm{2}\,\, \mathrm{T}\gamma_\mathrm{3}\\
&& \mathrm{T}\gamma_\mathrm{4}\,\, \mathrm{T}\gamma_\mathrm{5}\,\, \kappa\gamma_\mathrm{1}\,\, \kappa\gamma_\mathrm{2}\,\, \kappa\gamma_\mathrm{3}\,\, \kappa\gamma_\mathrm{4}\,\, \kappa\gamma_\mathrm{5}\,\, \delta_\mathrm{L}\gamma_\mathrm{1}\,\, \delta_\mathrm{L}\gamma_\mathrm{2}\,\, \delta_\mathrm{L}\gamma_\mathrm{3}\\
&& \left. \delta_\mathrm{L}\gamma_\mathrm{4}\,\, \delta_\mathrm{L}\gamma_\mathrm{5}\,\, \delta_\mathrm{C}\gamma_\mathrm{1}\,\, \delta_\mathrm{C}\gamma_\mathrm{2}\,\, \delta_\mathrm{C}\gamma_\mathrm{3}\,\, \delta_\mathrm{C}\gamma_\mathrm{4}\,\, \delta_\mathrm{C}\gamma_\mathrm{5} \right) \, .
\label{datavec_cross}
\end{eqnarray*}
We use the above described setup to perform Markov Chain Monte Carlo (MCMC) analyses using the \textsc{emcee}~\footnote{https://emcee.readthedocs.io/en/stable/} affine invariant ensemble sampler \cite{Goodman2010} using 96 walkers per run. We sample the combined likelihood given by Eq. (\ref{likelihood}) by varying the base cosmological parameters $\{h, \Omega_\mathrm{m}, \Omega_\mathrm{b}, n_s, \sigma_8\}$, the sum of the neutrino masses $\sum m_\nu$, as well as the nuisance parameters $\{m_\mathrm{T}\, ,m_\kappa\, ,m_{\gamma 1}\, , m_{\gamma 2}\, , m_{\gamma 3}\, , m_{\gamma 4}\, , m_{\gamma 5}\, , b_\mathrm{LOWZ}\, , b_\mathrm{CMASS}\}$, depending on the probes included. We use broad, flat priors for the cosmological parameters and for all nuisance parameters for our fiducial MCMC run.

The prior ranges for the varied parameters are given in the second row of Table \ref{prior_post_table}. For all our MCMC runs we set the optical depth to reionisation to the updated value of $\tau_\mathrm{reion}=0.051$ from the intermediate \textit{Planck} results~\cite{refId0} after the official \textit{Planck} 2018 results. We further set the mean temperature of the CMB to $\mathrm{T}_\mathrm{CMB} = 2.726$ and the primordial helium fraction to $\mathrm{YHe} = 0.2454$, which correspond to the parameter values inferred in \cite{Planck2018I} from \textit{Planck} CMB temperature, polarisation and lensing power spectra. For our analysis, we choose not to vary the IA amplitude and set it to the best fit value of $A_\mathrm{IA} = 0.973$, obtained from the KiDS-1000 Band Power analysis \cite{Asgari2021}.

Prior to our joint analysis combining the probes, we compute parameter constraints for the \textit{Planck} 2018, BOSS DR 12 and KiDS-1000 weak lensing data sets separately as consistency tests. The results are shown and discussed in Appendix \ref{sec:consistency}. Figure~\ref{figure_planck_consistency_TTTEEE} shows the parameter constraints for the probes TT, TE, EE obtained from our map-based CMB analysis, compared to when using the public MCMC package \textsc{CosmoMC}~\cite{PhysRevD.66.103511} with matching multipole cuts. We find in general a good agreement between both constraints for all sampled parameters. Due to the conservative multipole cuts chosen in our map-based CMB analysis, we obtain a higher value for $\Omega_m$ and therefore a lower value for $\Omega_\Lambda$ compared to the results from the official \textit{Planck} 2018 analysis (see Table~2 in~\cite{Aghanim:2018eyx}).
\section{\label{sec:constraints}Cosmological Constraints}
\subsection{Results from the Combined Analysis}
\begin{figure*}[htbp!]
\centering
\includegraphics[width=0.8\paperwidth]{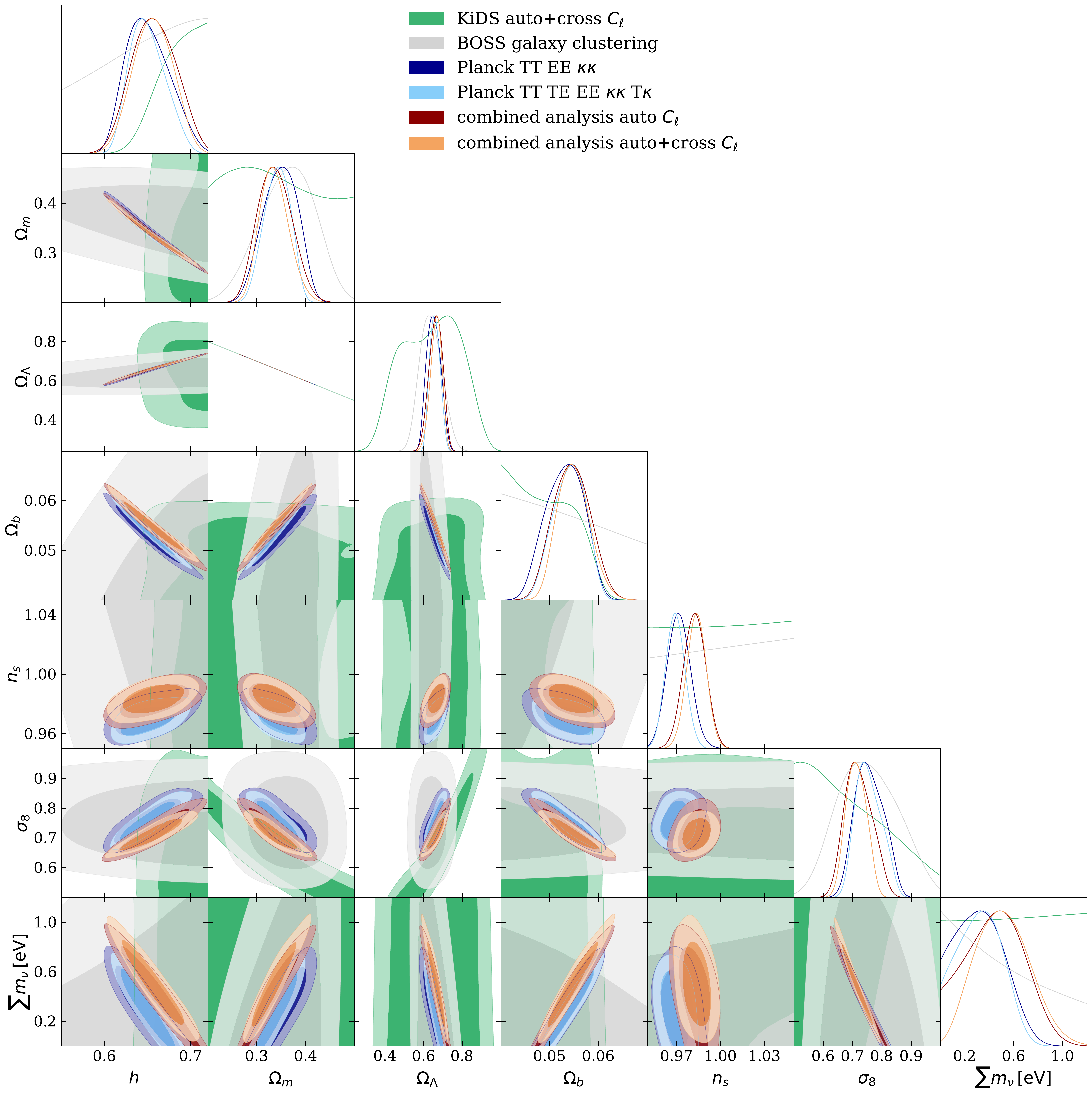}
\caption{Cosmological parameter constraints derived from the combined analysis using the 10 auto-spectra (in dark red) and using the 46 auto+cross-spectra (in orange), marginalised over all nuisance parameters described in Table \ref{prior_post_table}. The constraints are compared to the CMB-only results using the auto-spectra alone (in dark blue) as well as using the auto+cross-spectra (in light blue), the weak lensing shear (in green) and the galaxy clustering results (in grey). The inner (outer) contours show the 68\% (95\%) confidence limits.}
\label{figure_results_main}
\end{figure*}

In Figure \ref{figure_results_main}, we show the %fiducial 
%combined 
cosmological parameter constraints obtained from the combination of the 10 auto-power spectra and from the 46 auto- and cross-power spectra, which are shown in Figures \ref{cls_data_A}, \ref{cls_data_B} and \ref{cls_data_C}. The associated prior ranges, means and 68\% confidence limits (CL) are given in Table \ref{prior_post_table}. As depicted in the figure, the constraining power from including all auto-spectra (denoted “combined analysis auto $C_\ell$”) mainly comes from the TT, EE and $\kappa \kappa$ spectra. The resulting auto-$C_\ell$ parameter constraints only slightly exhibit smaller confidence limits than the TT, EE, $\kappa \kappa$ results. Adding cross-spectra between the different probes (denoted “combined analysis auto + cross $C_\ell$”) increases the constraining power and leads to smaller contours compared to the results using auto-$C_\ell$'s only, which is consistent with the findings of ~\cite{Sgier2021}. We note here that the observed increase in constraining power by adding the cross-correlations for the combined analysis is of the same order as for the addition of the TE and T$\kappa$ spectra for the CMB-only analysis.

\begin{table*}
\caption{Parameter priors and constraints for the fiducial joint analysis combining CMB temperature and polarisation, CMB lensing, BOSS spectroscopic tracers and KiDS-1000 weak lensing shear. The priors shown in the second column are given as uniform priors \(\mathcal{U}\) with lower and upper bound. The third to sixth column give the mean and the 68\% CL bounds for the cosmological and nuisance parameters for different probe-combinations.}
\begin{tabular}{ cccccc }
 \hline
 \\
 Parameter [unit] & Prior & CMB auto \quad\quad & CMB auto + cross\quad\quad & all probes auto\quad\quad & all probes auto + cross\\
 \\
 \hline
  \\
$h$ &\(\mathcal{U}\) (0.3,1.1) & $0.651^{+0.022}_{-0.029}$ & $0.650^{+0.018}_{-0.024}$ & $0.658^{+0.025}_{-0.025}$ & $0.656^{+0.022}_{-0.022}$\\[0.1cm]
$\Omega_\mathrm{m}$ & \(\mathcal{U}\)(0.1, 0.7) & $0.348^{+0.038}_{-0.033}$ & $0.342^{+0.029}_{-0.025}$ & $0.337^{+0.034}_{-0.034}$ & $0.336^{+0.026}_{-0.032}$\\[0.1cm]
$\Omega_\Lambda$ & - & $0.652^{+0.033}_{-0.038}$ & $0.658^{+0.025}_{-0.029}$ & $0.663^{+0.034}_{-0.034}$ & $0.664^{+0.032}_{-0.026}$\\[0.1cm]
$\Omega_\mathrm{b}$ & \(\mathcal{U}\)(0.02, 0.08) & $0.0531^{+0.0042}_{-0.0037}$ & $0.054^{+0.0033}_{-0.0033}$ & $0.0545^{+0.0037}_{-0.0037}$ & $0.0548^{+0.0030}_{-0.0036}$\\[0.1cm]
$n_s$ & \(\mathcal{U}\)(0.4, 1.2) & $0.9714^{+0.0078}_{-0.0078}$ & $0.9691^{+0.0065}_{-0.0065}$ & $0.9822^{+0.0074}_{-0.0075}$ & $0.9835^{+0.0065}_{-0.0065}$\\[0.1cm]
$\sigma_8$ & \(\mathcal{U}\)(0.4, 1.2) & $0.758^{+0.046}_{-0.054}$ & $0.755^{+0.038}_{-0.051}$ & $0.721^{+0.042}_{-0.052}$ & $0.713^{+0.034}_{-0.034}$\\[0.1cm]
$S_8$ & - & $0.813^{+0.032}_{-0.032}$ & $0.804^{+0.025}_{-0.025}$ & $0.761^{+0.022}_{-0.022}$ & $0.754^{+0.016}_{-0.016}$\\[0.1cm]
$\sum m_\nu\,\, [\mathrm{eV}]$ & \(\mathcal{U}\)(0.0, 5.0) &$0.27^{+0.23}_{-0.23}$ & $0.29^{+0.24}_{-0.2}$ & $0.46^{+0.23}_{-0.23}$ & $0.51^{+0.21}_{-0.24}$\\
\\
\hline \\
$m_\mathrm{T}$ & \(\mathcal{U}\)(-0.2,0.2) & $0.003^{+0.0032}_{-0.0032}$ & $0.0035^{+0.0028}_{-0.0027}$ & $0.0045^{+0.0032}_{-0.0032}$ & $0.0051^{+0.0027}_{-0.0027}$\\[0.1cm]
$m_\kappa$ & \(\mathcal{U}\)(-0.5,0.5) & $0.032^{+0.029}_{-0.029}$ & $0.025^{+0.026}_{-0.026}$ & $0.069^{0.027}_{-0.027}$ & $0.072^{+0.026}_{-0.026}$\\[0.1cm]
$b_\mathrm{LOWZ}$ & \(\mathcal{U}\)(0.0, 10.0) & - & - & $2.19^{+0.16}_{-0.16}$ & $2.24^{+0.14}_{-0.14}$\\[0.1cm]
$b_\mathrm{CMASS}$ & \(\mathcal{U}\)(0.0, 10.0) & - & - & $2.32^{+0.15}_{-0.15}$ & $2.29^{+0.10}_{-0.11}$\\[0.1cm]
$m_{\gamma 1}$ & \(\mathcal{U}\)(-0.2,0.2) & - & - & $0.001^{+0.041}_{-0.041}$ & $0.013^{+0.043}_{-0.030}$\\[0.1cm]
$m_{\gamma 2}$ & \(\mathcal{U}\)(-0.2,0.2) & - & - & $0.008^{+0.046}_{-0.039}$ & $-0.003^{+0.033}_{-0.033}$\\[0.1cm]
$m_{\gamma 3}$ & \(\mathcal{U}\)(-0.2,0.2) & - & - & $-0.001^{+0.041}_{-0.041}$ & $-0.011^{+0.027}_{-0.038}$\\[0.1cm]
$m_{\gamma 4}$ & \(\mathcal{U}\)(-0.2,0.2) & - & - & $-0.040^{+0.018}_{-0.032}$ & $-0.040^{+0.017}_{-0.028}$\\[0.1cm]
$m_{\gamma 5}$ & \(\mathcal{U}\)(-0.2,0.2) & - & - & $-0.029^{+0.024}_{-0.043}$ & $-0.032^{+0.023}_{-0.038}$\\
\\
 \hline
\end{tabular}
\label{prior_post_table}
\end{table*}

The reduced-$\chi^2$ statistics (and p-values) for the performed fits are given by 1.07 (p=0.35), 1.12 (p=0.28), 1.25 (p=0.2) and 1.71 (p=0.12) for CMB auto-, CMB auto- and cross-, all probes auto- and all probes auto- and cross-correlations, respectively. 
Using a threshold value of $\alpha=0.01$ for the p-value, we obtain good fits for all four analyses described above. Here, we use the DES convention for the threshold value $\alpha=0.01$, as described in Appendix D in~\cite{DES:2021wwk}. Using a more conservative threshold value of $\alpha=0.1$, we would also obtain good fits for all the combinations considered. The fit for the combination of all probes is less good compared to than using the CMB data alone. Including the cross-correlations further compromises the goodness of the fit. However, our results for the joint fit still represent a good fit that helps in reducing the tension between the individual data sets. We mainly attribute this to the inclusion of the nuisance parameters, which are self-calibrated in the analysis alongside varying the cosmological parameters. Combining all auto-correlations leads to a shift from the CMB-only towards the weak lensing shear-only constraints. The inclusion of the galaxy overdensity and weak lensing shear data results in higher values for $h$ and $n_s$, whereas the parameters $\Omega_\mathrm{m}$ and $\Omega_\mathrm{b}$ favour lower values compared to the CMB-only results. Such parameter shifts are in agreement with the findings from other works (e.g. \cite{Nicola2016,Nicola2017} and references therein).

\subsection{\label{sec:nuisance}Constraints on the Calibration Parameters}
\begin{figure*}[htbp!]
\centering
\includegraphics[width=0.8\paperwidth]{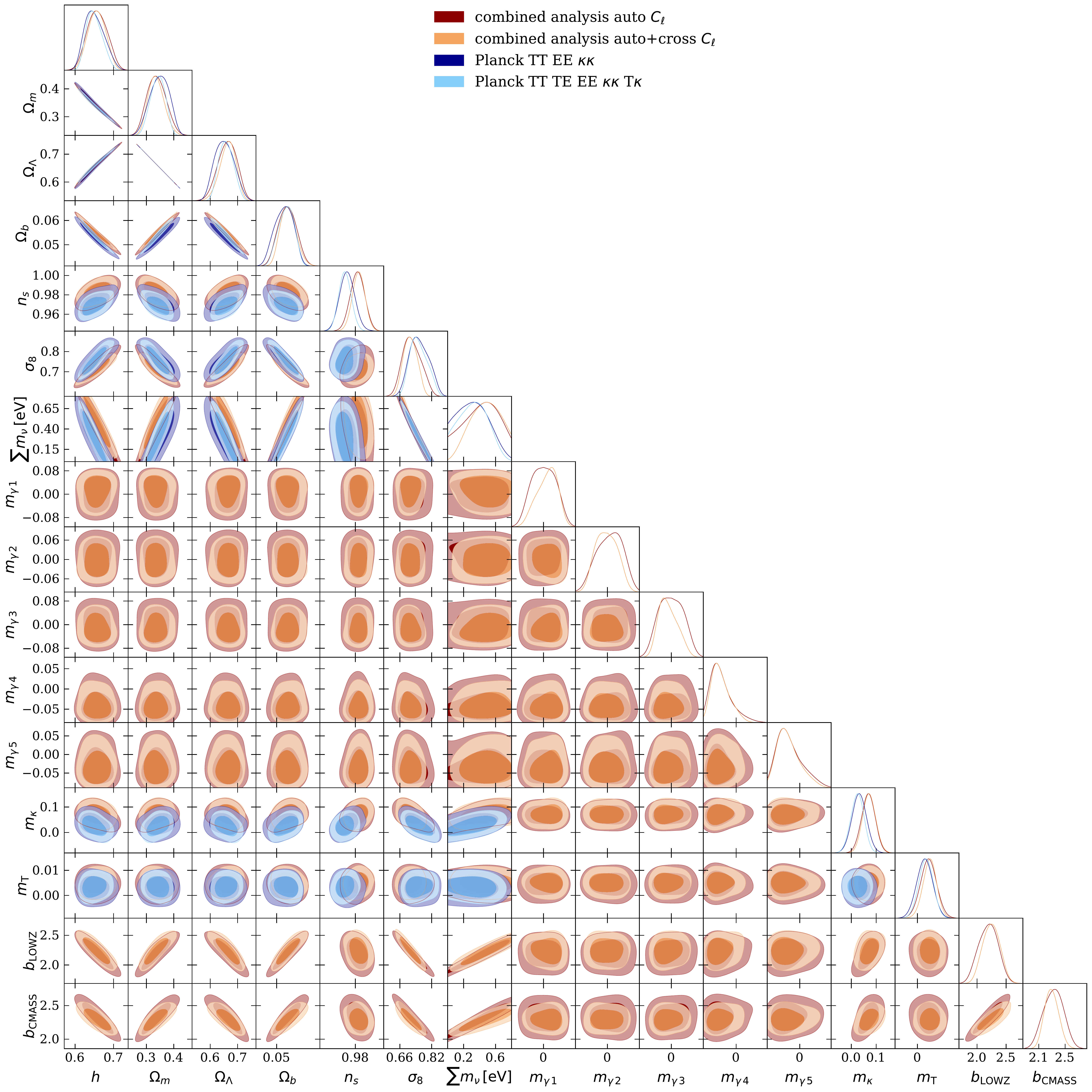}
\caption{Constraints for the 6 cosmological and the 9 nuisance parameters derived from the combined analysis using the 10 auto-spectra (dark red) and using the 46 auto- + cross-spectra (orange). The inner (outer) contours show the 68\% (95\%) confidence limits. \label{figure_results_main_nuisance}}
\end{figure*}

As explained in Section~\ref{sec:systematics}, we vary seven multiplicative bias and two linear galaxy bias parameters in our combined analysis, which allows for a calibration of all probes at the map level. Therefore, varying these nine nuisance parameters allows for a self-calibration of the auto- and the cross-power spectra during the inference process. 
In Figure~\ref{figure_results_main_nuisance}, we show the constraints on the six cosmological and the nine nuisance parameters for the different data combinations. The priors and the central values for the nuisance parameters are given in Table \ref{prior_post_table}. In Table \ref{nuisance_table}, we additionally show the expected values for the nuisance parameters and the measured differences between the expected and obtained values.

\begin{table*}
\caption{Constraints on the nuisance parameters for the analysis combining all 46 power spectra. The results are compared to the expected values from~\cite{Doux2018} for the galaxy bias parameters $b_\mathrm{LOWZ}$ and $b_\mathrm{CMASS}$ and from~\cite{Asgari2021} for the multiplicative shear bias parameters $m_{\gamma i}$.\label{nuisance_table}}
\begin{tabular}{c @{\vline} c @{\vline} cc @{\vline} cc @{\vline} cc}
 \hline
 &  &  &  &  &  &  & \\
 &  & \multicolumn{2}{c|}{Broad Priors for $m_{\gamma i}$ (fiducial)} & \multicolumn{2}{c|}{Tight Priors for $m_{\gamma i}$}  & \multicolumn{2}{c}{Fixed $\sum m_\nu$}\\
 &  &  &  &  &  &  & \\
Nuisance Parameter \quad & \, Expected Value\footnote{Note that the CMB temperature, polarisation and CMB lensing measurements also contain calibration uncertainties~\cite{Aghanim:2019ame}. In this analysis we do not express these uncertainties in terms of the expected values for the multiplicative bias parameters $m_\mathrm{T}$ and $m_\kappa$.} \quad & \quad Obtained Value & $\Delta \sigma$ & \quad Obtained Value & $\Delta \sigma$ & \quad Obtained Value & $\Delta \sigma$\\[0.1cm]
 &  &  &  &  &  &  & \\
\hline
\hline
 &  &  &  &  &  &  & \\
 $m_\mathrm{T}$ & 0.0 & $0.0051^{+0.0027}_{-0.0027}$ & $+1.8 \sigma$ & $0.0018^{+0.0055}_{-0.0055}$ & $+0.3 \sigma$ & $0.0047^{+0.0021}_{-0.0021}$ & $+2.2 \sigma$ \\[0.1cm]
$m_\kappa$ & 0.0 & $0.072^{+0.026}_{-0.026}$ & $+2.7 \sigma$ & $0.067^{+0.041}_{-0.041}$ & $+1.6 \sigma$ &  $0.036^{+0.027}_{-0.027}$ & $+1.3 \sigma$ \\[0.1cm]
$b_\mathrm{LOWZ}$ & $1.837^{+0.033}_{-0.033}$ & $2.24^{+0.14}_{-0.14}$ & $+2.8 \sigma$ & $2.11^{+0.22}_{-0.18}$ & $+1.5 \sigma$ & $1.914^{+0.064}_{-0.053}$ & $+1.2 \sigma$\\[0.1cm]
$b_\mathrm{CMASS}$ & $2.086^{+0.032}_{-0.032}$ & $2.29^{+0.10}_{-0.11}$ & $+1.8 \sigma$ & $2.05^{+0.15}_{-0.15}$ & $-0.3 \sigma$ & $2.026^{+0.05}_{-0.05}$ & $-1.2 \sigma$ \\[0.1cm]
$m_{\gamma 1}$ & $-0.009^{+0.02}_{-0.02}$ & $0.013^{+0.043}_{-0.030}$ & $+0.7 \sigma$ & $-0.009^{+0.021}_{-0.021}$ & $+0.0 \sigma$ & $0.018^{+0.048}_{-0.037}$ & $+0.56 \sigma$ \\[0.1cm]
$m_{\gamma 2}$ & $-0.011^{+0.02}_{-0.02}$ & $-0.003^{+0.033}_{-0.033}$ & $+0.24 \sigma$ & $-0.008^{+0.02}_{-0.02}$ & $+0.12 \sigma$ & $-0.001^{+0.032}_{-0.032}$ & $+0.3 \sigma$\\[0.1cm]
$m_{\gamma 3}$ & $-0.015^{+0.02}_{-0.02}$ & $-0.011^{+0.027}_{-0.039}$ & $+0.1 \sigma$ & $-0.022^{+0.02}_{-0.02}$ & $-0.33 \sigma$ & $-0.014^{+0.018}_{-0.036}$ & $+0.06 \sigma$ \\[0.1cm]
$m_{\gamma 4}$ & $0.002^{+0.02}_{-0.02}$ & $-0.040^{+0.017}_{-0.028}$ & $-1.5 \sigma$ & $-0.006^{+0.02}_{-0.02}$ & $-0.36 \sigma$ & $-0.039^{+0.014}_{-0.029}$ & $-1.4 \sigma$ \\[0.1cm]
$m_{\gamma 5}$ & $0.007^{+0.02}_{-0.02}$ & $-0.032^{+0.023}_{-0.038}$ & $-1.05 \sigma$ & $-0.001^{+0.02}_{-0.02}$ & $-0.36 \sigma$ & $-0.035^{+0.017}_{-0.035}$ & $-1.2 \sigma$ \\[0.1cm]
 \hline
\end{tabular}
\end{table*}

As shown in Figure \ref{figure_results_main_nuisance}, we find that the multiplicative bias parameters for the higher weak lensing shear redshift bins, $m_{\gamma 3}$, $m_{\gamma 4}$ and $m_{\gamma 5}$, are shifted towards negative values, whereas the recovered distributions for $m_{\gamma 1}$, $m_{\gamma 2}$ $m_\mathrm{T}$ and $m_\kappa$ move in the opposite direction towards positive values. This is expected, as the nuisance parameters $m_{\gamma i}$, $m_\mathrm{T}$ and $m_\kappa$ recalibrate the spectra and help to resolve the tension between CMB and weak lensing shear measurements. The observed shifts for the parameters $m_\kappa$ at 2.7$\sigma$ and $m_T$ at 1.8$\sigma$ could be related to the $A_\mathrm{lens}$ anomaly. A shift of $A_\mathrm{lens}\neq1$ at $2.8\sigma$ has been found in~\cite{Aghanim:2018eyx} for \textit{Planck} 2018 temperature and polarisation data.

This is affecting the parameters $m_{\gamma 3}
$, $m_{\gamma 4}$ and $m_{\gamma 5}$, since the weak lensing kernels for the redshift bins 3, 4 and 5 show a more dominant overlap with the CMB lensing kernel, and the two lowest redshift bins have smaller constraining power. The posterior distributions of the two linear galaxy bias parameters $b_\mathrm{LOWZ}$ and $b_\mathrm{CMASS}$ experience are slightly narrowed when combining all auto- and cross-spectra, compared to when using only the auto-spectra. Our results are shifted by $\sim 2.8 \sigma$ for $b_\mathrm{LOWZ}$ and by $\sim 1.8 \sigma$ for $b_\mathrm{CMASS}$ towards higher values compared to the results in~\cite{Doux2018}. 
%%%%%%%%%%%%%%%%%%%%%%%%%%%%%%%%%%%%%%%%%%%%%%%%%%%%%%%%%%%%%%%%%%%%%%%%%%%%%%%%%%%%%%%%%%%%%%%%%%%%%%%%%%%%%%%%%%%%%%%%%%%%%%%%%%%%%%%%%%%%%%%%%%%%%%%%%%%%%%
\subsection{\label{sec:constraints_mnu} Constraints on the Neutrino Mass Sum}

\begin{figure*}[htbp!]
\centering
\includegraphics[width=0.8\paperwidth]{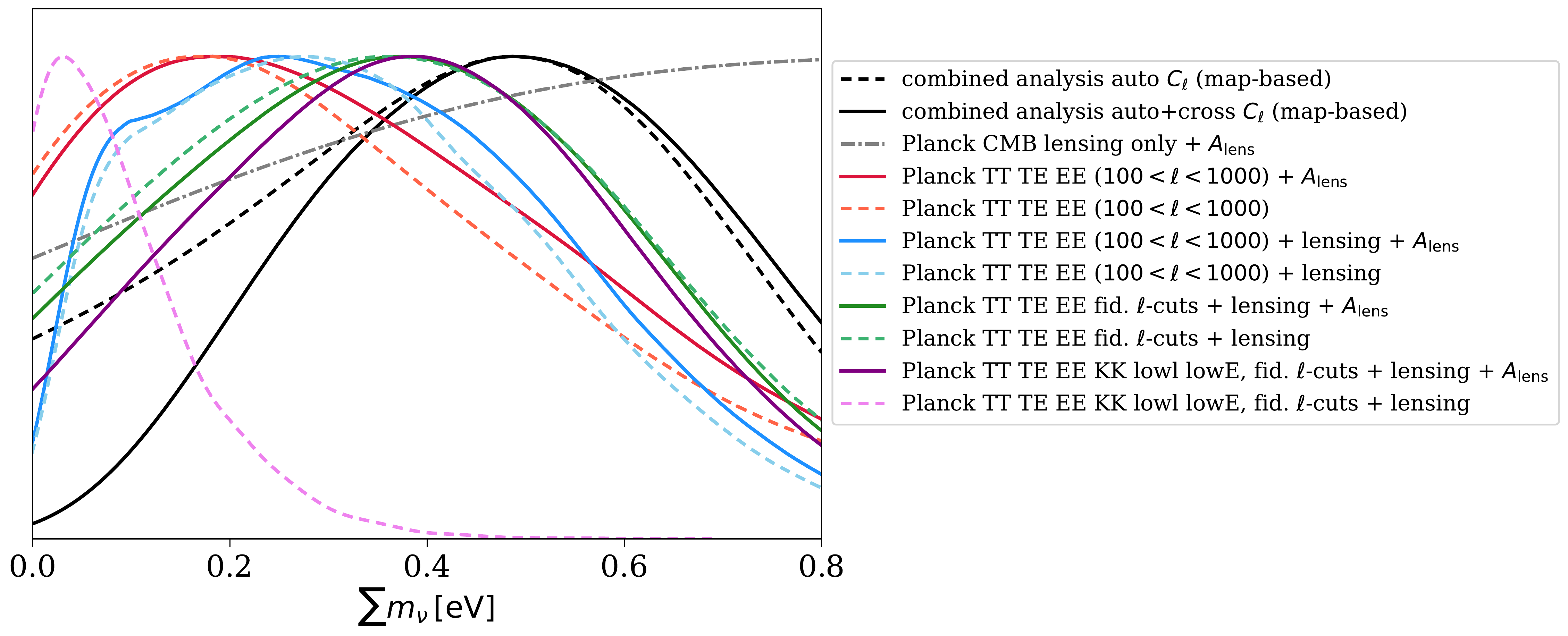}
\caption{One-dimensional posterior distribution for $\sum m_\nu$. The results from the (map-based) combined analysis using auto-spectra only (black dashed) and auto- and cross-spectra (black solid) are compared to results obtained from using \textsc{CosmoMC} with various \textit{Planck} data-configurations. Solid lines include the variation of the parameter $A_\mathrm{lens}$, whereas the dashed lines use $A_\mathrm{lens}=1$. Fid. $\ell$-cuts denote the fiducial multipole range used by the \textit{Planck} collaboration given by $30 \leq \ell \leq 2508$ for TT, $30 \leq \ell \leq 1996$ for TE and EE and $8 \leq \ell \leq 400$ for CMB lensing. \label{figure_results_main_mnu}}
\end{figure*}

In Figure~\ref{figure_results_main_mnu}, we show the one-dimensional posterior distribution of the neutrino mass sum $\sum m_\nu$ for different data combinations, and in Table~\ref{mnutable} the corresponding upper limits for $\sum m_\nu$. We show both our results from our analysis for the combination of all auto- and auto- + cross-correlations, respectively, as well as several runs considering only CMB and CMB lensing data. For the latter, we show the results both for fixing the $A_\mathrm{lens}$ parameter (see Section~\ref{sec:systematics}), as well as for varying it as a free parameter in the analysis. These results have been obtained using the public Monte-Carlo Markov chain package \textsc{CosmoMC}~\cite{Bridle2007}, which uses the Einstein-Boltzmann code~\textsc{CAMB}~\cite{Lewis:1999bs}. Note that for all \textsc{CosmoMC} runs we leave the optical depth to reionization, $\tau_\mathrm{reion}$, to vary as a free parameter, whereas for the runs in our main analysis it is fixed to $\tau_\mathrm{reion}=0.051$. In our runs, we vary instead the CMB calibration parameter $m_T$. This could potentially have a similar effect on the resulting parameter constraints as varying $\tau_\mathrm{reion}$ in the inference, as both $m_T$ and $\tau_\mathrm{reion}$ affect the normalisation of the CMB temperature power spectrum. 

\begin{table*}
\caption{Neutrino mass constraints from different data combinations. All presented limits are upper limits at 95\% CL. \label{mnutable}}
\begin{tabular}{l|c}
 \hline
 \\
 Probes \quad & $\sum m_\nu$ [eV]
 \\
 \hline
 \hline
  $A_\mathrm{lens}=1$ & \\
 \textit{Planck} 2018 TTTEEE + lowl + lowE + lensing & 0.25 \\
 \textit{Planck} 2018 TTTEEE + lensing & 0.80 \\
\textit{Planck} 2018 TTTEEE ($100\leq\ell<1000$) + lensing & 0.69 \\
\textit{Planck} 2018 TTTEEE ($100\leq\ell<1000$) & 0.90 \\
 \hline
 $A_\mathrm{lens}$ free & \\
\textit{Planck} 2018 TTTEEE + lowl + lowE + lensing & 0.76\\
\textit{Planck} 2018 TTTEEE + lensing & 0.79 \\
\textit{Planck} 2018 TTTEEE ($100\leq\ell<1000$) + lensing & 0.71 \\
\textit{Planck} 2018 TTTEEE ($100\leq\ell<1000$) & 0.91 \\
\hline 
$m_T$ and $m_\kappa$ free & \\
All probes auto-correlations & 0.82 \\
All probes auto- + cross-correlations & 0.93 \\
 \hline
\end{tabular}
\end{table*}

The baseline run for this comparison is the combination of the full \textit{Planck} 2018 TT, TE, EE \textit{lowl}, \textit{lowE}, and lensing data set (the pink dashed posterior distribution), resulting in an upper limit of $\sum m_\nu<0.25$ eV (95\% CL). When the $A_\mathrm{lens}$ parameter is a free parameter in the analysis (the violet posterior), the upper limit of the total neutrino mass sum is weakened significantly to $\sum m_\nu<0.76$ eV (95\% CL). This result is consistent with previous findings in the literature~\cite{Aghanim:2018eyx,RoyChoudhury:2019hls,Renzi:2017cbg}. As a next step, we reduce the multipole range for the \textit{Planck} 2018 primary CMB %and CMB lensing 
data and remove the \textit{lowl} and \textit{lowE} part of the likelihood. Then the constraint of $\sum m_\nu$ is weakened to $\sum m_\nu<0.69$ eV (95\% CL), and shifted towards higher values (the blue dashed posterior compared to the pink dashed posterior). By reducing the multipole range, we exclude parts of the CMB data set where the impact of the $A_\mathrm{lens}$ parameter is the most relevant~\cite{Aghanim:2018eyx}. Furthermore, by removing the \textit{lowE} dataset, the constraint of the optical depth, $\tau_\mathrm{reion}$, is weakened, and because of the three-dimensional degeneracy between $\sum m_\nu$, $\tau_\mathrm{reion}$ and $A_s$ then also the neutrino mass constraint~\cite{Allison:2015qca,Calabrese:2016eii}. 

Similarly, we also expect a higher neutrino mass bound in our analysis, compared to the upper limit of $\sum m_\nu<0.24$ eV (95\% CL) found by the official \textit{Planck} 2018 analysis~\cite{Aghanim:2018eyx} for the combination of the full \textit{Planck} primary CMB and CMB lensing data set. This is due to the conservative multipole cuts used in this analysis ($100\leq \ell \leq 1000$ for TTTEEE and $50 \leq \ell \leq 400$ for CMB lensing), as well as to the inclusion of the two multiplicative CMB bias parameters $m_T$ and $m_\kappa$, which are similar to the $A_\mathrm{lens}$ parameter. 

In our combined analysis, we find a high upper bound for the neutrino mass sum, 
\begin{align}
    \sum m_\nu&<0.82 \hspace{0.2em}\text{eV} \hspace{2em}\text{Auto} \nonumber\\
    \sum m_\nu&<0.93 \hspace{0.2em}\text{eV} \hspace{2em}\text{Auto + Cross} \hspace{1em} \text{(95\% CL)}.
\end{align}
Here, `Auto' corresponds to the combination of the auto-correlations of the individual probes and `Auto + Cross' to the combination of all auto- and cross-correlations. 
The corresponding mean and standard deviations are given by
\begin{align}
    \sum m_\nu&= 0.46^{+0.23}_{-0.23} \hspace{0.2em}\text{eV} \hspace{1em}\text{Auto} \nonumber\\
    \sum m_\nu&=0.51^{+0.21}_{-0.24} \hspace{0.2em}\text{eV} \hspace{1em}\text{Auto + Cross} \hspace{0.5em} \text{(68\% CL)}.
\end{align}

These results are in line with previous results in the literature that hint towards higher neutrino mass constraints using late Universe probes (see e.g.~\cite{Battye:2013xqa,Wyman:2013lza,Beutler:2014yhv,Poulin:2018zxs,Muir:2020puy,Lorenz:2021alz}), in particular when combining cosmic microwave background and weak lensing measurements. Furthermore, several low-redshift measurements of $\sum m_\nu$, independent from CMB measurements, allow high neutrino masses~\cite{Palanque-Delabrouille:2019iyz,Ivanov:2019pdj,Ivanov:2021zmi}. Our result is also in the parameter region allowed by terrestrial neutrino experiments~\cite{Aker:2021gma,10.1093/ptep/ptaa104}, and would be consistent with a potential direct neutrino mass detection at the Karlsruhe Tritium Neutrino Experiment (KATRIN)~\cite{Drexlin:2013lha}.

In principle, this result might be affected by the tension between the KiDS and the \textit{Planck} data sets. Previously, the inclusion of massive neutrinos in the $\Lambda$CDM model had been discussed as a potential solution of this tension~\cite{Wyman:2013lza,Battye:2013xqa}, however the current consensus is that massive neutrinos increase the CMB contours along the degeneracy lane in the $\sigma_8$-$\Omega_\mathrm{m}$ plane, but do not decrease the discrepancy between the CMB and weak lensing measurements~\cite{Joudaki:2016kym,DiValentino:2018gcu,Lorenz:2018fzb}. 

Combining all auto-correlations results in a $\sim 2.0$~$\sigma$ constraint of the total neutrino mass sum at a mean value of $\sum m_\nu = 0.46$ eV, which is still consistent with massless neutrinos. Adding cross-correlations improves to a $\sim 2.3$~$\sigma$ constraint at $\sum m_\nu = 0.51$ eV.
%%%%%%%%%%%%%%%%%%%%%%%%%%%%%%%%%%%%%%%%%%%%%%%%%%%%%%%%%%%%%%%%%%%%%%%%%%%%%%%%%%%%%%%%%%%%%%%%%%%%%%%%%%%%%%%%%%%%%%%%%%%%%%%%%%%%%%%%%%%%%%%%%%%%%%%%%%%%%%%%%%%%%%%%%%%%%%%%%%%%%%%%%%
\subsection{Constraints on the Growth of Structure}
\begin{figure}[htbp!]
\centering
\includegraphics[width=\linewidth]{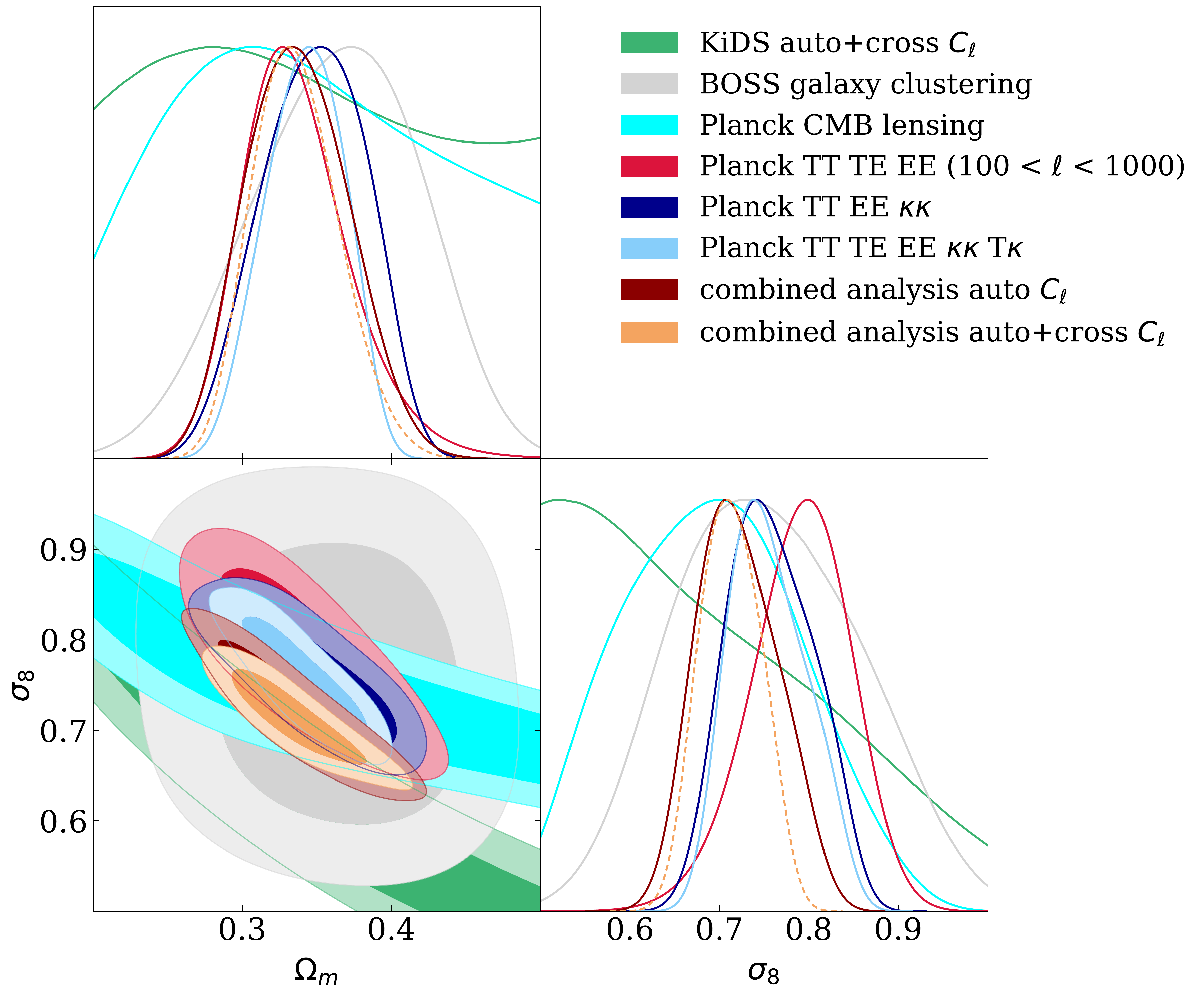}
\caption{Cosmological parameter constraints for $\Omega_m$ and $\sigma_8$ derived from the combined analysis using the 10 auto-spectra (dark red) and using the 46 auto- + cross-spectra (orange), marginalised over all nuisance parameters described in Table \ref{prior_post_table}. The constraints are compared to the CMB-only results using the auto-spectra (dark blue), the auto+cross-spectra (light blue), TT TE EE spectra (red), CMB lensing only (cyan), the weak lensing shear (green) and the galaxy clustering results (grey).  The inner (outer) contours show the 68\% (95\%) confidence limit. \label{figure_results_om_sigma8}}
\end{figure}

\begin{figure}[htbp!]
\centering
\includegraphics[width=\linewidth]{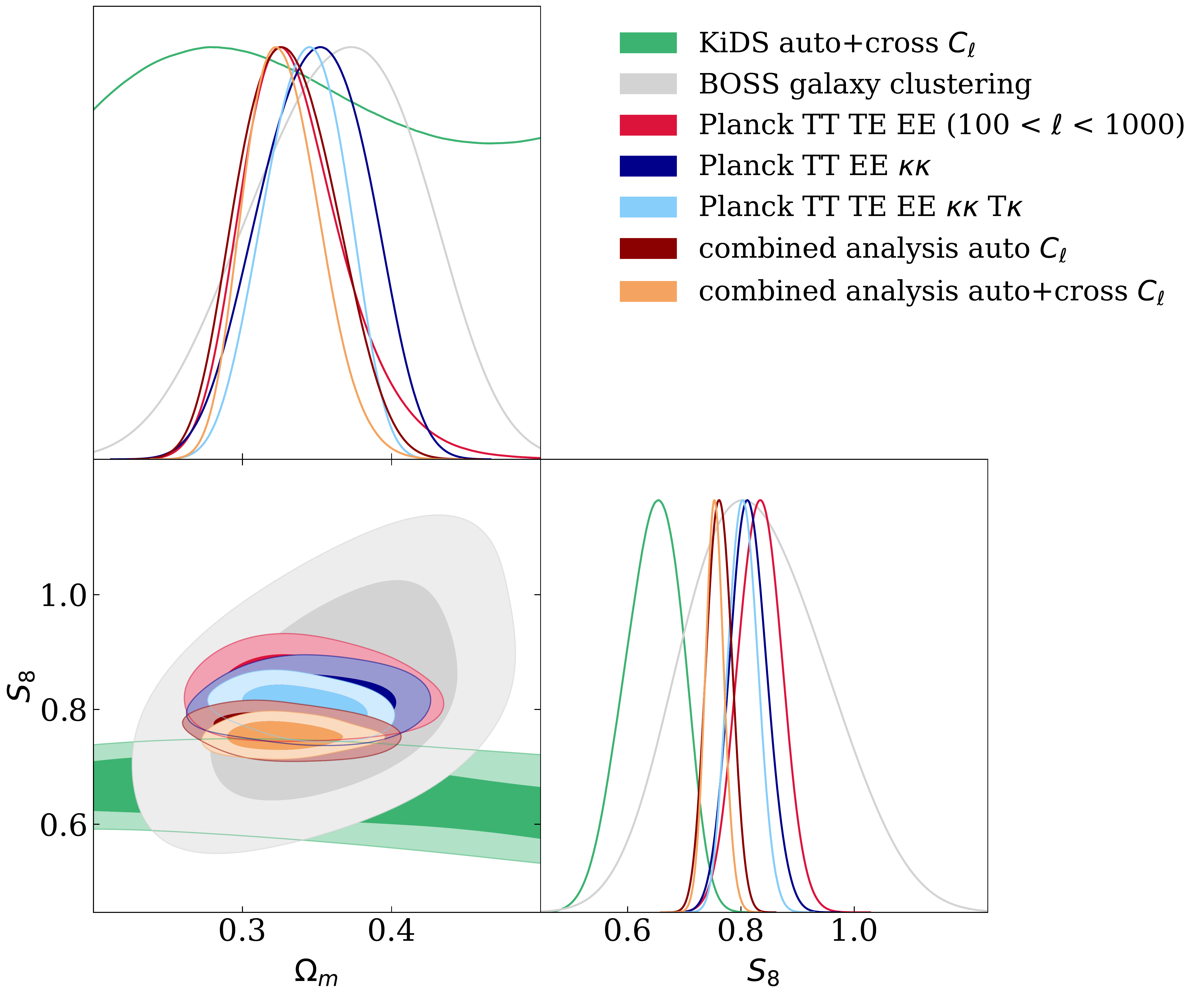}
\caption{Cosmological parameter constraints for $\Omega_m$ and $S_8 = \sigma_8 \sqrt{\Omega_m / 0.3}$ derived from the combined analysis using the 10 auto-spectra (dark red) and using the 46 auto- + cross-spectra (orange), marginalised over all nuisance parameters described in Table \ref{prior_post_table}. The constraints are compared to the CMB-only results using the auto-spectra (dark blue), the auto- + cross-spectra (light blue), the TT TE EE spectra (red), the weak lensing shear (green) and the galaxy clustering results (grey). The inner (outer) contours show the 68\% (95\%) confidence limit. \label{figure_results_om_S8}}
\end{figure}

In Figure \ref{figure_results_om_sigma8}, we show the constraints in the $\Omega_m - \sigma_8$ plane obtained from the combined analysis, compared to various different data combinations. Similarly, Figure \ref{figure_results_om_S8} shows our results in the $\Omega_m - S_8$ plane, using the definition of
$S_8 = \sigma_8 \sqrt{\Omega_m / 0.3}$. In general, we observe a shift towards lower values for $\sigma_8$ or $S_8$ when adding the CMB lensing $\kappa \kappa$ to the primary CMB spectra TT TE EE. This effect is even more pronounced when adding the cross-correlation T$\kappa$. Combining all auto-spectra or all auto- and cross-spectra further shifts the obtained constraints in the same direction.

Our fiducal result for $S_8$ is given by 
\begin{equation}
    S_8 = 0.754^{+0.016}_{-0.016} \hspace{1em} \text{(68\% CL)}. 
\end{equation} 
This result can be compared to the cosmic shear results from the DES Y3 analysis, $S_8 = 0.759^{+0.023}_{-0.025}$~\cite{Amon:2021kas}, the KiDS-1000 Band Power analysis~\cite{Asgari2021}, $S_8 = 0.765^{+0.018}_{-0.024}$, and to a recent result obtained in a combined analysis of current galaxy and quasar clustering, cosmic shear and CMB lensing data ~\cite{Garcia-Garcia:2021unp}, $S_8 = 0.7769 \pm 0.0095$. Our result lies below the \textit{Planck} 2018 TT TE EE+\textit{lowE}+lensing value given by $S_8 = 0.830^{+0.013}_{-0.013}$~\cite{Aghanim:2018eyx}.
\newpage

\subsection{\label{sec:consistency_tests}Additional Tests}

In the following, we describe three tests that we performed to cross-check our results presented in the previous sections: using tight Gaussian priors for the multiplicative bias parameters $m_{\gamma i}$ for cosmic shear, fixing the neutrino mass sum to a low value of $\sum m_\nu=0.06$ eV and fixing all multiplicative bias parameters and the two galaxy bias parameters to their fiducial values depicted in Table~\ref{nuisance_table}. 

\begin{figure*}[htbp!]
\centering
\includegraphics[width=0.8\paperwidth]{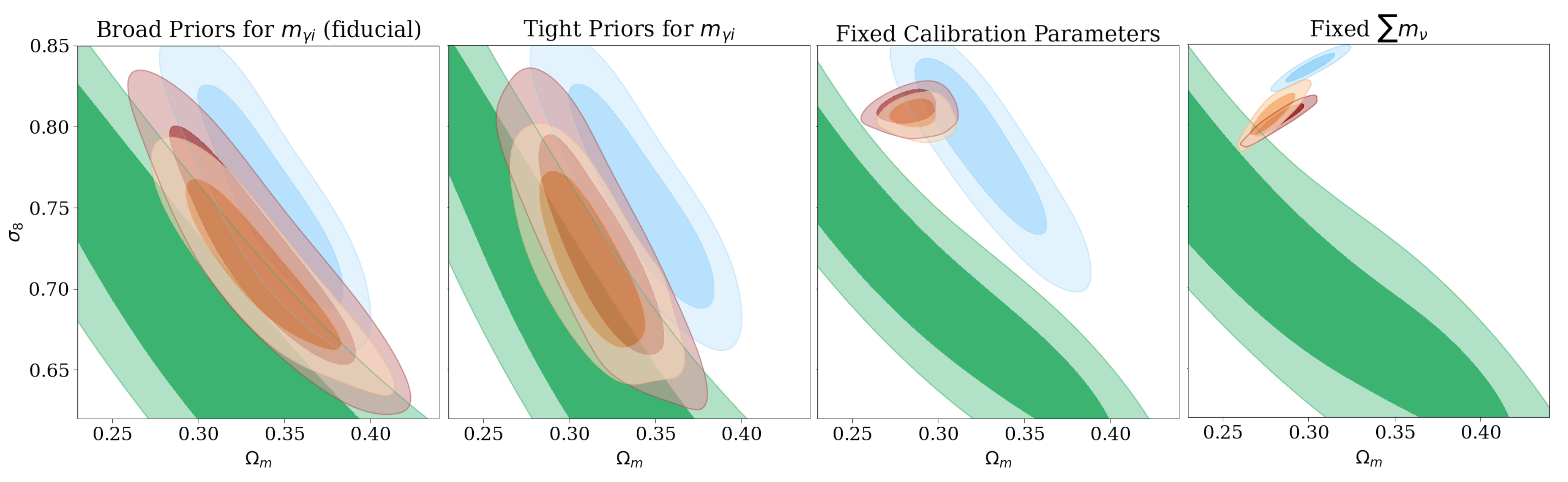}
\caption{Parameter constraints in the $\Omega_m-\sigma_8$ plane for the fiducial run, using broad priors for the multiplicative bias parameters $m_{\gamma i}$ for cosmic shear, and for the three tests: using tight priors for $m_{\gamma i}$, fixing the calibration parameters, and fixing the neutrino mass sum to $\sum m_\nu=0..06$ eV. The results from using the 10 auto-spectra (dark red) and using all 46 auto- + cross-spectra (orange) are compared to the CMB-only results using the auto- + cross-spectra (light blue) and the weak lensing shear results (green). \label{Om_sigma8_consistency}}
\end{figure*}

\begin{figure*}[htbp!]
\centering
\includegraphics[width=0.83\paperwidth]{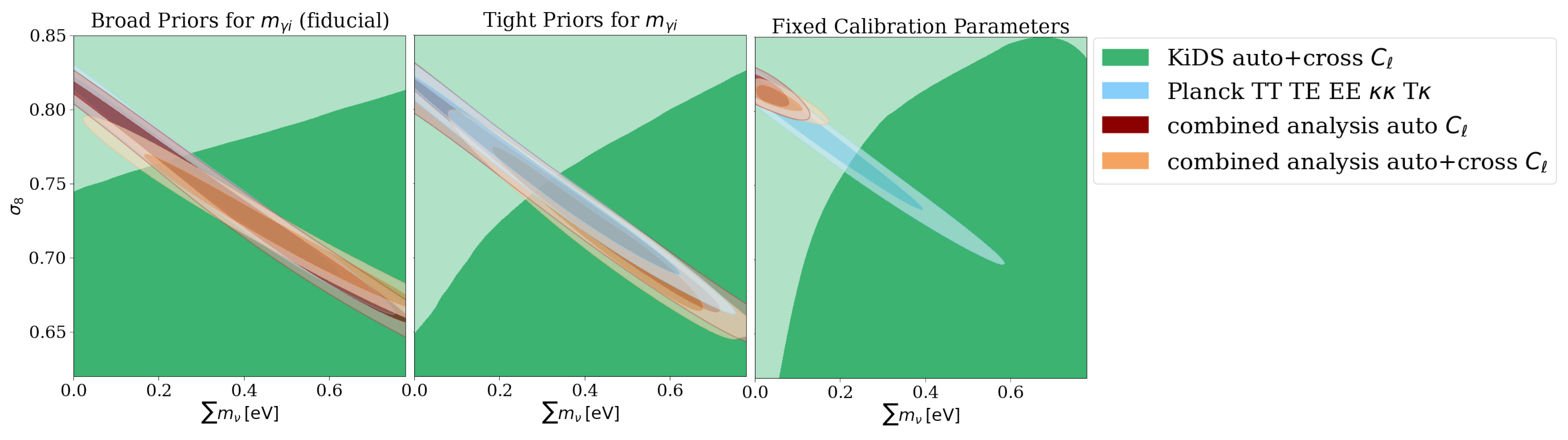}
\caption{Parameter constraints in the $\sum m_\nu-\sigma_8$ plane for the fiducial run, using broad priors for the multiplicative bias parameters $m_{\gamma i}$ for cosmic shear, and for the two tests using tight priors for $m_{\gamma i}$ and for fixing the calibration parameters. The results from using the 10 auto-spectra (dark red) and using all 46 auto+cross-spectra (orange) are compared to the CMB-only results using the auto- + cross-spectra (light blue) and the weak lensing shear results (green).\label{mnu_sigma8_consistency}}
\end{figure*}

In Figure~\ref{Om_sigma8_consistency} we show the obtained parameter constraints in the $\Omega_m-\sigma_8$ plane for the fiducial run, using broad, flat priors for the multiplicative bias parameters $m_{\gamma i}$ for cosmic shear, as well as for the three test runs described above. Similarly, Figure~\ref{mnu_sigma8_consistency} shows the obtained constraints in the $\sum m_\nu - \sigma_8$ plane. The reduced $\chi^2$ values and corresponding p-values for the fiducial and the additional runs are given in Table~\ref{chi2table}. Note that the tests using fixed nuisance parameters (both for the combination of all auto- and all auto- and cross-correlations) and the one with a fixed neutrino mass sum (for the combination of all auto- and cross-correlations) would not satisfy a stronger threshold value of $\alpha=0.1$ for the p-value.

In Table~\ref{nuisance_table}, we show a comparison between the expected and the observed values of the nuisance parameters and the associated shifts for all runs including calibration parameters. 

\begin{enumerate}
    \item \textbf{Tight priors for cosmic shear calibration.} 
In the first test, we use the tighter Gaussian priors for the multiplicative shear bias parameters $m_{\gamma i}$ of the KiDS-1000 analysis~\cite{Asgari2021}, with mean and standard deviation given by the expected values in Table~\ref{nuisance_table}. As can be inferred from Table~\ref{chi2table}, this case results in a slightly worse fit compared to when using broader flat priors for $m_{\gamma i}$ used in our fiducial analysis. The fit is however still acceptable. As expected, choosing these tighter priors leads to lower shifts in the multiplicative shear bias parameters $m_{\gamma i}$. The calibration parameters associated to the CMB data, $m_\mathrm{T}$ and $m_\kappa$, experience lower shifts compared to the shifts for the fiducial case. Moreover, choosing Gaussian priors leads to a slightly lower sum of the neutrino masses than for the fiducial case. 

\item \textbf{Fixed neutrino mass sum.} For the second test, we choose the flat broad priors for $m_{\gamma i}$ (as for the fiducial run), but fix the neutrino mass sum to a low value given by $\sum m_\nu=0.06$ eV. This run results in a worse fit compared to the runs with varied $\sum m_\nu$, resulting in a p-value of 0.09. Furthermore, the constraints for the nuisance parameters $m_{\gamma i}$ are similar as those derived in the fiducal run. Interestingly, we observe a lower value for $m_\kappa$ in this case. As shown in Fig.~\ref{Om_sigma8_consistency}, the $\sigma_8-\Omega_m$ constraints are tighter in this case, which is consistent with the findings of Refs.~\cite{Joudaki:2016kym,Lorenz:2018fzb}.

\item \textbf{Fixed calibration parameters.} For the third test, we fix all calibration parameters to the expected values given in Table~\ref{nuisance_table}. This results in a relatively high reduced $\chi^2$ value (and low p-value) and therefore in a bad fit. We can therefore conclude that the inclusion of a  set of calibration parameters is a possibility to reduce the tension between the different data sets. Based on the obtained bad fit, we caution against the combination of all probes without the inclusion of calibration parameters. 
Moreover, we note that the sum of the neutrino masses is forced to a low value by fixing the nuisance parameters, $\sum m_\nu<0.105$ eV (95\% CL) for the combination of all auto-correlations and $\sum m_\nu<0.140$ eV (95\% CL) for the combination of all auto- and cross-correlations. The relation between $\sum m_\nu$ and the two CMB calibration parameters $m_\kappa$ and $m_T$ can be compared to the relation between $\sum m_\nu$ and the $A_\mathrm{lens}$ parameter, which is discussed in Section~\ref{sec:constraints_mnu}. Fixing $m_\kappa$ and $m_\mathrm{T}$ would correspond to fixing $A_\mathrm{lens}$, which also results in low neutrino mass bounds, as shown in Fig.~\ref{figure_results_main_mnu}. 
\end{enumerate}

We conclude that our fiducial run with varying all nuisance parameters using broad, flat priors, as well as varying the total neutrino mass simultaneously, resulted in the best fit compared to these different test cases. 

\begin{table*}
\caption{Reduced $\chi^2$ and p-values.\label{chi2table}}
\begin{tabular}{ lcccc }
 \hline
 \\
 Probes \quad & \multicolumn{2}{ c }{reduced $\chi^2$} \quad & \multicolumn{2}{ c }{p-value}\\[0.1cm]
  & auto & \quad auto+cross & \quad auto & \quad auto+cross
 \\
 \hline
 \hline
 \\
Flat priors on $m_{\gamma i}$ (fiducial) & 1.25 & \quad 1.7 & \quad 0.2 & \quad 0.12 \\[0.05cm]
Tight priors for $m_{\gamma i}$ & 1.32 & \quad 1.84 & \quad 0.17 & \quad 0.08 \\[0.05cm]
Fixed $\sum m_\nu$ & 1.46 & \quad 1.93 & \quad 0.15 & \quad 0.09 \\[0.05cm]
Fixed nuisance parameters & 1.65 & \quad 2.12 & \quad 0.07 & \quad 0.025 \\[0.05cm]
 \hline
\end{tabular}
\end{table*}
\section{\label{sec:discussion}Conclusions}
In this paper, we presented a joint analysis combining the latest CMB measurements from \textit{Planck} with spectroscopic galaxy samples from  BOSS DR12 and the latest KiDS-1000 weak lensing shear data release. We used the CMB temperature and polarisation anisotropies and CMB lensing data from the \textit{Planck} 2018 release. Concerning late-Universe probes, we included the two BOSS galaxy samples LOWZ and CMASS, as well as weak lensing shear data from the five tomographic bins from the latest KiDS-1000 data release. We performed our analysis on a map level, i.e., we extracted the data and computed pixelised maps of all considered cosmological probes in order to include the auto- and cross-correlations between the probes in our analysis.

We described the framework for our multi-probe likelihood analysis based on auto- and cross-spherical harmonic power spectra. Our analysis is based on a single-parameter extension of a flat $\Lambda$CDM cosmological model, including the total neutrino mass sum $\sum m_\nu$ for degenerate neutrinos. The analytical predictions used for the parameter inference are based on the pseudo-$C_\ell$ method, which consists of transforming the analytical full-sky power spectra using mode-coupling matrices, taking into account the effect of the finite sky covered by the surveys. 

The analytical power spectra have been computed using a hybrid approach: The CMB spectra TT, TE, EE and $\kappa\kappa$ are predicted using the code \textsc{Class}~\cite{Lesgourgues2011, Blas2011}, whereas all remaining LSS spectra and cross-correlations with the CMB fields are computed using \textsc{PyCosmo}~\cite{pycosmo2018,Tarsitano2020}. We use the lightcone generation code \textsc{UFalcon}~\cite{Sgier2019,Sgier2021} to construct full-sky maps for all LSS probes considered (including ISW maps) by post-processing 20 \textsc{PKDGrav3}~\cite{Stadel2001} $N$-Body simulation realisations. A shell randomisation procedure is subsequently used to increase the number of quasi-independent realisations up to 1000 to obtain a well-converged estimate of the multi-probe covariance matrix used for the statistical inference.

We thus performed a $13\times2$-point multi-probe likelihood analysis based on spherical harmonic power spectra by constraining 6 cosmological parameters and the probe-specific calibration parameters. The latter are given by multiplicative bias parameters for the probes T, $\kappa$ and $\gamma$ and linear galaxy bias parameters for the two galaxy samples. The variation of these calibration parameters allows for a self-calibration of the different auto- and cross-spectra included in the combined analysis and can thus give a possible interpretation of the tensions between various data sets.

First, we used the data from the individual surveys alone, and found good agreement between our constraints and the results reported by the experimental teams for each probe. Our main analysis consisted of varying six cosmological and nine calibration parameters simultaneously for the 10 auto- and additional 36 cross-power spectra using conservative multipole cuts. We observe that the constraining power from our combined analysis is dominated by the CMB spectra and that using all 46 auto- and cross-spectra for the analysis further tightens the obtained constraints. Compared to the CMB-only parameter constraints, we observed a shift of our combined analysis constraints towards the cosmic shear only results. We obtain relatively good fits when combining all auto- or all auto- and cross-spectra in our analysis. The corresponding reduced-$\chi^2$ statistics (and p-values) are given by 1.25 (p=0.20) and 1.7 (p=0.12) for all probes auto- and all probes auto- and cross-spectra, respectively. These good fits are due in part to the inclusion of nine calibration parameters in our joint analysis.

We further examined the impact of combining the different data sets on the constraints for the sum of the neutrino mass. In our multi-probe analysis, we find a $2.0\sigma$ constraint of $\sum m_\nu=0.46^{+0.23}_{-0.23}$ eV (68\% CL) for using the auto-correlations and 
a $2.3\sigma$ constraint of $\sum m_\nu=0.51^{+0.21}_{-0.24}$ eV (68\% CL) when additionally including the cross-correlations. 
These results are compatible with current particle physics limits. We compared our findings to the results obtained for different \textit{Planck} 2018 data-configurations (shown in Fig.~\ref{figure_results_main_mnu}). 

We further performed three tests by using tight Gaussian priors for the multiplicative bias parameters $m_{\gamma i}$ for cosmic shear, by fixing the neutrino mass sum to a low value, and by fixing all the calibrating nuisance parameters to their expected values. The resulting reduced $\chi^2$ and p-values indicate worse fits for all these tests compared to our fiducial run, with the worst fit for the case when all calibration parameters are fixed. Moreover, fixing the nuisance parameters $m_\kappa$ associated to CMB lensing results in a low value for the neutrino mass sum. 
Our results highlight the interplay between the different cosmological and calibration parameters and its impact on the $S_8$-tension and the resulting constraints on the neutrino mass sum.

The work presented in this paper can be compared to other map-based joint analyses using spherical harmonic power spectra, such as the combination of CMB data, photometric data from SDSS and DES and geometric probes in~\cite{Nicola2016, Nicola2017} in the context of a flat $\Lambda$CDM model. In addition, our results for $\sum m_\nu$ are consistent with other studies showing that combining late-Universe with CMB data can lead to higher upper limits for the neutrino mass sum, compared to when using CMB data alone~\cite{Battye:2013xqa,Wyman:2013lza,Beutler:2014yhv,Poulin:2018zxs,Muir:2020puy}.

Our results demonstrate the power of multi-probe cosmological analyses including a large number of statistically correlated data sets. The inclusion of the cross-correlations between these data sets proves particularly fruitful in combination with a suitable set of calibration parameters. Varying these parameters in the inference process allows for a self-calibration of the used correlations and can help to alleviate possible tensions between the data sets. 

 Our map-based analysis has the potential for several extensions. In particular, choosing less conservative scale cuts for the spherical harmonic power spectra and incorporating more data sets could further improve the constraints on cosmological parameter. This would require a non-Limber integration~\cite{Fang:2019xat,Schoneberg:2018fis} in order to correctly resolve large scales and a treatment of baryonic effects on small angular scales~\cite{Mead2015,Schneider:2019snl,Chisari:2019tus}. 

As highlighted in this work, testing for possible extensions to the $\Lambda$CDM model, such as including neutrino masses, within such a holistic analysis proves to be very promising. Future surveys are expected to push the boundaries in terms of precision and will cover larger overlapping regions of the sky. Combining them using the integrated approach presented in this work and probing a wider range of angular scales could lead to a significant increase in constraining power and possibly confirm the necessity for new physics.

\begin{acknowledgments}
We thank Cyrille Doux for the help concerning the construction of the BOSS completeness maps. Furthermore, we thank Uwe Schmitt for his help with the computing implementation, and Devin Crichton and Beatrice Moser for useful discussions. We thank Andrina Nicola and Adam Amara for early discussions about multi-probe analyses. The \textsc{PKDgrav3} simulations have been run on the Piz Daint supercomputer (CSCS~\footnote{\url{https://www.cscs.ch/}}, Switzerland). Based on observations made with ESO Telescopes at the La Silla Paranal Observatory under programme IDs 177.A-3016, 177.A-3017, 177.A-3018 and 179.A-2004, and on data products produced by the KiDS consortium. The KiDS production team acknowledges support from: Deutsche Forschungsgemeinschaft, ERC, NOVA and NWO-M grants; Target; the University of Padova, and the University Federico II (Naples).

Funding for SDSS-III has been provided by the Alfred P. Sloan Foundation, the Participating Institutions, the National Science Foundation, and the U.S. Department of Energy Office of Science. The SDSS-III web site is \url{http://www.sdss3.org/}.

SDSS-III is managed by the Astrophysical Research Consortium for the Participating Institutions of the SDSS-III Collaboration including the University of Arizona, the Brazilian Participation Group, Brookhaven National Laboratory, Carnegie Mellon University, University of Florida, the French Participation Group, the German Participation Group, Harvard University, the Instituto de Astrofisica de Canarias, the Michigan State/Notre Dame/JINA Participation Group, Johns Hopkins University, Lawrence Berkeley National Laboratory, Max Planck Institute for Astrophysics, Max Planck Institute for Extraterrestrial Physics, New Mexico State University, New York University, Ohio State University, Pennsylvania State University, University of Portsmouth, Princeton University, the Spanish Participation Group, University of Tokyo, University of Utah,Vanderbilt University, University of Virginia, University of Washington, and Yale University.

Based on observations obtained with Planck (\url{http://www.esa.int/Planck}), an ESA science mission with instruments and contributions directly funded by ESA Member States, NASA, and Canada.
\end{acknowledgments}

\appendix
\section{\label{sec:lightcones}Projected and Integrated Lightcones}
The lightcone generation code \textsc{UFalcon} generally adopts the Born approximation to construct the CMB lensing and weak lensing convergence lightcones, which holds when the change in comoving separation between deflected light rays is small compared to the comoving separation between undeflected lightrays. The expression for the convergence value of a given pixel is then given by \cite{teyssier2009}, \cite{Sgier2019}, \cite{Sgier2021}:
\begin{equation}
\kappa (\theta_\mathrm{pix}) \approx \frac{3}{2} \Omega_m \sum_b W_b \left( \frac{H_0}{c} \right)^3 \frac{N_\mathrm{pix}}{4 \pi} \frac{V_\mathrm{sim}}{N_\mathrm{p}} \frac{n_p (\theta_\mathrm{pix}, \Delta \chi_b)}{\mathcal{D}^2(z_b)} \, ,
\label{kappaproj}
\end{equation}
where $n_p$ is the number of particles in shell $b$ with thickness $\Delta \chi_b$ and $N_\mathrm{pix}$ is the total number of pixels on the sky. Note that the above equation uses the dimensionless comoving distance denoted by $\mathcal{D} (z) = (H_0 / c) \chi(z)$ and the sum runs over all timesteps with $N$-Body simulation output.

For an arbitrary redshift distribution $n(z)$, the weak lensing convergence weights are given by
\begin{eqnarray*}
W_b^{n(z)} &=& \left( \int_{\Delta z_b} \frac{\mathrm{d} z}{E(z)} \int_z^{z_s} \mathrm{d} z' n(z') \frac{\mathcal{D}(z) \mathcal{D}(z, z')}{\mathcal{D}(z')} \frac{1}{a(z)} \right) \\
&\times& \left( \int_{\Delta z_b} \frac{\mathrm{d} z}{E(z)} \int_{z_0}^{z_s} \mathrm{d} z' n(z')\right)^{-1} \, ,
\label{wnz}
\end{eqnarray*}
with $\mathcal{D}(z, z') = \mathcal{D}(z') - \mathcal{D}(z)$. The CMB lensing convergence weights are written as
\begin{eqnarray*}
W_b^{z_s} &=& \left( \int_{\Delta z_b} \frac{\mathrm{d} z}{ E(z)} \frac{\mathcal{D}(z) \mathcal{D}(z, z_s)}{\mathcal{D}(z_s)} \frac{1}{a(z)} \right) \\
&\times& \left( \int_{\Delta z_b} \frac{\mathrm{d} z}{ E(z)} \right)^{-1} \, ,
\end{eqnarray*}
with a single source located at $z_s = z_\ast$. The pixelised expression for the galaxy overdensity is given by
\begin{equation}
\delta_g ( \theta_\mathrm{pix}) \approx \sum_b W_b^{\delta_g} \left( \frac{H_0}{c} \right)^2 \frac{N_\mathrm{pix}}{4 \pi} \frac{V_\mathrm{sim}}{N_p} \frac{n_p (\theta_\mathrm{pix}, \Delta \chi_b)}{\mathcal{D}^2(z_b)} \, ,
\end{equation}
with the associated weights
\begin{equation}
W_b^{\delta_g} = \left( \int_{\Delta z_b} \frac{\mathrm{d}z}{E(z)} H(z) b(z) n(z) \right) \cdot \left( \int_{\Delta z_b} \frac{\mathrm{d}z}{E(z)} \right)^{-1} \, .
\end{equation}
The expression for the temperature anisotropies from the linear ISW effect is given as a sum over discrete steps in comoving radial distance
\begin{equation}
\Delta T_\mathrm{ISW} (\theta_\mathrm{pix}) = \frac{2}{c^2} \sum_{b,\delta z} \dot{\Phi}(\chi \hat{n}_\mathrm{pix}, z)\, a\, \Delta \chi \, ,
\label{iswsum}
\end{equation}
where the sum runs over finer redshift steps of $\delta z = 0.01$ within each shell $\Delta z_b$ related to the timesteps of the simulation. The time derivative of the gravitational potential is directly determined by the overdensity field through
\begin{equation}
\dot{\Phi} (\vec{k}, t) = \frac{3}{2} \left(\frac{H_0}{k}\right)^2 \Omega_m \frac{\dot{a}}{a^2} \delta(\vec{k}, t) [1 - \beta(t)] \, ,
\label{psinum}
\end{equation}
where $\beta(t) \equiv d\, \mathrm{ln}\, D(t) / d\, \mathrm{ln}\, a$ is the linear growth rate. \textsc{UFalcon} constructs the quantity $\dot{\Phi}$ as follows:
\begin{enumerate}
\item Cloud-in-cell mass assignment scheme (CIC) to obtain the overdensity field $\delta (\vec{x})$ on a 3D cubic grid with $1024^3$ cells.
\item Fast Fourier transform to obtain $\dot{\Phi}(\vec{k})$ in Fourier space using Eq. (\ref{psinum}).
\item Inverse Fourier transform to obtain $\dot{\Phi}(\vec{x})$ and interpolate linearly.
\end{enumerate}

\section{\label{sec:consistency} Parameter Constraints for Individual Surveys}
\begin{figure*}[htbp!]
\centering
\includegraphics[width=0.75\paperwidth]{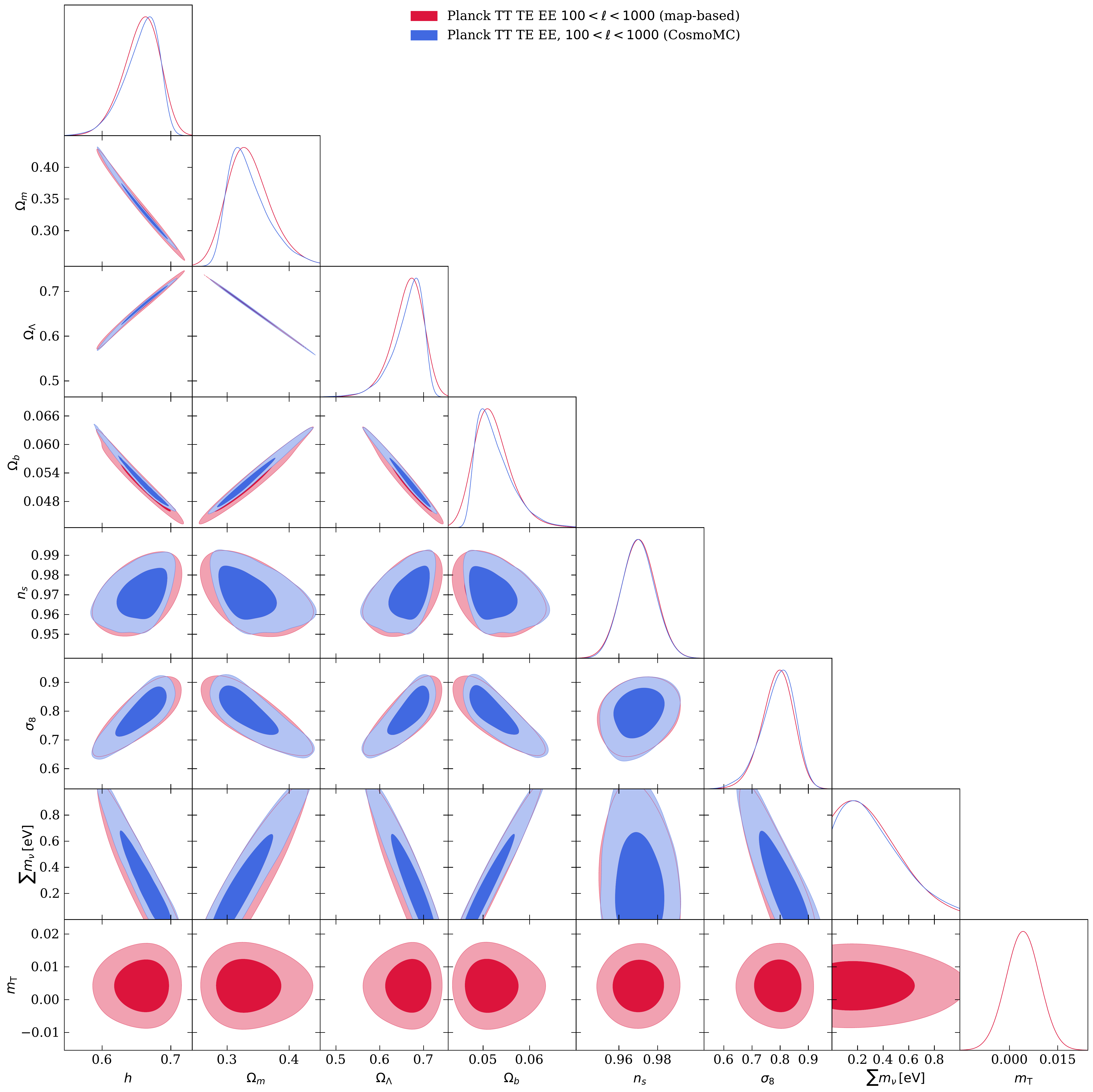}
\caption{Parameter constraints based on CMB temperature and polarisation data (TT, TE, TT) obtained from our map-based analysis (blue) and from the \textsc{CosmoMC} analysis for a multipole range of $100 \leq \ell \leq 1000$. The inner (outer) contours show the 68\% (95\%) confidence limit.}
\label{figure_planck_consistency_TTTEEE}
\end{figure*}

\begin{figure*}[htbp!]
\centering
\includegraphics[width=0.8\paperwidth]{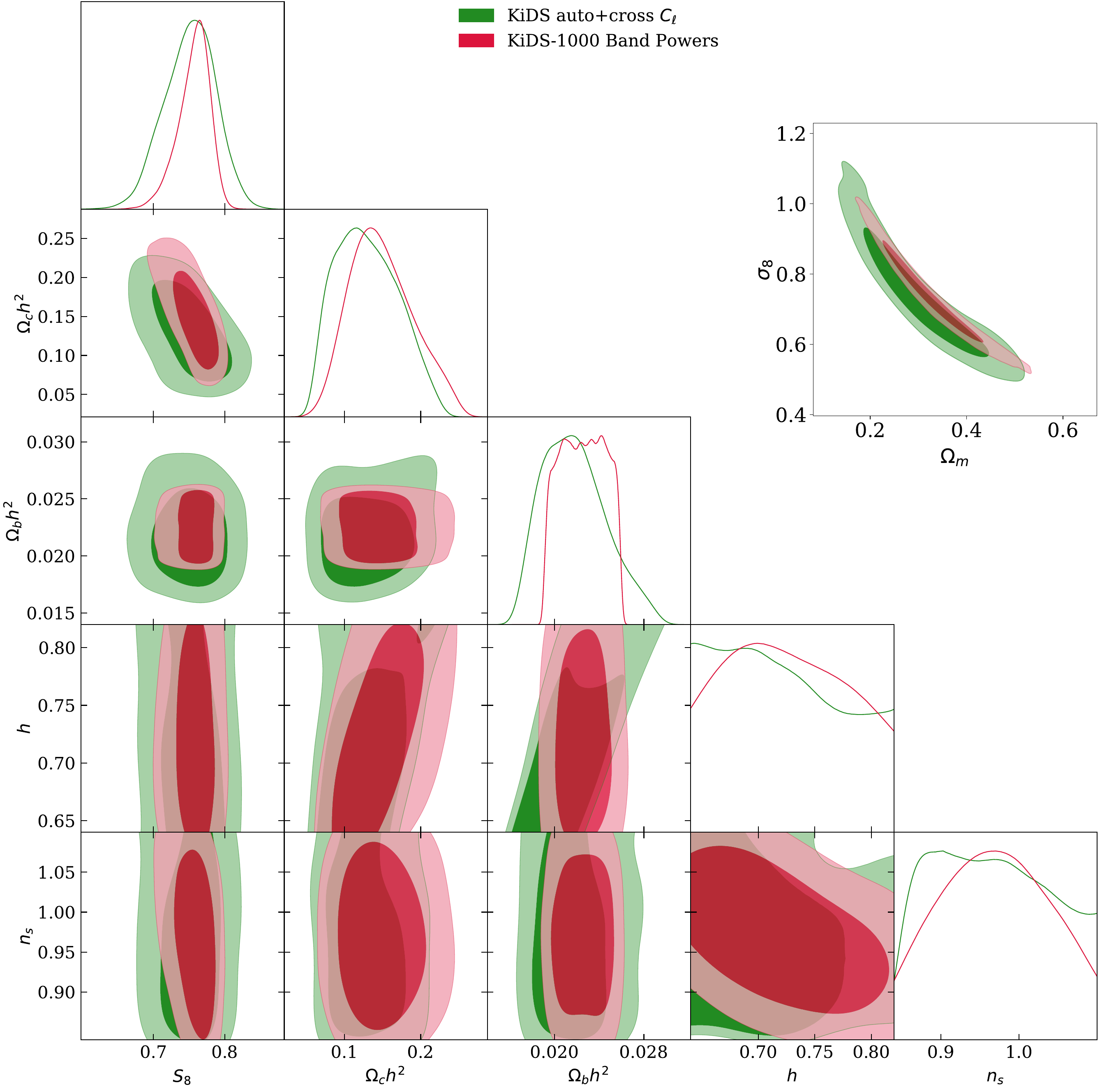}
\caption{Parameter constraints based on the KiDS-1000 weak lensing gold sample using the 13 auto- and cross-spherical harmonic power spectra for the multipole range $100 \leq \ell \leq 1000$ (green) and from the official band power analysis for the multipole range $100 \leq \ell \leq 1500$ (red). The inner (outer) contours show the 68\% (95\%) confidence limit.\label{figure_kids_consistency}}
\end{figure*}

For our CMB data analysis we use a multipole range between $100 \leq \ell \leq 1000$ for temperature and polarisation and $50 \leq \ell \leq 400$ for CMB lensing convergence. We can therefore not directly compare our findings to the \textit{Planck} 2018 "TT, TE, EE + lensing" baseline cosmological results, since they use the multipole range $30 \leq \ell \leq 2508$ for TT, $30 \leq \ell \leq 1996$ for TE and EE and $8 \leq \ell \leq 400$ for CMB lensing. Furthermore, as discussed in \ref{sec:systematics}, we use a multiplicative bias parameter $m_\kappa$ for the CMB lensing convergence map instead of using a scaling of the lensing spectrum by $A_\mathrm{lens}$ as done in the \textit{Planck} 2018 analysis. We therefore use the Markov Chain Monte Carlo package \textsc{CosmoMC} \cite{PhysRevD.66.103511}, which is based on the publicly available Einstein-Boltzmann code \textsc{CAMB} \cite{Lewis_2000}. We run \textsc{CosmoMC} by using the "TT, TE, EE + lensing" data sets and adapt the multipole range for the CMB temperature and polarisation to match our $\ell$-cuts. In Figure \ref{figure_planck_consistency_TTTEEE} we show the parameter constraints obtained from our map-based analysis and from using \textsc{CosmoMC} for the multipole range $100 \leq \ell \leq 1000$. Our results show good overall agreement with the \textsc{CosmoMC} results for all samples parameters. We attribute the differences between the two set of contours to using a different $\ell$-binnig scheme (our analysis uses logarithmic bins) or to using a different covariance matrix estimation.
We further compute parameter constraints using the KiDS-1000 weak lensing shear data for a $\Lambda$CDM model by varying 5 cosmological parameters $\{h, \Omega_\mathrm{m}, \Omega_\mathrm{b}, n_s, \sigma_8\}$ and 5 multiplicative bias parameters $\{m_{\gamma 1}\, , m_{\gamma 2}\, , m_{\gamma 3}\, , m_{\gamma 4}\, , m_{\gamma 5}\}$, where the priors are given in Table \ref{prior_post_table}. We further set the IA amplitude to the best fit value $A_\mathrm{IA} = 0.973$ obtained from the KiDS-1000 Band Power analysis done in \cite{Asgari2021}. 
Figure \ref{figure_kids_consistency} shows our results when using the 13 auto- and cross-spherical harmonic power spectra between the five tomographic bins and the parameter constraints from the band power analysis performed in \cite{Asgari2021}. The parameter constraints obtained from our analysis have significantly larger confidence regions that the official band power results, which we mainly attribute to the larger multipole range $100 \leq \ell \leq 1500$ used by the latter.
\begin{figure*}[htbp!]
\centering
\includegraphics[width=0.9\paperwidth]{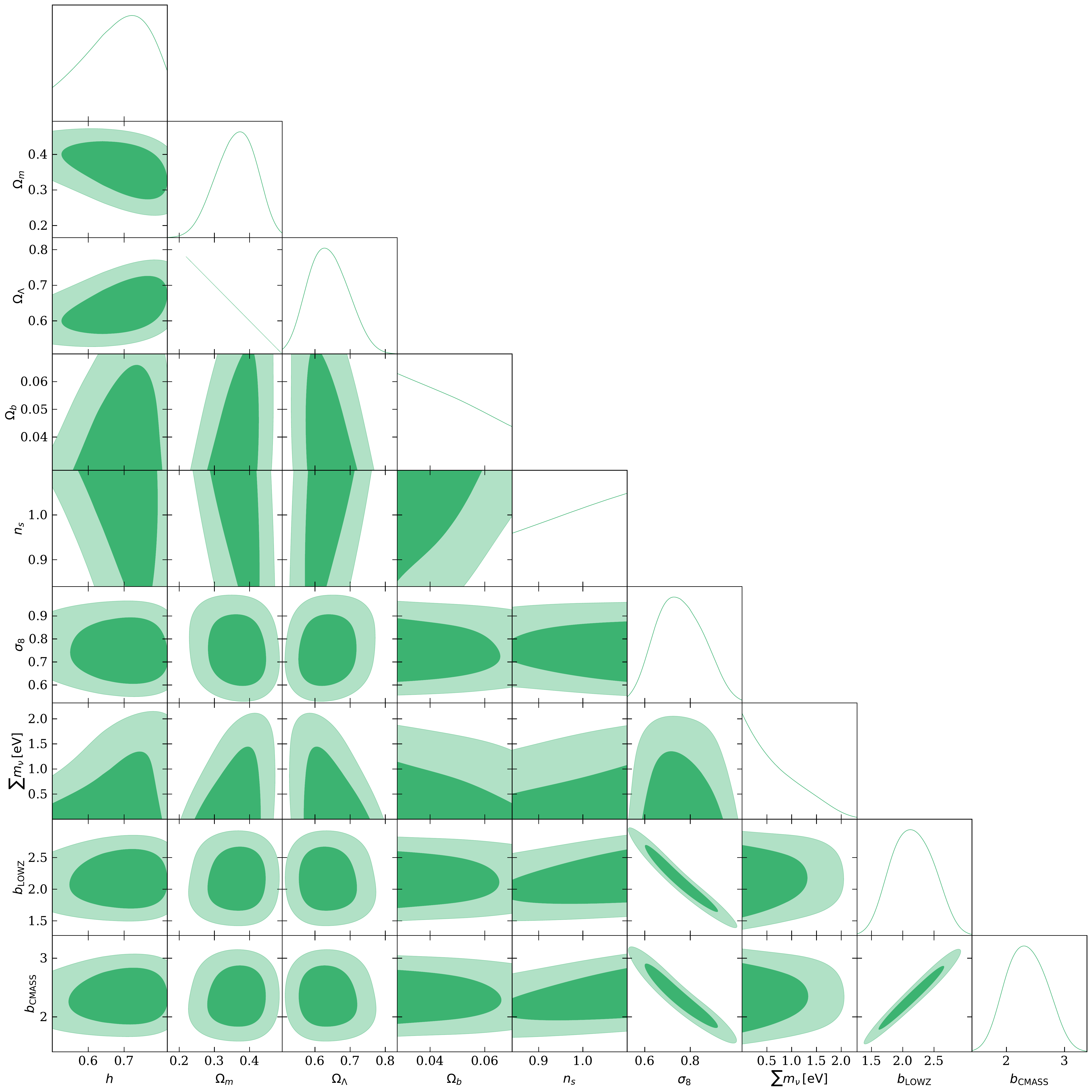}
\caption{Parameter constraints based on the two BOSS DR12 galaxy samples LOWZ and CMASS using the 2 auto-spherical harmonic power spectra for the multipole range $50 \leq \ell \leq 200$ for a $\nu \Lambda$CDM model. The inner (outer) contours show the 68\% (95\%) confidence limit.}
\label{figure_BOSS_consistency}
\end{figure*}

\section{\label{sec:wig_mat_check}Compatibility of the Mode Coupling Matrix and the Limber Approximation}
\begin{figure}[htbp!]
\centering
\includegraphics[width=\linewidth]{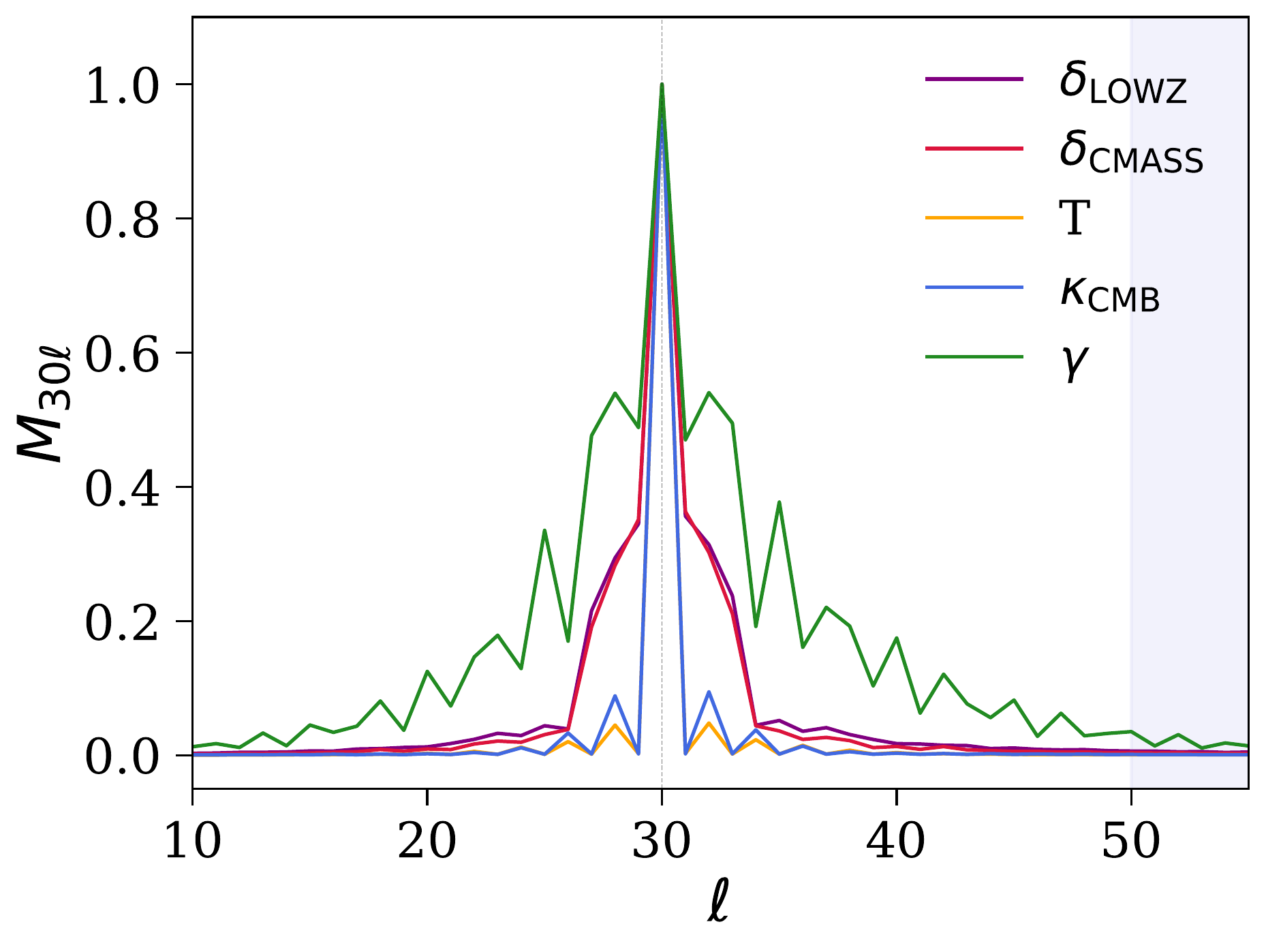}
\caption{Mode coupling matrix normalised to one as a function of $\ell$ with one angular scale fixed to $\ell' = 30$, where we assume the Limber approximation to fail. The lines correspond to the survey masks of the probes $\delta_\mathrm{LOWZ}$ (purple), $\delta_\mathrm{CMASS}$ (red), T (orange), $\kappa_\mathrm{CMB}$ (blue) and $\gamma$ (green). The blue shaded area starts at $\ell_\mathrm{min} = 50$, which is the lower angular scale cut used for all spectra in our analysis. \label{wig_mat_check}}
\end{figure}

As shown in Table \ref{cl_table}, we chose a lower cut in angular scales of $\ell_\mathrm{min} = 50$ for all spectra considered. In principle, the application of the mode-coupling matrix to the Limber approximated full-sky spectra could lead to mixing of miscalculated modes $\ell < 30$ with modes associated to smaller scales, and therefore introducing errors in our pseudo-$C_\ell$ calculation. In Fig. \ref{wig_mat_check} we show the normalised mode-coupling matrices $M_{\ell' \ell}$ for the different survey masks and fix $\ell' = 30$, which corresponds approximately to the maximum scale at which the Limber approximation fails. We see that for the masks corresponding to LOWZ, CMASS, CMB temperature and CMB lensing (the CMB polarisation mask is very similar to the CMB temperature one), the mode-coupling matrix only mixes modes within $\pm 5$ around $\ell' = 30$. For the smaller KiDS-1000 survey mask, the modes within $\sim \pm 20$ are coupled around $\ell' = 30$. We therefore conclude that using a lower cut of $\ell_\mathrm{min} = 50$ for all spectra in our analysis does not bias our calculation of the pseudo-$C_\ell$'s. The coupling matrices corresponding to the combined masks exhibit a very similar behavior and never mix modes within more than $\sim \pm 20$ around $\ell' = 30$.
\clearpage
\bibliography{bibliography}% Produces the bibliography via BibTeX.

\end{document}